\documentclass[10pt,journal,compsoc]{IEEEtran}

\usepackage{cite}
\usepackage{amsmath,amssymb,amsfonts}
\usepackage[linesnumbered,ruled,vlined]{algorithm2e}
\usepackage[noend]{algpseudocode}
\usepackage{graphicx}
\usepackage{textcomp}
\usepackage{xcolor}
\usepackage[section]{placeins}
\usepackage{float}
\usepackage{epstopdf}
\usepackage{subfigure}
\usepackage{mathrsfs}
\usepackage{enumitem}

\usepackage{times}
\usepackage{booktabs} 
\usepackage{bbm}
\usepackage{graphicx}
\usepackage{epstopdf}
\usepackage{booktabs}
\usepackage{url}
\usepackage{multirow}

\pdfoutput=1

\newtheorem{defn}{Definition}
\newenvironment{proof}{\quad{\it Proof:}}{\hfill $\square$\par}

\newtheorem{Lemma}{Lemma}
\newtheorem{example}{Example}
\newtheorem{theo}{Theorem}

\newcommand{\kw}[1]{{\ensuremath {\mathsf{#1}}}\xspace}

\newcommand{\stitle}[1]{\vspace{0.5ex} \noindent{\bf #1}}
\long\def\comment#1{}
\newcommand{\eop}{\hspace*{\fill}\mbox{$\Box$}}

\newcommand{\etal}{\emph{et~al.}\xspace}

\newcommand\figref[1]{Fig.~\ref{#1}}
\newcommand\lamref[1]{Lemma~\ref{#1}}

\newcommand\tabref[1]{Table~\ref{#1}}

\newcommand\secref[1]{Section~\ref{#1}}

\newcommand{\heuristic}{\kw{CalHeurOrd}}
\newcommand{\wforder}{\kw{CalColorOD}}
\newcommand{\colorful}{\kw{ColorfulCore}}
\newcommand{\baseweak}{\kw{BaseWeak}}
\newcommand{\basestrong}{\kw{BaseStrong}}

\newcommand{\baserelative}{\kw{BaseRelative}}

\newcommand{\weak}{\kw{WFCEnum}}
\newcommand{\strong}{\kw{SFCEnum}}
\newcommand{\backtrack}{\kw{BackTrack}}
\newcommand{\cnt}{\kw{cnt}}
\newcommand{\group}{\kw{Group}}

\newcommand{\connectedcpn}{\kw{ConnectedGraph}}

\newcommand{\fairness}{\kw{FairnessCore}}
\newcommand{\strongbacktrack}{\kw{StrongBackTrack}}
\newcommand{\checkmaximality}{\kw{IsMaximal}}

\newcommand{\relativeweak}{\kw{RFCRefineEnum}}
\newcommand{\relativestrong}{\kw{RFCAlterEnum}}
\newcommand{\rstrongenum}{\kw{DeepRFCAlter}}
\newcommand{\rweakenum}{\kw{DeepRFCRefine}}

\newcommand{\enhancedcolordeg}{\kw{EnhancedColCal}}
\newcommand{\enhancedcolor}{\kw{EnhancedColorCore}}

\newcommand{\fairdeg}{\kw{FairDegCal}}

\newcommand{\agroupone}{\kw{OA1Group}}
\newcommand{\agrouptwo}{\kw{OA2Group}}
\newcommand{\agroupmix}{\kw{MixGroup}}
\newcommand{\agroupmixs}{\kw{MixGroups}}

\newcommand{\colororder}{\kw{ColorOD}}
\newcommand{\bfsorder}{\kw{BfsOD}}
\newcommand{\idorder}{\kw{VidOD}}
\newcommand{\frorder}{\kw{FairOD}}
\newcommand{\heurorder}{\kw{HeurOD}}

\newcommand{\wiki}{\kw{WikiTalk}}
\newcommand{\flixster}{\kw{Flixster}}
\newcommand{\themarker}{\kw{Themarker}}
\newcommand{\webwiki}{\kw{Wikipedia}}
\newcommand{\slashdot}{\kw{Slashdot}}
\newcommand{\dblp}{\kw{DBLP}}

\newcommand{\dbcase}{\kw{DBCS}}

\begin{document}

\title{Fairness-aware Maximal Clique in Large Graphs: Concepts and Algorithms}

\author{Qi~Zhang, Rong-Hua~Li, Minjia~Pan, Yongheng~Dai,  Qun~Tian, and~Guoren~Wang
\IEEEcompsocitemizethanks{
\IEEEcompsocthanksitem Q. Zhang, R.-H. Li,  M. Pan, and G. Wang are with the School of Computer Science and Technology, Beijing Institute of Technology, Beijing 100081, China.\protect\\
E-mail: qizhangcs@bit.edu.cn, lironghuabit@126.com, panminjia\_cs@163.com, wanggrbit@126.com.

\IEEEcompsocthanksitem Y. Dai and Q. Tian are with the Diankeyun Technologies Ltd, Beijing 100043, China.\protect\\
E-mail: toyhdai@163.com, tianqun@cetccloud.com.

}
}

\markboth{IEEE TRANSACTIONS ON KNOWLEDGE AND DATA ENGINEERING, ~Vol.~XX, No.~X, May~2021}%
{Shell \MakeLowercase{\textit{et al.}}: Bare Advanced Demo of IEEEtran.cls for IEEE Computer Society Journals}

\IEEEtitleabstractindextext{%
\begin{abstract}
Cohesive subgraph mining on attributed graphs is a fundamental problem in graph data analysis. Existing cohesive subgraph mining algorithms on attributed graphs do not consider the \emph{fairness} of attributes in the subgraph. In this paper, we, for the first time, introduce fairness into the widely-used clique model to mine fairness-aware cohesive subgraphs. In particular, we propose three novel fairness-aware maximal clique models on attributed graphs, called weak fair clique, strong fair clique and relative fair clique, respectively. To enumerate all weak fair cliques, we develop an efficient backtracking algorithm called \weak~equipped with a novel colorful $k$-core based pruning technique. We also propose an efficient enumeration algorithm called \strong to find all strong fair cliques based on a new attribute-alternatively-selection search technique. To further improve the efficiency, we also present several non-trivial ordering techniques for both weak and strong fair clique enumerations. To enumerate all relative fair cliques, we design an enhanced colorful $k$-core based pruning technique for 2D attribute, and then develop two efficient search algorithms: \relativeweak and \relativestrong based on the ideas of \weak and \strong for arbitrary dimension attribute. The results of extensive experiments on four real-world graphs demonstrate the efficiency, scalability and effectiveness of the proposed algorithms.
\end{abstract}

}

\maketitle
\IEEEdisplaynontitleabstractindextext

%
\IEEEpeerreviewmaketitle

\IEEEraisesectionheading{\section{Introduction}\label{sec:introduction}}
\IEEEPARstart{C}{omplex} networks in the real world, such as social networks, communication networks and biological networks, can be modeled as graphs. Graph analysis techniques have been extensively studied to help understand the features of networks. Community detection, which aims at finding cohesive subgraph structures in networks, is a fundamental problem in graph analysis that has attracted much attention for decades \cite{li2015influential, papadopoulos2012community, huang2014querying}. As an elementary model, clique has been widely used to reveal dense community structures of graphs \cite{friedrich2015cliques,li2019improved}. Mining cliques in a graph has a wide range of applications, including mining overlapping communities in social networks \cite{18tkdekpc}, identifying protein complexes in protein networks \cite{yu2006predicting}, and finding groups with abnormal transactions in financial networks \cite{boginski2006mining}.

Many real-life networks are often attributed graphs where vertices or edges are associated with attribute information. There are a large number of studies that focus on finding communities on attributed graphs \cite{li2018community, tong2007fast, fang2016effective, khan2020compact, pizzuti2018genetic, wu2014graph,yang2013community, xu2012model}. However, those works either require a high correlation of attributes in a community or aim to find communities satisfying some attribute constraints. None of them takes into account the \emph{fairness} of attributes in the community.

Recently, the concept of fairness is mainly considered in the machine learning community \cite{verma2018fairness, hardt2016equality, dwork2012fairness}. Many studies reveal that a rank produced by a biased machine learning model can result in systematic discrimination and reduce visibility for an already disadvantaged group (e.g., incorporations of gender and racial and other biases) \cite{zehlike2017fa, serbos2017fairness, beutel2019fairness}. Therefore, many different definitions of fairness, such as individual fairness, group fairness \cite{verma2018fairness}, and related algorithms were proposed to generate a fairness ranking. Some other studies focus on the fairness in classification models, such as demographic parity \cite{dwork2012fairness} and equality of opportunity \cite{hardt2016equality}. All these studies suggest that the concept of fairness is very important in machine learning models.

Motivated by the concept of fairness in machine learning, we introduce fairness for an important graph mining task, i.e., mining cliques in a graph. Mining \emph{fair cliques} has a variety of applications. For example, consider an online social network where each user has an attribute denoting his/her gender. We may want to find a clique community in which both the number of males and females reach a certain threshold, or the number of males is exactly the same or slightly different from the number of females. Compared to the traditional clique communities, the fair clique communities can overcome gender bias. In a collaboration network, each vertex has an attribute representing his/her research topic. The fair cliques can be used to identify research groups who work closely and also have diverse research topics, because the fair cliques have already considered the fairness over different research topics. Finding such fair cliques can help identify the groups of experts from diverse research areas to conduct a particular task.


In this paper, we focus on the problem of finding fairness-aware cliques in attributed graphs where each vertex in the graph has one attribute. We propose three new models to characterize the fairness of a clique, called weak fair clique, strong fair clique and relative fair clique, respectively. A weak fair clique is a maximal subgraph which 1) is a clique, and 2) requires the number of vertices of every attribute value is no less than a given threshold $k$, thus it can guarantee the fairness over all attributes to some extent. A strong fair clique is a maximal subgraph in which 1) the vertices form a clique, and 2) the number of vertices for each attribute value is no less than $k$ and exactly the same, thus it can fully guarantee the fairness over all attributes. A relative fair clique is a maximal subgraph in which 1) the vertices form a clique, 2) the number of vertices for each attribute value is no less than $k$, and 3) the difference in the number of vertices for all attributes is no larger than a given threshold $\delta$. Thus, the relative fair clique is a compromise model between the weak and strong fair cliques, which not only guarantees the coverage of each attribute, but also implements a more flexible balance between all attributes. We show that finding all weak, strong and relative fair cliques is NP-hard. Furthermore, the problem of enumerating all strong and relative fair cliques is often much more challenging than the problem of enumerating all weak fair cliques. To solve our problems, we first propose a backtracking enumeration algorithm called \weak with a novel colorful $k$-core based pruning technique to find all weak fair cliques. Then, we propose a \strong algorithm to enumerate all strong fair cliques based on a new attribute-alternatively-selection search strategy. We also develop several non-trivial ordering techniques to further speed up the \weak and \strong algorithms. Additionally, to enumerate all relative fair cliques, we design an enhanced colorful $k$-core based pruning technique for 2D attribute, and present two efficient search algorithms, i.e., \relativeweak and \relativestrong, to handle any dimension attribute. Below, we summarize the main contributions of this paper.

\underline{\kw{New~models}.} We propose a weak fair clique, a strong fair clique and a relative fair clique to characterize the fairness of a cohesive subgraph. To the best of our knowledge, we are the first to introduce the concept of fairness for cohesive subgraph models.

\underline{\kw{Novel~algorithms}.} We first propose a novel concept called colorful $k$-core and develop a linear-time algorithm to compute the colorful $k$-core. We show that the weak fair cliques, strong fair cliques and relative fair cliques must be contained in the colorful $k$-core, thus we can use it to prune unpromising vertices before enumerating weak, strong or relative fair cliques. Then, we propose a backtracking algorithm \weak to find all weak fair cliques with a colorful $k$-core induced ordering. To enumerate all strong fair cliques, we further develop a novel fairness $k$-core based pruning technique which is more effective than the colorful $k$-core pruning. We also present a backtracking algorithm \strong with a new attribute-alternatively-selection search strategy to enumerate all strong fair cliques. In addition, a heuristic ordering method is also proposed to further improve the efficiency of the strong fair clique enumeration algorithm. For the problem of relative fair clique enumeration, we develop two efficient algorithms, i.e., \relativeweak based on a weak fair clique refinement technique and \relativestrong equipped with attribute-alternatively-selection strategy. We also design an enhanced colorful $k$-core based pruning technique for 2D attributes which can also be used to find all weak fair cliques.

\underline{\kw{Extensive~experiments}.} We conduct extensive experiments to evaluate the efficiency and effectiveness of our algorithms using four real-world networks. The results indicate that the colorful $k$-core based pruning technique is very powerful which can significantly prune the original graph. The results also show that the \weak, \strong, \relativeweak and \relativestrong algorithms are efficient in practice. These algorithms can enumerate all fair cliques on a large graph with 2,523,387 vertices and 7,918,801 edges in less than 3 hours. In addition, we conduct a case study on \dblp to evaluate the effectiveness of our algorithms. The results illustrate that the proposed fair clique enumeration algorithms, i.e., \weak, \strong, \relativeweak and \relativestrong, can find fair communities with different research areas. Moreover, \strong can further keep balance of attribute values in the subgraph, and \relativeweak and \relativestrong can explore the communities which not only cover each attribute, but also appropriately avoid the imbalance of attributes.

{\underline{\kw{Reproducibility}.} The source code of this paper is released at Github: \url{https://github.com/honmameiko22/fairnessclique} for reproducibility purpose}.

\section{Preliminaries} \label{sec:preliminaries}
Let $G = (V, E, A)$ be an undirected, unweighted attributed graph with $n = |V|$ and $m = |E|$. Each vertex $u$ in $G$ has an attribute $A$ and we denote its value as $u.val$. Let $A_{val}$ be the set of all possible values of attribute $A$, namely, $A_{val} = \{u.val | u \in V\}$. The cardinality of $A_{val}$ is denoted by $A_n$, i.e., $A_n = | A_{val}|$. For brevity, we also represent $A_{val}$ as $A_{val} = \{a_i| 0 \le i < A_n\}$. We denote the set of neighbors of a vertex $u$ by $N(u)$, and the degree of $u$ by $d(u)=|N(u)|$. For a vertex subset $S \subseteq V$, the subgraph induced by $S$ is defined as $G_{S}=(S, E_{S}, A)$, where $E_{S} = \{(u, v) | (u, v) \in E, u, v \in S \}$ and $A$ is the vertex attribute in $G$.

\begin{figure}[t]
	\centering
	\subfigure[$G$]{
		\label{fig:expgraph1}
		\begin{minipage}{3cm}
			\centering
			\raisebox{0.1\height}{
				\includegraphics[width=\textwidth]{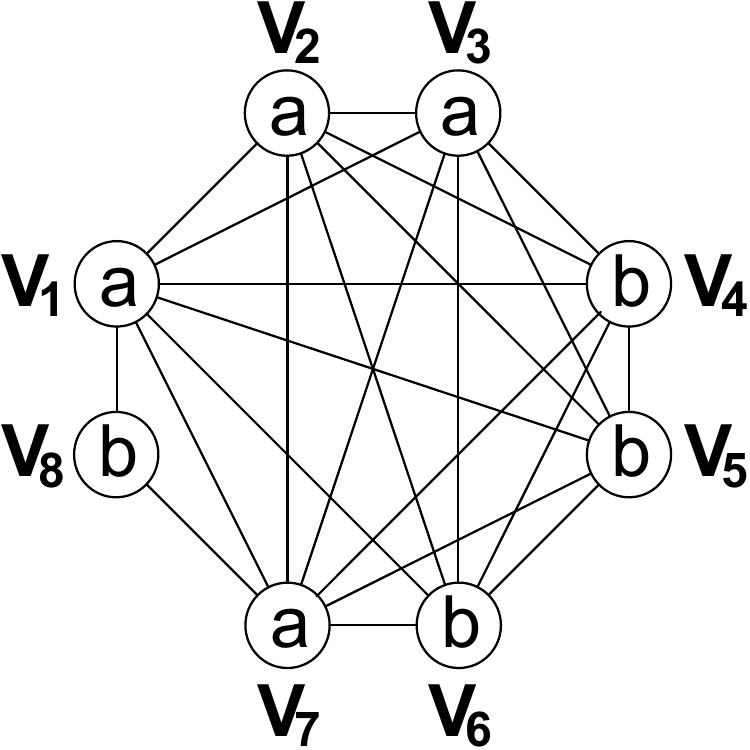}
			}
		\end{minipage}
	}
	\subfigure[colorful $G$]{
		\label{fig:expgraph2}
		\begin{minipage}{3cm}
			\centering
			\raisebox{0.1\height}{
				\includegraphics[width=\textwidth]{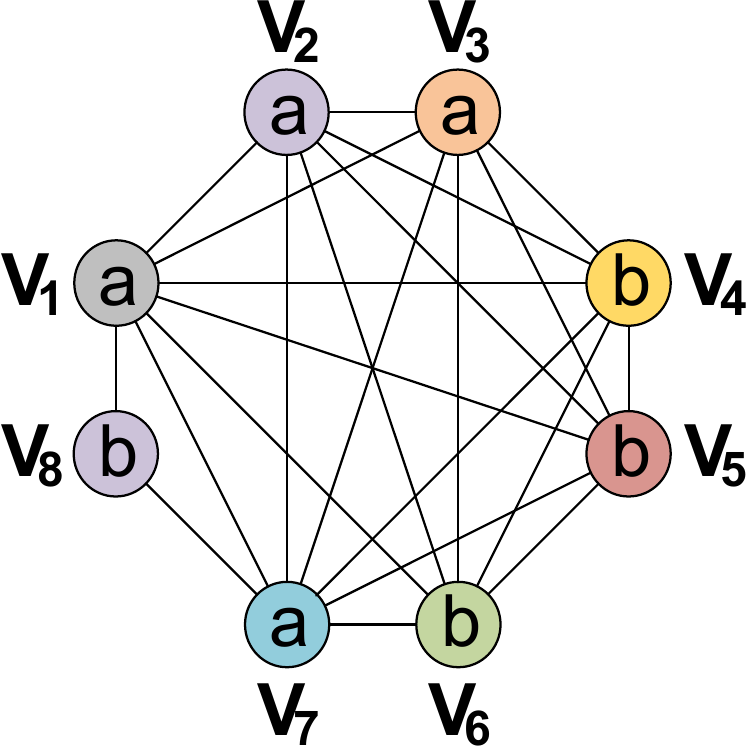}
			}
		\end{minipage}
	}
	\vspace*{-0.2cm}
	\caption{Running example}
	\vspace*{-0.4cm}
	\label{fig:expgraph}
\end{figure}

In a graph $G$, a clique $C$ is a complete subgraph where each pair of vertices in $C$ is connected. Based on the concept of clique, we present three fairness-aware clique models as follows.

\begin{defn}\label{def:weakfairclique}
	(\kw{Weak~fair~clique}) Given an attributed graph $G$ and an integer $k$, a clique $C$ of $G$ is a weak fair clique of $G$ if (1) for each value $a_i \in  A_{val}$ , the number of vertices whose value equals $a_i$ is no less than $k$; (2) there is no clique $C' \supset C$ satisfying (1).
\end{defn}

\begin{example}
	Consider a graph $G = (V, E, A)$ with $A_{val}=\lbrace a, b \rbrace$ in \figref{fig:expgraph1}. Suppose that $k = 3$. By Definition~\ref{def:weakfairclique}, we can see that the subgraph $C$ induced by the vertex set $\{ v_1, v_2, v_3, v_4, v_5, v_6, v_7 \}$ is a weak fair clique. This is because the number of vertices with attribute value $a$ in $C$ is 4 ($\geqslant k = 3$), and with attribute $b$ is 3 ($\geqslant k = 3$). Moreover, there does not exist a subgraph $C^\prime$ that contains $C$ and also satisfies the condition (1) in Definition~\ref{def:weakfairclique}. \eop
\end{example}

Clearly, by Definition~\ref{def:weakfairclique}, the weak fair clique model exhibits the \emph{fairness property} over all types of vertices (with different attribute values), as it requires the number of vertices for each attribute in the subgraph must be no less than $k$. However, the weak fair clique model may not strictly guarantee fairness for all attributes because there may be an excessive number of nodes with some attributes. Below, we propose a strong fair clique model which strictly requires the subgraph has the same number of vertices for each attribute.

\begin{defn}\label{def:strongfairclique}
	(\kw{Strong~fair~clique}) Given an attributed graph $G$ and an integer $k$, a clique $C$ of $G$ is a \emph{strong fair clique} of $G$ if (1) for each $a_i \in  A_{val}$, the number of vertices whose value equals $a_i$ is no less than $k$; (2) the number of vertices for each $a_i$ is exactly the same; (3) there is no clique $C' \supset C$ satisfying (1) and (2).
\end{defn}

\begin{example} \label{example:strongclique}
	Reconsider the attributed graph $G$ in \figref{fig:expgraph1}. Again, we assume that $k=3$. According to Definition~\ref{def:strongfairclique}, we can easily check that the subgraph induced by $\lbrace v_1, v_2, v_3, v_4, v_5, v_6\rbrace$ is a strong fair clique. Note that the subgraph induced by $\{ v_1, v_2, v_3, v_4, v_5, v_6, v_7 \}$ is a weak fair clique, but it is not a strong fair clique, as it violates the condition (2) in Definition~\ref{def:strongfairclique}. \eop
\end{example}
		
With Definition~\ref{def:strongfairclique}, the strong fair clique model requires the subgraph has the strictly same number of vertices for each attribute. Thus, it can overcome the imbalance between attributes in a clique caused by the excessive number of vertices for some attributes in the weak fair clique. However, the strong fair clique model guaranteeing fairness for all attributes is too strict to work in some real-life applications flexibly. For example, in an online social network with gender as the attribute, we only want to find a clique community in which the number of males and females is roughly equal rather than strictly equal. To this end, we propose a relative fair clique to achieve a good compromise, which absorbs the advantages of the weak and strong fair clique models.
		
\begin{defn}\label{def:relativefairclique}
	(\kw{Relative~fair~clique}) Given an attributed graph $G$ and two integers $k, \delta$, a clique $C$ of $G$ is a relative fair clique of $G$ if (1) for each value $a_i \in  A_{val}$, the number of vertices whose value equals $a_i$ is no less than $k$; (2) for arbitrary two attribute $a_i$ and $a_j$, the difference of the number of vertices with $a_i$ and $a_j$ in $C$ is no larger than $\delta$, i.e., $|cnt_C(a_i)-cnt_C(a_j)| \le \delta$; (3) there is no clique $C' \supset C$ satisfying (1) and (2).
\end{defn}

\begin{figure}[t]
	\centering
	\label{fig:expgraphnew}
	\includegraphics[width=0.28\textwidth]{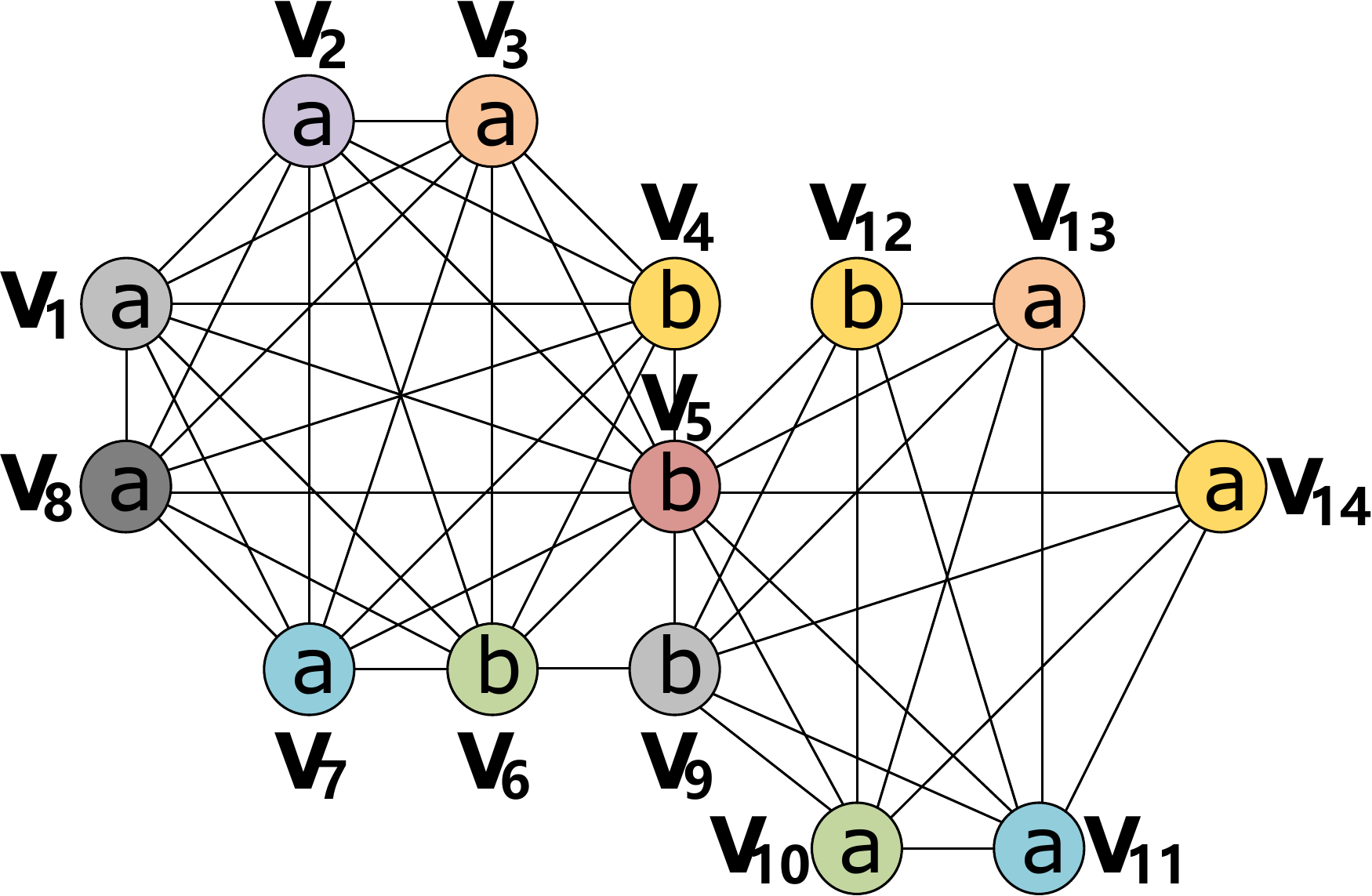}	\vspace*{-0.2cm}
	\caption{Running example: colorful $G$}
	\vspace*{-0.4cm}
\end{figure}

\begin{example} \label{example:relativeclique}
	Consider the attributed graph $G$ in \figref{fig:expgraphnew}. We suppose that $k=3$ and $\delta=1$. By Definition \ref{def:relativefairclique}, we can easily derive that the clique $C_1$ induced by $\lbrace v_1, v_2, v_3, v_4, v_5, v_6, v_7\rbrace$ is a relative fair clique that involves 3 vertices with $a$ and 4 vertices with $b$. While the clique $C_2$ induced by $\lbrace v_1, v_2, v_3, v_4, v_5, v_6, v_7, v_8 \rbrace$ is not a relative fair clique since it contains 5 vertices with $a$ and 3 vertices with $b$, which violates the condition (2) of Definition \ref{def:relativefairclique}. The clique $C_3$ induced by $\lbrace v_1, v_2, v_3, v_4, v_5, v_6 \rbrace$ is also not because $C_1$ is a larger clique that contains $C_3$, which violates the condition (3) of Definition \ref{def:relativefairclique}. Clearly, $C_2$ is a weak fair clique and $C_3$ is a strong fair clique, we have $C_2 \supset C_1 \supset C_3$. Thus, the relative fair clique $C_2$ is indeed a compromise clique between $C_1$ and $C_3$. \eop
\end{example}

\stitle{Remark.} According to Definition~\ref{def:weakfairclique}, Definition~\ref{def:strongfairclique}, and Definition~\ref{def:relativefairclique}, the parameter $k$ in our fair clique models provides a lower bound on the size of a clique. There are at least $k\times A_n$ vertices in a weak/strong/relative fair clique. Note that the guarantee of fairness in our models lies in that no matter how large a clique is, every attribute owns at least $k$ vertices. The weak fair clique model is suitable to the applications which require a lower-bound guarantee of fairness. The strong fair clique, however, aims at finding absolutely fair cliques, which can be applied in the scenarios like finding a group of people where the number of females equals that of males. In comparison, the relative fair clique achieves a compromise between the weak fair clique and strong fair clique models. Specifically, when $\delta = \infty$, a relative fair clique degenerates to a weak fair clique, and it evolves into a strong fair clique in the case of $\delta = 0$. Hence, a relative fair clique must be contained in weak fair cliques, and a strong fair clique must be contained in relative fair cliques.

Note that in the relative fair clique model, we also require the number of vertices for each attribute in the clique must be no less than $k$. This is because if we only guarantee that the difference of the number of each attribute is below a given threshold $\delta$, we may miss fairness in some cases. For example, suppose that we have three attributes: A, B and C, and the given difference threshold is $\delta=5$. Then, we may find a 5-clique that has 5 vertices with A, 0 vertex with B, and 0 vertex with C which is clearly unfair for the attributes B and C. Hence, all our definitions of fairness-aware cliques need to guarantee that each attribute has at least $k$ vertices.


\stitle{Problem statement.} Given an attributed graph $G$ and two integers $k$ and $\delta$, our goal is to enumerate all weak fair cliques and strong fair cliques with $k$, and enumerate all relative fair cliques in $G$ with $k$ and $\delta$.

\begin{example}
	Consider the attributed graph $G$ in \figref{fig:expgraph1}. Suppose that $k$ equals 2. We aim to find all 2-weak fair cliques and 2-strong fair cliques in $G$. The answer of 2-weak fair clique enumeration is $C = \lbrace v_1, v	_2, v_3, v_4, v_5, v_6, v_7\rbrace$ because it is the maximal clique satisfying {Definition \ref{def:weakfairclique}}. We can also find that there are three 2-strong fair cliques in $G$, i.e., $C_1 = \lbrace v_1, v_2, v_3, v_4, v_5, v_6\rbrace$, $C_2 = \lbrace v_1, v_2, v_7, v_4, v_5, v_6\rbrace$, and $C_3 = \lbrace v_2, v_3, v_7, v_4, v_5, v_6\rbrace$, thus they are the answers for 2-strong fair clique search. Clearly, all 2-strong fair cliques are subgraphs of the 2-weak fair clique. Let us consider the attributed graph $G$ in \figref{fig:expgraphnew}. Assume that $k=3$ and $\delta=2$, and we want to find all relative fair cliques in $G$. The answer of $(3, 2)$-relative fair clique enumeration problem are the subgraphs induced by $V_C^1=\{ v_1, v_2, v_3, v_4, v_5, v_6, v_7, v_8\}$ and $V_C^2=\lbrace v_5, v_9, v_{10}, v_{11}, v_{12}, v_{13}\rbrace$. They are also two weak fair cliques. While when $\delta$ equals $1$, there are $C_5^4$ and $1$ relative fair cliques in the subgraphs induced by $V_C^1$ and $V_C^2$, respectively. In the case of $\delta=0$, we can also find $C_5^3$ and $1$ relative fair cliques (i.e., strong fair cliques) in the subgraphs induced by $V_C^1$ and $V_C^2$. Obviously, all $(3, \delta)$-relative fair cliques are contained in all $3$-weak fair cliques, and all $3$-strong fair cliques are included in all $(3, \delta)$-relative fair cliques. \eop
\end{example}

\stitle{Challenges.} We first discuss the hardness of the weak fair clique enumeration problem. Considering a special case: $k = 0$. Clearly, the weak fair clique enumeration problem degenerates to the traditional maximal clique enumeration problem which is NP-hard. Thus, finding all weak fair cliques is also  NP-hard. Enumerating strong fair cliques is more challenging than enumerating all weak fair cliques for the following reasons. (1) The number of strong fair cliques is often much larger than that of weak fair cliques. By definition, we can see that a strong fair clique is always contained in a weak fair clique. On the contrary, a weak fair clique is not necessarily a strong fair clique. (2) Each weak fair clique must be a traditional maximal clique, but the strong fair clique may not be a traditional maximal clique (see Example~\ref{example:strongclique}), which means that it is difficult to check the maximality of strong fair cliques. For relative fair clique enumeration problem, when $\delta = \infty$, it degenerates to the weak clique enumeration problem which is NP-hard. Moreover, like the strong fair clique model, the number of relative fair cliques is also much larger than that of weak fair cliques and it is also difficult to check the maximality.

Unlike traditional maximal cliques, our fair clique models have an additional attribute value constraint, thus a potential solution is to apply attribute information to prune the search space. The challenges of our problems are (1) how can we efficiently prune unpromising vertices, and (2) how to maintain the fair clique property during the search procedure. To tackle the above challenges, we will propose the \weak algorithm with a new colorful $k$-core based pruning technique for weak fair clique enumeration; propose the \strong algorithm with a novel attribute-alternatively-selection strategy for enumerating all strong fair cliques; and propose a \relativeweak algorithm based on a weak fair clique refinement technique and a \relativestrong algorithm with an attribute-alternatively-selection strategy to enumerate all relative fair cliques. All the proposed algorithms are able to correctly find all fair cliques and significantly improve the efficiency compared to the baseline enumeration algorithm.


\section{Weak fair clique enumeration} \label{sec:wfcsearch}
\label{sec:wfcsearch}

In this section, we present the \weak algorithm to enumerate all weak fair cliques. The key idea of \weak is that it first prunes the vertices that are not contained in any weak fair clique based on a novel concept called colorful $k$-core. Then, it performs a carefully-designed backtracking search procedure to enumerate all results. Below, we first introduce the concept of colorful $k$-core, followed by a heuristic search order and the \weak algorithm.


\subsection{The colorful $k$-core pruning} \label{sec:colorfulcore}
Before introducing the colorful $k$-core based pruning technique, we first briefly review the problem of vertex coloring for a graph. The goal of vertex coloring is to color the vertices such that no two adjacent vertices have the same color~\cite{matula1972graph, jensen2011graph}. Given a graph $G = (V, E)$, we denote by $color(u)$ the color of a vertex $u \in V$. Based on the vertex coloring, we define the \emph{colorful degree} of a vertex as follows.

\begin{defn}
	\label{def:colorfuldeg}
	(\kw{Colorful} \kw{degree}) Given an attributed graph $G = (V, E, A)$ and an attribute value $a_i \in A_{val}$. The colorful-degree of vertex $u$ based on $a_i$, denoted by $D_{a_i}(u, G)$, is the number of colors of $u$'s neighbors whose attribute value is $a_i$, i.e., $D_{a_i}(u, G) = | \lbrace color(v) |v \in N(u), v.val = a_i \rbrace |$.
\end{defn}

Clearly, each vertex $u$ has $A_n$ colorful degrees. Let $D_{\min}(u, G)$ denotes the minimum colorful degree of a vertex $u$, i.e., $D_{\min}(u, G) = \min \lbrace D_{a_i}(u, G) | a_i \in A_{val}\rbrace$. We omit the symbol $G$ in $D_{a_i}(u, G)$ and $D_{\min}(u, G)$ when the context is clear. Below, we give the definition of \emph{colorful $k$-core}.

\begin{defn}
	\label{def:colorfulcore}
	(\kw{Colorful} \kw{k}-\kw{core}) Given an attributed graph $G = (V, E, A)$ and an integer $k$, a subgraph $H = (V_H, E_H, A)$ of $G$ is a colorful $k$-core if: (1) for each vertex $u \in V_H$, $D_{\min}(u, H) \ge k$; (2) there is no subgraph $H' \subseteq G$ that satisfies (1) and $H \subset H'$.
\end{defn}

Based on Definition~\ref{def:colorfulcore}, we have the following lemma.


\begin{Lemma}
	\label{lem:kccorekwfc}
	Given an attributed graph $G = (V, E, A)$ and a parameter $k$, any weak fair clique must be contained in the colorful ($k$-1)-core of $G$.
\end{Lemma}

\begin{proof}
	Assume that $C$ is a weak fair clique and consider a vertex $u \in C$. Based on Definition \ref{def:weakfairclique}, for each $a_i \in A_{val}$, $u$ has at least $k-1$ neighbors in $C$ whose attribute value is $a_i$. Since the vertices with the same color must not be adjacent, we have $D_{a_i}(u, C) \ge D_{\min}(u, C) \ge k-1$ for each $a_i \in A_{val}$. Thus, if a subgraph $g \subseteq G$ satisfies $D_{\min}(u, g) < k-1$, $C$ must not be included in $g$.
\end{proof}	


Equipped with \lamref{lem:kccorekwfc}, we propose a novel algorithm, called \colorful, to compute the colorful-$k$-core of $G$, which can be used to prune unpromising vertices in the weak fair clique enumeration procedure. The pseudo-code of \colorful is shown in Algorithm \ref{alg:colorcoreprune}. The algorithm computes the colorful-$k$-core of $G$ by iteratively peeling vertices from the remaining graph based on their colorful degrees, which is a variant of the classic core decomposition algorithm \cite{coresdecom2003, coresdecomMatulaB83} (lines 8-20). Specifically, it first performs greedy coloring on $G$ which colors vertices based on the order of degree \cite{mitchem1976various,14spaacolororder} (line 1). Note that finding the optimal coloring is an NP-hard problem \cite{jensen2011graph, matula1972graph}, thus we use a greedy algorithm to compute a heuristic coloring which is sufficient for defining the colorful $k$-core. A priority queue $Q$ is employed to maintain the vertices with smaller $D_{\min}$ which will be removed during the peeling procedure (line 2). \colorful computes the colorful degrees of all vertices to initialize $Q$ (lines 3-10). $M_u$ records the number of $u$'s neighbors whose attribute values and colors are the same. After that, the algorithm computes the colorful $k$-core of $G$ by iteratively peeling vertices from the remaining graph based on their colorful degrees (lines 11-20). Finally, \colorful returns the remaining graph $\hat{G}$ as the colorful $k$-core. Below, we analyze the complexity of Algorithm~\ref{alg:colorcoreprune}.

\begin{algorithm}[t]
	\scriptsize
	\caption{\colorful}
	\label{alg:colorcoreprune}
	\KwIn{$G = (V, E, A)$, an integer $k$}
	\KwOut{The colorful $k$-core $\hat {G}$}
	Color all vertices by invoking a degree-based greedy coloring algorithm\;
	Let ${\mathcal Q} $ be a priority queue; ${\mathcal Q} \leftarrow \emptyset $\;
	\For{$u \in V$}
	{
		\For{$v \in N(u)$}
		{
			{\bf {if}} $M_u(v.val, color(v)) = 0$ {\bf {then}} $D_{v.val}(u)$++\;
			$M_u(v.val, color(v))\text{++}$\;
		}
		$D_{\min}(u) \leftarrow \min \lbrace D_{a_i}(u) | a_i \in A_{val} \rbrace$\;
	}
	\For{$u \in V$}
	{
		\If{$D_{\min}(u) < k$}
		{
			${\mathcal Q}.push(u)$; Remove $u$ from $G$\;
		}
	}
	\While{${\mathcal Q} \neq \emptyset$}
	{
		$u \leftarrow {\mathcal Q}.pop()$\;
		\For{$v \in N(u)$}
		{
			\If{$v$ is not removed}
			{
				$M_v(u.val, color(u)){-}{-}$\;
				\If{$M_v(u.val, color(u)) \le 0$}
				{
					$D_{u.val}(v) \leftarrow D_{u.val}(v) - 1$\;
					$D_{\min}(v) \leftarrow \min \lbrace D_{a_i}(v) | a_i \in A_{val} \rbrace$\;
					
					\If{$D_{\min}(v) < k$}
					{
						${\mathcal Q}.push(v)$; Remove $v$ from $G$\;
					}
				}
			}
		}
	}
	The colorful $k$-core $\hat {G} \leftarrow$ the remaining graph of $G$\;
	{\bf return} $\hat {G}$;
\end{algorithm}

\begin{example}
	Consider the graph $G = (V, E, A)$ in \figref{fig:expgraph1}. Assume that we want to search all $2$-weak fair cliques. By \lamref{lem:kccorekwfc}, we invoke \colorful to calculate the colorful-$1$-core of $G$. Specifically, we first color the vertices of $G$ using the greedy method. Then, we obtain a colored graph which is illustrated in \figref{fig:expgraph2} with seven different colors. Take the vertex $v_8$ as an example. $v_8$ connects to $v_1$ and $v_7$ in $G$ and both of them have attribute value $a$, thus $D_a(v_8) = 2$ and $D_b(v_8) = 0$ hold. Due to $D_{\min}(v_8) = D_b(v_8) = 0 < 1$, $v_8$ is not contained in any $2$-weak fair clique. Thus, \colorful removes $v_8$ from $G$. The removal of $v_8$ subsequently updates the colorful-degrees of $v_1$ and $v_7$. \colorful repeatedly removes vertices until all the remaining vertices satisfying $D_{\min} \ge 1$. Finally, we can obtain a subgraph induced by the vertex set $V-\{v_8\}$ which is a colorful-$1$-core with $D_{\min} = 2$. \eop
\end{example}

\begin{theo}
	Algorithm~\ref{alg:colorcoreprune} consumes $O(E+V)$ time using $O(V \times A_n \times {\kw {color}})$ space, where ${\kw {color}}$ denotes the total number of colors.
\end{theo}

\begin{proof}
	In line~1, the greedy coloring procedure takes $O(E+V)$ time \cite{14spaacolororder}. In lines~2-7, we can easily derive that the algorithm takes $O(E+V)$ time. In lines~{11}-20, the algorithm can update $M_v$ for each $v\in N(u)$ in $O(1)$ time. For each edge $(u, v)$, the update operator only performs once, thus the total time complexity is bounded by $O(E+V)$. For the space complexity, the algorithm needs to maintain the structure $M_v$ for each vertex which takes at most $O(V\times A_n\times {\kw {color}})$ space in total.
\end{proof}



\subsection{The colorful $k$-core based ordering} \label{sec:onlineupdate}	
\weak finds all weak fair cliques by performing a backtracking search procedure. Hence, the search order of vertices is vital as the search spaces with various orderings are significantly different. Below, we propose a heuristic order based on the colorful $k$-core, called \colororder, which can significantly improve the performance of \weak as confirmed in our experiments.

\begin{algorithm}[t]
	\scriptsize
	\caption{\wforder}
	\label{alg:colorcoreorder}
	\KwIn{A connected graph $G = (V, E)$}
	\KwOut{The \colororder ordering $\mathcal{O}$}
	
	Let $B$ be an array with $B(i) = false, 1 \le i \le |V|$\;
	$\mathcal{O} \leftarrow \emptyset$; $H \leftarrow \emptyset$; $\cnt \leftarrow 0$\;
	\For{$u \in V$}
	{
		Calculate $D_{\min}(u)$ as lines 4-7 in Algorithm \ref{alg:colorcoreprune}\;
		$H.push(u, D_{\min}(u))$\;
	}
	\While{$H \neq \emptyset$}
	{
		$(u, D_{\min}(u)) \leftarrow H.pop()$\;
		${\mathcal {O}}[u]=\cnt$; $B(u) \leftarrow true$; $\cnt$++\;
		\For{$v \in N(u)$}
		{
			\If{$B(v) = false$}
			{
				$M_v(u.val, color(u)){-}{-}$\;
				\If{$M_v(u.val, color(u)) \le 0$}
				{
					$D_{u.val}(v){-}{-}$; $dif \leftarrow D_{\min}(v)-D_{u.val}(v) $\;
					\If{$dif \neq 0$}
					{
						$D_{\min}(v) \leftarrow D_{u.val}(v)$; $H.update(v, dif)$\;
					}
				}
			}
		}
	}
	{\bf return} $\mathcal {O}$;
\end{algorithm}

Consider a vertex $u$ and its neighbor $v$ with $D_{\min}(u, G) \ge (k-1) > D_{\min}(v, G)$. According to \lamref{lem:kccorekwfc}, $u$ may be contained in a weak fair clique but $v$ is impossible. Thus, we can construct a smaller subgraph induced by $u$'s neighbors whose $D_{\min}$ values are no less than $D_{\min}(u, G)$ to search weak fair cliques. Inspired by this, we design a search order denoted by \colororder; and we propose an algorithm, called \wforder, to calculate such an order. Similar to the idea of \colorful, \wforder iteratively removes a vertex with the minimum $D_{\min}$ from the remaining graph. The vertices-removal ordering by this procedure is the \colororder.

Algorithm \ref{alg:colorcoreorder} outlines the pseudo-code of \wforder. For each vertex $u$, we use $\mathcal{O} (u)$ to indicate the rank of $u$ in our order $\mathcal{O}$. A heap-based structure $H$ is employed to maintain the vertices with their $D_{\min}$ values, which always pops out the pair $(u, D_{\min}(u))$ with minimum $D_{\min}$. \wforder first calculates $D_{\min}(u)$ for every vertex $u$ and pushes $(u, D_{\min}(u))$ into $H$ (lines 3-5). Then, \wforder iteratively pops out the vertex with minimum $D_{\min}$ from $H$ and records its rank in $\mathcal{O}$ (lines 6-15). As a vertex is removed, we maintain the $D_{\min}$ values for its neighbors and update $H$ (lines 9-15). It is easy to check that the time and space complexities of Algorithm \ref{alg:colorcoreorder} are the same as those of Algorithm~\ref{alg:colorcoreprune}.


The reason why \colororder works is that the search procedure beginning with vertices that have low ranks in \colororder tends to be less possible to form weak fair cliques. Note that the main searching time of the enumeration algorithm is spent on the vertices that have a dense and large neighborhood. \colororder can guarantee that the unpromising vertices are explored first, thus reducing the number of candidates of the vertices that have a dense and large neighborhood.


\subsection{The weak fair clique enumeration algorithm} \label{sec:weakfairclique}	

The main idea of \weak is to prune the unpromising vertices first, and then perform the backtracking procedure to find all weak fair cliques. Unlike the traditional maximal clique enumeration, \weak is equipped with a colorful $k$-core-based pruning rule and a carefully-designed \colororder ordering technique, which can significantly reduce the search space. The pseudo-code of \weak is outlined in Algorithm \ref{alg:weakenumeration}.

\begin{algorithm}[t]
	\scriptsize
	\caption{\weak}
	\label{alg:weakenumeration}
	\KwIn{$G = (V, E, A)$, an integer $k$}
	\KwOut{The set of weak fair cliques $Res$}
	$Res \leftarrow \emptyset$; $R \leftarrow \emptyset$; $X \leftarrow \emptyset$; $C \leftarrow \emptyset$\;
	$\hat{G}=(\hat{V}, \hat{E}) \leftarrow \colorful(G, k-1)$\;
	Initialize an array $B$ with $B(i) = false, 1 \le i \le |\hat{V}|$\;
	\For{$u \in {\hat V}$}
	{
		\If{$B(u) = false$}
		{	
			$C \leftarrow \connectedcpn(u, B)$\;
			${\mathcal{O}} \leftarrow \text{\wforder}(C)$\;
			$R \leftarrow \emptyset$; $X \leftarrow \emptyset$; $\backtrack(R, C, X, {\mathcal {O}})$;
		}
	}
	{\bf return} $Res$\;
	
	\vspace*{0.1cm}
	{\bf Procedure} $\backtrack(R, C, X, {\mathcal O})$\\
	{\bf {if}} $C = \emptyset$ and $X = \emptyset$ {\bf {then}} $Res \leftarrow Res \cup R$\;
	\For{$u \in C$ in non-descending \colororder order}
	{
		$\hat{R} \leftarrow R \cup u$; ${\hat C} \leftarrow \emptyset$; $flag \leftarrow false$\;
		Let ${{\hat C}_{\cnt}}, {{\hat R}_{\cnt}}$ be the arrays of size $A_n$\;
		\For{$v \in C$}
		{
			\If{$v \in N(u)$ and ${\mathcal O}(v) > {\mathcal O}(u)$}
			{
				${\hat C} \leftarrow$ ${\hat C} \cup v$; ${{\hat C}_{\cnt}}(v.val)$++\;
			}
		}	
		{\bf {if}} $|{\hat C}| + |{\hat R}| < k\times A_n$ {\bf then continue}\;
		{\bf {for}} $v \in {\hat R}$ {\bf {do}} ${{\hat R}_{\cnt}}(v.val)$++\;
		
		\For{$a_i \in A_{val}$}
		{
			\If{${{\hat R}_{\cnt}}(a_i) + {{\hat C}_{\cnt}}(a_i) < k$}
			{
				$flag \leftarrow true$; {\bf break};\
			}
		}
		{\bf if} $flag = true$ {\bf then continue}\;
		${\hat X} \leftarrow X \cap N(u)$\;
		$\backtrack({\hat R}, {\hat C}, {\hat X}, {\mathcal O})$\;
		$X \leftarrow X \cup u$\;
	}
\end{algorithm}

The \weak algorithm works as follows. It first initializes four sets $R$, $X$, $C$, and $Res$ (line 1). The set $R$ represents the currently-found clique which may be extended to a weak fair clique. $X$ is the set of vertices in which every vertex can be used to expand the current clique $R$ but has already been visited in previous search paths. $C$ is the candidate set that can be used to extend the current clique $R$ in which each vertex must be neighbors of all vertices in $R$. After initialization, \weak performs \colorful to prune the vertices that are definitely not contained in any weak fair clique (line 2). The algorithm invokes the \backtrack procedure to find all weak fair cliques in the pruned graph $\hat G$ (lines 4-9). Note that $\hat G$ may have several connected colorful $(k-1)$-cores, so \backtrack should be performed on each connected component in $\hat G$. An array $B$ is used to indicate whether a vertex $u$ has been searched, and it is initialized as false for each vertex. For an unvisited vertex $u$, \weak identifies the connected colorful-$(k-1)$-core $C$ containing $u$ and sets $B$ as true for all vertices within $C$ to denote that $C$ will not be searched again (line 6). \weak then calls \wforder to derive the search order \colororder of vertices in $C$, and performs the \backtrack procedure on $C$ to enumerate all weak fair cliques (lines 7-8).

The workflow of \backtrack is depicted in lines 10-26 of Algorithm \ref{alg:weakenumeration}. It first identifies whether the current $R$ is a weak fair clique (line 11). $R$ is an answer if and only if $C = \emptyset$ and $X = \emptyset$. $C$ is empty means that no vertex can be added into $R$. In addition, the set $X$ must be empty, otherwise any vertex in $X$ can be added into $R$ and makes $R$ non-maximal. If $R$ is not a weak fair clique, we add each vertex $u \in C$ into $R$ and start the next iteration of \backtrack (lines 12-26). Note that each candidate in $C$ is a neighbor of all vertices in $R$, therefore after adding $u$ into $R$, $C$ must be updated to keep out those vertices that are not adjacent with $u$ (lines 15-17). Here, we only consider the vertices whose rank is larger than $u$'s rank to avoid finding the same clique repeatedly. After obtaining the updated sets $\hat C$ and $\hat R$, if $|{\hat C}| + |{\hat R}| < k\times A_n$ holds, \backtrack terminates as the sets cannot reach the minimum size of a weak fair clique (line 18). On the other hand, we use ${{\hat R}_{\cnt}}$ and ${{\hat C}_{\cnt}}$ to denote the number of vertices whose attribute value is $a_i$ in $\hat {R}$ and $\hat {C}$, respectively (line 17 and line 19). By checking the count for each $a_i \in A_{val}$, we can quickly determine whether the current/next clique is promising. For any $a_i \in A_{val}$, if ${{\hat R}_{\cnt}}(a_i) + {{\hat C}_{\cnt}}(a_i) < k$ holds, we cannot obtain a weak fair clique even if we add the whole set $C$ into $R$. This is because the condition (1) of Definition \ref{def:weakfairclique} is not satisfied, thus \backtrack terminates (lines 20-23). Otherwise, the procedure derives the set $\hat X$ by adding $u$'s neighbors into $X$, and then performs the next iteration (lines 24-25). After exploring the vertex $u$, \backtrack adds it into $X$ because $u$ has already been searched in the current search path and cannot be processed in the following recursions (line 26).

\section{Strong fair clique Enumeration} \label{sec:sfcsearch}
In this section, we first develop an efficient strong fair clique enumeration algorithm with a novel pruning technique for the two-dimensional (2D) case, where the attributed graph has only two types of attributes (i.e., $|A_n|=2$). Then, we will show how to extend our enumeration algorithm to handle the high-dimensional case ($|A_n| > 2$).

\subsection{The pruning technique for 2D case} \label{sec:fairnessreduction}
Suppose that the attributed graph $G = (V, E, A)$ has two types of attributes, i.e., $A_{val} = \{a_1, a_2\}$.  The neighbors of a vertex $u$ can be divided into $h_u$ groups by coloring where each group contains vertices with the same color. Clearly, by the property of coloring, only one vertex can be selected from a group to form a clique with $u$. Below, we give a new definition of fairness degree of a vertex.

\begin{defn}
	\label{def:fairnessdeg}
	(\kw{Fairness} \kw{degree}) Given a colored attributed graph $G = (V, E, A)$ with $A_{val} = \{a_1, a_2\}$, the fairness degree of $u$, denoted by $FD(u)$, is the largest number of groups from which we select vertices so that the number of vertices with attribute $a_1$ is the same as the number of vertices with attribute $a_2$.
\end{defn}

By Definition~\ref{def:fairnessdeg}, we can easily verify that the fairness degree of a vertex $u$, i.e., $FD(u)$, is an upper bound of the size of the strong fair clique containing $u$. Therefore, for any vertex $u$, if $FD(u) < 2\times (k-1)$, then $u$ cannot be contained in any strong fair clique, because any vertex in a strong fair clique must have a fairness degree no less than $2\times (k-1)$ by Definition~\ref{def:strongfairclique}. As a consequence, we can safely prune the vertex whose fairness degree is less than $2\times (k-1)$.

A remaining question is how can we efficiently compute the fairness degree for a vertex $u$. Below, we develop an efficient approach to answer this question.

Based on the attribute values, the $h_u$ color groups can be divided into three categories: (1) \agroupone: is a group that involves vertices of attribute $a_1$ only; (2) \agrouptwo: is a group that contains vertices of attribute $a_2$ only; (3) \agroupmix: is a group that contains vertices of both $a_1$ and $a_2$. Let $c_1$, $c_2$, and $c_m$ be the number of the \agroupone groups, the \agrouptwo groups, and the \agroupmix groups respectively. Suppose without loss of generality that $c_1 \le c_2$. Then, if $c_m \le (c_2-c_1)$ holds, we can easily derive that $FD(u)=2\times (c_m+c_1)$. Otherwise, we have $FD(u) = 2\times ((c_m-(c_2-c_1))/2+c_2)$. Based on these results, we can calculate the fairness degree for each vertex by using the three quantities $c_1$, $c_2$, and $c_m$. The pseudo-code of our \fairdeg algorithm to compute the fairness is given in lines 17-29 of Algorithm \ref{alg:fairnessreduction}.



\begin{algorithm}[t]
	\scriptsize
	\caption{\fairness}
	\label{alg:fairnessreduction}
	\KwIn{$G = (V, E, A)$, an integer $k$}
	\KwOut{The reduced graph $\hat G$}
	${\overline {G}} = ({\overline {V}}, {\overline {E}}) \leftarrow \colorful(G, k)$\;
	Let $FD$ be an array of size $|{\overline {V}}|$; Let ${\mathcal {Q}}$ be a queue\;
	\For {$u \in {\overline{V}}$}{
		\For{$v \in N(u)$}
		{
			$\group(u, color(v), v.val)$++;
		}
		$FD(u) \leftarrow \fairdeg(u, \group)$\;
		\If{$FD(u) < 2\times k$}
		{
			Remove $u$ from $\overline G$; $Q.push(u)$\;
		}
	}
	\While{$Q \neq \emptyset$}
	{
		$u \leftarrow Q.pop() $\;
		\For{$v \in N(u)$}
		{
			{\bf if} {$v$ is removed} {\bf then} {\bf continue}\;
			$\group(v, color(u), u.val) - -$\;
			Calculate $FD(v)$ and update $Q$ as lines 6-8\;
		}
	}
	$\hat {G} \leftarrow$ the remaining graph of $\overline G$\;
	{\bf return} $\hat {G}$\;
	
	\vspace*{0.1cm}
	{\bf Procedure} $\fairdeg(u, \group)$\\
	{
		$c_1 \leftarrow 0$; $c_2 \leftarrow 0$; $c_m \leftarrow 0$\;
		\For{each color $cr$}
		{
			{\bf if} $\group(u, cr, a_1) \ge 1$ and $\group(u, cr, a_2)=0$ {\bf then } $c_1\leftarrow c_1+1$\;
			{\bf if} $\group(u, cr, a_2) \ge 1$ and $\group(u, cr, a_1)=0$ {\bf then } $c_2\leftarrow c_2+1$\;
			{\bf if} $\group(u, cr, a_1) \ge 1$ and $\group(u, cr, a_1)\ge 1$ {\bf then } $c_m\leftarrow c_m+1$\;
		}
		\If{$c_1 \le c_2$}
		{
			{\bf if} $c_m \geqslant (c_2 - c_1)$ {\bf then} $FD(u) \leftarrow 2\times((c_m - (c_2 - c_1))/2 + c_2)$\;
			{\bf else} $FD(u) \leftarrow 2\times(c_m + c_1)$\;
		}
		\Else
		{
			{\bf if} $c_m \geqslant (c_1-c_2)$ {\bf then} $FD(u) \leftarrow 2\times((c_m-(c_1-c_2))/2+c_1$)\;
			{\bf else} $FD(u) \leftarrow 2\times(c_m + c_2)$\;
		}
		{\bf return } $FD(u)$\;
	}
\end{algorithm}

With the fairness degree, we can iteratively prune the vertices with fairness degrees smaller than $2\times (k-1)$. Below, we introduce a concept called fairness $k$-core to characterize the reduced subgraph after iteratively peeling the unqualified vertices.

\begin{defn}
	\label{def:fairnesscore}
	(\kw{fairness} \kw{k}-\kw{core}) Given an attributed graph ${G}= (V, E, A)$ with $A_{val} = \{a_1, a_2\}$ and an integer $k$, a subgraph $H = (V_H, E_H, A)$ of $G$ is a fairness $k$-core if: (1) for each $u \in V_H$, $FD(u) \ge 2k$; (2) there is no subgraph $H' \subseteq G$ that satisfies (1) and $H \subset H'$.
\end{defn}

By Definition~\ref{def:fairnesscore}, we can show that any strong fair clique must be contained in the fairness $k$-core.

\begin{Lemma}
	\label{lem:faircoreksfc}
	Given an attributed graph $G = (V, E, A)$ with $A_{val} = \{a_1, a_2\}$ and a parameter $k$, any strong fair clique must be contained in the fairness $(k-1)$-core of $G$.
\end{Lemma}

\begin{proof}
	Consider a strong fair clique $C$. According to Definition \ref{def:strongfairclique}, assume there are $k$ vertices of attribute $a_1$ and $k$ vertices of attribute $a_2$ in $C$. For an arbitrary vertex $u$ in $C$, we suppose that $u.{val} = a_1$. There are $k-1$ vertices of attribute $a_1$ and $k$ vertices of attribute $a_2$ in $u$'s neighbors. Therefore, after performing \fairdeg for $u$, we have $c_1 = k-1$, $c_2 = k$ and $c_m = 0$. Further, $FD(u)$ is equal to $2(k-1)$. Due to the arbitrariness of $u$, the fairness degree of each vertex in $C$ must reach $2(k-1)$, too. Hence, $C$ must be contained in the fairness-$(k-1)$-core of $G$.
\end{proof}


\begin{example}
	Reconsider the attributed graph in \figref{fig:expgraph2}. Suppose that $k=3$. By \lamref{lem:faircoreksfc}, we consider the fairness 2-core of $G$. For vertex $v_8$, $v_8$ has two neighbors $v_1$ and $v_7$, and both of them have attribute value $a$. Clearly, we have $FD(v_8) = 0 < 2\times 2$, thus $v_8$ is not contained in the fairness 2-core. For vertex $v_1$, the initial value of $c_1$, $c_2$ and $c_m$ are $2, 3, 1$. Obviously, $c_m+c_1 = c_2$, thus we have $FD(v_1) = 6 > 4$. Similarly, the fairness degrees of the other vertices are all equal to $6$. Therefore, the subgraph induced by $V\setminus\{v_8\}$ is a fairness $2$-core. Clearly, such a subgraph contains the strong fair clique as illustrated in Example~\ref{example:strongclique}. \eop
\end{example}


Similar to the colorful $k$-core computation algorithm, we can also devise a peeling algorithm to compute the fairness $k$-core by iteratively removing the vertices that have fairness degrees smaller than $2k$. The pseudo-code of our algorithm is outlined in Algorithm \ref{alg:fairnessreduction}. Note that a strong fair clique is always contained in a weak fair clique, thus we can first invoke \colorful to prune vertices that are definitely not included in the weak fair cliques before computing the fairness $k$-core of $G$ (line 1).

\begin{theo}
	Algorithm~\ref{alg:fairnessreduction} consumes $O((E+V)\times {\kw{color}})$ time using $O(V\times \kw{color})$ space.
\end{theo}

\begin{proof}
	In line~1, Algorithm~\ref{alg:fairnessreduction} invokes Algorithm~\ref{alg:colorcoreprune} which takes $O(V+E)$ time and $O(V\times \kw{color})$ space (since $A_n=2$). The \fairdeg procedure takes at most $O(\kw{color})$ time for each vertex. Therefore, the total time overhead taken in lines~3-8 is O$(V\times {\kw{color}}+E)$. In lines~9-14, for each edge $(u, v)$, the update cost is bounded by $O({\kw{color}})$, thus the total time complexity is $O((E+V)\times {\kw{color}})$. For the space complexity, the algorithm takes $O(V\times \kw{color})$ space to maintain the $\group$ structure.
\end{proof}


\stitle {Fairness $k$-core ordering.} Similar to the \colororder, we can derive an ordering based on the fairness $k$-core, called \frorder, for strong fair clique enumeration. In particular, \frorder is derived by iteratively removing the vertex with the minimum fairness degree which is very similar to the computational procedure of \colororder. We omit the details for brevity.

\subsection{The enumeration algorithm for 2D case} \label{sec:strongenumeration}


Armed with the fairness $k$-core based pruning technique and the \frorder ordering, we propose the \strong algorithm which alternatively picks a vertex of a specific attribute in the backtracking procedure to enumerate all strong fair cliques. The \strong is shown in Algorithm \ref{alg:strongenumeration}. We use $R$ to represent the currently-found clique and $C$ to denote the candidate set. Similar to \weak, \strong first applies \fairness to prune the vertices that are definitely not contained in strong fair cliques (line 2) and then performs the \strongbacktrack procedure for each connected fairness $(k-1)$-core in $\hat G$ to find all results (lines 4-8).

The pseudo-code of \strongbacktrack is outlined in lines 10-27 of Algorithm \ref{alg:strongenumeration}. Since a strong fair clique requires that the numbers of vertices for each attribute $a_i$ are exactly the same, we develop a novel attribute-alternatively-selection mechanism to select vertices in each iteration. That is, \strongbacktrack admits an input parameter $a_\phi$, which is initialized to $a_0$ (line 8), to indicate the attribute value of the vertices to be selected in the current iteration. In the next iteration, we pick the vertices with the attribute value $a_{\phi+1}$ to construct strong fair cliques (line 27). \strongbacktrack divides the candidates in $C$ into $A_n$ sets, where the attribute values of vertices in each set are the same, i.e., $C_A(a_i) = \{u | u \in C, u.val = a_i\}$ (line 14). For each candidate $u$ in $C_A(a_{\phi})$, we pick one vertex at a time as a part of the currently-found clique and update the candidate set based on the \frorder ordering (lines 16-27).

\begin{algorithm}[t]
	\scriptsize
	\caption{\strong}
	\label{alg:strongenumeration}
	\KwIn{$G = (V, E, A)$, an integer $k$}
	\KwOut{The set of all strong fair cliques $Res$}
	$Res \leftarrow \emptyset$; $R \leftarrow \emptyset$; $C \leftarrow \emptyset$\;
	$\hat{G}=(\hat{V}, \hat{E}) \leftarrow \fairness(G, k-1)$\;
	Initialize an array $B$ with $B(i) = false, 1 \le i \le |\hat{V}|$\;
	\For{$u \in {\hat V}$ }
	{
		\If{$B(u) = false$}
		{
			$C \leftarrow \connectedcpn(u, B)$\;
			${\mathcal{O}} \leftarrow$ \frorder($C$)\;
			$R \leftarrow \emptyset$; $C \leftarrow \emptyset$; $\strongbacktrack(R, C, a_0, {\mathcal O})$\;
		}
	}
	{\bf return} $Res$\;
	
	\vspace*{0.1cm}
	{\bf Procedure} $\strongbacktrack(R, C, a_{\phi}, {\mathcal O})$\\
	{
		\If{$|R|\% A_n = 0$ and $|R| \ge k\times A_n$}
		{	
			\If{$\checkmaximality(C)$}
			{
				$Res \leftarrow Res \cup R$; {\bf return};\
			}
			
		}
		{\bf for} $u \in C$ {\bf then} $C_A(u.val) \leftarrow C_A(u.val) \cup u$\;
		\For{$u \in C_A(a_{\phi})$}
		{
			${\hat R} \leftarrow R \cup u$\;
			\For{$v \in C$}
			{
				\If{$v \in N(u)$ and $\mathcal{O}(v) > \mathcal{O}(u)$}
				{
					${\hat C} \leftarrow {\hat C} \cup v$; ${\hat C}_A(v.val) \leftarrow {\hat C}_A(v.val) \cup v$\;
				}
			}
			$c_{\min} \leftarrow \min(|{\hat C}_A(a_i)|)$; $a_{\min} \leftarrow \mathop{\arg\min}_{a_i} |{\hat C}_A(a_i)|$\;
			{\bf if} $|{\hat R}|\%A_n = 0$ {\bf then} $R_c \leftarrow c_{\min} \times A_n + |{\hat R}|$\;
			\Else
			{
				
				\If{$a_{\min} \in \{a_{0}, a_{1}, ..., a_{\phi}\}$}
				{
					$R_c \leftarrow c_{\min} \times A_n + (|{\hat R}|/A_n +1)\times A_n$\;
				}
				{\bf else} $R_c \leftarrow (c_{\min}-1) \times A_n + (|{\hat R}|/A_n +1)\times A_n$\;
			}
			{\bf if} $R_c < k\times A_n$ {\bf then continue}\;
			$\strongbacktrack({\hat R}, {\hat C}, a_{\phi+1}, {\mathcal {O}}$);
		}
	}
\end{algorithm}

After adding $u$ into the current clique, we can combine the set $\hat R$ and $\hat C$ to determine whether to call \strongbacktrack for a more in-depth search (lines 16-27). Specifically, we classify the candidates in $\hat C$ according to their attribute values and record $a_{\min}$ as the attribute value with the minimum number of vertices (denoted by $c_{\min}$) (line 20). Note that if there are multiple attribute values satifying $|{\hat C}_A(a_i)| =c_{\min}$, we pick $a_i$ with the largest $i$ as $a_{\min}$. Clearly, $c_{\min}$ determines how large a strong fair clique can be. We use $R_c$ to denote the largest size of possible strong fair cliques. If $|{\hat R}|\%A_n = 0$, the numbers of vertices with various attribute values are the same in the current set $\hat{R}$, thus there are at most $c_{\min}\times A_n$ vertices can be added into ${\hat R}$, and further we have $R_c = c_{\min}\times A_n +|{\hat R}|$ (line 21). Otherwise, we calculate $R_c$ and try to search a larger clique (lines 22-27). By the attribute-alternatively-selection strategy, in the current iteration with $a_{\phi}$, the number of vertices with attribute value  $a_f$ ($a_f \in \{a_0, ..., a_{\phi}\}$) is always one more than that of vertices with  $a_b$ ($a_b \in \{a_{\phi+1}, ..., a_{n-1}\}$) in $R$. If $a_{min} = a_f$ , we can add one vertex, for each $a_b$, into $R$ to obtain a clique with size $(|\hat {R}|/A_n+1)\times A_n$, which is denoted by $R_M$. Note that there are still $c_{min}\times A_n$ vertices that may form a larger clique with $R_M$. Therefore, we calculate $R_c$ as shown in line 24. Similarly, when $a_{min} = a_b$, we have at most $(c_{min}-1)\times A_n$ vertices that may add into $R_M$ to construct a strong fair clique with size $R_c$ (line 25). After calculating $R_c$, we can terminate the search procedure early if $R_c < k \times A_n$, because it violates the definition of strong fair clique in this case. Otherwise, we recursively perform \strongbacktrack  with the attribute value $a_{\phi+1}$ (line 27).


\stitle{Maximality checking.} The results of all traditional maximal cliques and our weak fair cliques lie in the leaves of the backtracking enumeration tree. We can check whether a weak fair clique is found by $C = \emptyset$ and $X = \emptyset $ (see line 11 of Algorithm \ref{alg:weakenumeration}). However, such a maximality checking method cannot be used for strong fair cliques. The reasons are twofold: (1) an empty candidate set $C$ does not mean that we find a strong fair clique because the number of vertices in $R$ corresponding to each attribute value may not be the same; (2) even if $X$ is not empty, $R$ can be a strong fair clique. That is to say, strong fair cliques can appear in the intermediate nodes of the backtracking enumeration tree. Therefore, we need to develop new solution to check the maximality for strong fair cliques. We propose a maximality checking technique as follows.

\begin{algorithm}[t]
	\scriptsize
	\caption{ $\checkmaximality(C)$}
	\label{alg:checkmaximality}
	{
		{\bf if $|C| < A_n$}
		{\bf then return true};\\
		\Else
		{
			\For{each $a_i \in A_{val}$}{
				$C_i \leftarrow \{u | u \in C, u.val = a_i\}$\;
				{\bf if} $|C_i|=0$ {\bf return true};
			}
			$Record \leftarrow C_0$\;
			\For{each $a_i \in \{ A_{val} - \{a_0\} \}$}{
				$SwapRecord \leftarrow \emptyset$\;
				\For{$v_i \in C_i$}{
					\For{$r \in Record$}{
						\If{$v_i$ is a neighbor of all vertices in $r$}
						{
							$SwapRecord \leftarrow SwapRecord \cup \{r \cup v_i\}$\;
						}
					}
				}
				$Record \leftarrow SwapRecord$\;
			}
			{\bf if $Record \neq \emptyset$}
			{\bf return false}\;
		}
	}
	
\end{algorithm}

Once the \strongbacktrack procedure finds a clique whose size is equal to $k^\prime \times A_n$ with $k^\prime \ge k$, we need to check the maximality according to Definition \ref{def:strongfairclique}. Since the vertices in $C$ are neighbors of all vertices in $R$, if we find any clique in $C$ with every attribute, $R$ is definitely not a strong fair clique as it violates the constraint (3) in Definition \ref{def:strongfairclique}. Based on this, we propose a verification method, called \checkmaximality, which is shown in Algorithm \ref{alg:checkmaximality}. Specifically, if the size of $C$ is less than $A_n$, which means adding all vertices in $C$ will destroy the fairness property of $R$, $R$ is maximal and thus the algorithm returns \emph{true} (line 1). Otherwise, we need to explore the common neighbors to find if there exist cliques with size at least $A_n+|R|$ that are also strong fair cliques. The \checkmaximality algorithm uses $C_i$ to represent the vertices in $C$ with the attribute value $a_i$. Clearly, if $|C_i| = 0$ holds for an arbitrary attribute $a_i$, the attribute constraint will not be satisfied and the procedure outputs \emph{true}, indicating $R$ is maximal (lines 3-5). Otherwise, \strongbacktrack tries to construct cliques from $C$. The variables $Record$ and $SwapRecord$ are used to maintain the current partial cliques. Finally, if $Record$ is not empty, we can find a clique with size at least $A_n+|R|$. In such case, $R$ is not a strong fair clique and the \strongbacktrack procedure returns \emph{false} (lines 6-14).

\subsection{Handling the high-dimensional case} \label{sec:onlineupdate}
We note that the idea of the fairness degree based pruning rule is not easy to extend to the high-dimensional case, because there may be $2^{A_n}-1-A_n$ \agroupmixs in the worst case. Therefore, it is very difficult to compute the exact fairness degree for each vertex when $A_n > 2$. To circumvent this problem, we propose a heuristic greedy algorithm to calculate an approximation of the fairness degree for each vertex $u$, instead of deriving the exact fairness degree.

\begin{algorithm}[t]
	\scriptsize
	\caption{\heuristic}
	\label{alg:heuristic}
	\KwIn{A connected graph $G = (V, E)$}
	\KwOut{The \heurorder ordering $\mathcal{O}$}
	$\mathcal{O} \leftarrow \emptyset$; ${\mathcal Q} \leftarrow \emptyset$\;
	Let $B$ be an array with $B(i) = false, 1 \le i \le |V|$\;
	\For{$u \in V$}
	{
		\For{$v \in N(u)$}
		{
			$S_u(color(v), v.val)\leftarrow S_u(color(v), v.val) +1$\;
		}
		Let $\cnt$ be an array with cnt(i) = 0, $0 \le i < A_n$\;
		\For{each color $cr$}
		{
			\For{$a_i \in A_{val}$ }
			{
				\If{$S_u(cr, a_i) \ge 1 $}
				{
					$a_m = \mathop{\arg\min}_{a_i\in S_u(cr, a_i)} \cnt(a_i)$\;
				}
			}
			$\cnt(a_m)\leftarrow \cnt(a_m)+1$\;
		}
		$GD(u) = \min \{\cnt(a_i), a_i \in A_{val}\}$\;
		${\mathcal Q}.push(u, GD(u))$;
	}
	\While{${\mathcal Q} \ne \emptyset$}
	{
		$u \leftarrow {\mathcal Q}.pop()$; $\mathcal{O}.push(u)$; $B(u) \leftarrow true$\;
		\For{$v \in N(u)$}
		{
			\If{$B(v) = false$}
			{
				$S_v(color(u), u.val) \leftarrow S_v(color(u), u.val)-1$\;
				Calculate $GD(v)$ and update ${\mathcal Q}$ as lines 6-13\;
			}
		}
	}
	\bf{return} $\mathcal{O}$;
\end{algorithm}

Specifically, we let $GD(u)$ be the approximate fairness degree computed by our greedy algorithm. By coloring, the neighbors of a vertex $u$ can be classified into $h_u$ color groups. For each color $cr$, we have a group, denoted by $\group(cr)$.
For a color group $\group(cr)$, we let $S(cr)$ be the set of attributes of the vertices in $\group(cr)$. For an attribute $a_i$, if $a_i\in S(cr)$ and $|S(cr)|=1$ hold, we know that the group $\group(cr)$ only contains the vertices with the attribute $a_i$. For each attribute $a_i$, we maintain a counter $\cnt(a_i)$ to record the number of color groups that only contain vertices with  $a_i$. Clearly, $|S(cr)| > 1 $ indicates a mix group $\group(cr)$. The greedy algorithm greedily assigns $\group(cr)$ to the attribute with the minimum number of color groups. In other words, the algorithm increases the counter of $a_m$ by 1 where $a_m=\arg \min_{a_j\in S(cr)} \cnt(a_j)$. Finally, $GD(u)$ is obtained by taking the minimum counter over all attributes, i.e.,  $GD(u) = \min \{\cnt(a_i), a_i \in A_{val}\}$.

It is easy to see that the approximate fairness degree $GD(u)$ of a vertex $u$ is always no larger than the exact fairness degree of $u$, thus it cannot be directly used to prune vertices for strong fair clique enumeration. This is because $GD(u)$ is not an upper bound of the size of the strong fair cliques containing $u$. However, we can use the approximate fairness degrees to derive a good heuristic ordering, because the vertices with high exact fairness degrees tend to have high approximate fairness degrees. Such a heuristic ordering can be applied to reduce the search space for strong fair clique enumeration, as confirmed in our experiments. Specifically, to obtain the heuristic ordering denoted by \heurorder, we can iteratively delete the vertex with the minimum $GD$ (similar to the procedure of computing \colororder and \frorder). The pseudo-code of our greedy algorithm to generate \heurorder is given in Algorithm~\ref{alg:heuristic}. 

\begin{theo}
	Algorithm~\ref{alg:heuristic} takes $O((V+E)\times A_n\times {\kw{color}})$ using $O(V\times A_n\times {\kw{color}})$ space.
\end{theo}

\begin{proof}
	It is easy to derive that the time complexity to compute $GD$ for all vertices is $O(E+V\times {\kw{color}}\times A_n)$ (lines~3-13). The total cost to update the $GD$ in line~19 is  $O(E\times {\kw{color}}\times A_n)$. Therefore, the total time complexity is $O((V+E)\times A_n\times {\kw{color}})$. For the space complexity, the algorithm takes $O(V\times {\kw{color}}\times A_n)$ space to maintain all $S_u$, and $O(V)$ to maintain all $GD$s. Thus, the total space overhead of the algorithm is  $O(V\times A_n\times {\kw{color}})$.
\end{proof}

\stitle{The enumeration algorithm.} Algorithm~\ref{alg:strongenumeration} can be easily extended to handle the  high-dimensional case. Note that \fairness and \frorder in Algorithm \ref{alg:strongenumeration} do not work for the high-dimensional case. However, we can use \colorful (Algorithm \ref{alg:colorcoreprune}), which is designed for pruning unpromising vertices in weak fair clique enumeration, to reduce search space because a strong fair clique is always contained in a weak fair clique. In addition, we use the ordering \heurorder computed by Algorithm~\ref{alg:heuristic} for strong fair clique enumeration with $A_n > 2$. Clearly, the \strongbacktrack procedure with the attribute-alternatively-selection strategy in Algorithm \ref{alg:strongenumeration} can be directly applied to handle the $A_n > 2$ case. Therefore, we only need to slightly modify Algorithm~\ref{alg:strongenumeration} to enumerate strong fair cliques for the high-dimensional attributes. Specifically, in Algorithm \ref{alg:strongenumeration}, we use \colorful instead of \fairness to prune the unpromising vertices (line 2), and invoke Algorithm~\ref{alg:heuristic} to obtain the \heurorder ordering to reduce the search space (line 7).

\section{Relative fair clique enumeration} \label{sec:rfcsearch}
In this section, we first develop an enhanced pruning technique for the case of two-dimensional (2D) attributes to prune the unpromising vertices in the original graph. Then, two search frameworks with different strategies, namely, \relativeweak and \relativestrong, are proposed to enumerate relative fair cliques for both 2D and high-dimensional attributes. 

\subsection{The enhanced pruning technique for 2D case} \label{sec:enhancedpruning}
Suppose that the attributed graph $G = (V, E, A)$ with $A_{val} = \{a_1, a_2\}$, and we also divide the neighbors of a vertex $u$ into $h_u$ groups where each group contains vertices with the same color. Below, we define the enhanced colorful degree as follows.

\begin{defn}
	\label{def:relativefairdeg}
	(\kw{Enhanced} \kw{colorful} \kw{degree}) Given a colored attributed graph $G = (V, E, A)$ with $A_{val} = \{a_1, a_2\}$, the enhanced colorful degree of $u$, denoted by $ED(u)$, is the minimum number of groups that assigned to either to attribute $a_1$ or to attribute $a_2$.
\end{defn}

For a vertex $u$, as only one vertex in a group can be selected to form a clique with $u$, the number of groups assigned to an arbitrary attribute is no greater than the number of $u$'s neighbors with this attribute. And further, the enhanced colorful degree is no larger than the minimum colorful degree, thus it determines a tighter upper bound of the size of the relative fair clique containing $u$. By Definition~\ref{def:relativefairclique}, the enhanced colorful degree of any vertex in a relative fair clique is no less than $(k-1)$. Consequently, we can safely prune the vertex whose enhanced colorful degree is less than $(k-1)$. Below, we introduce an algorithm, called \enhancedcolordeg, to compute the enhanced colorful degree for a vertex $u$.

The pseudo-code of \enhancedcolordeg is outlined in Algorithm \ref{alg:enhancedcolorcal}. Similar to \fairdeg, we divide $h_u$ color groups into three categories, i.e., \agroupone, \agrouptwo and \agroupmix, and denote the number of the groups in these three categories by $c_1$, $c_2$, and $c_m$. The main idea of \enhancedcolordeg is to assign each color group in the \agroupmix to \agroupone or \agrouptwo when $c_1$ or $c_2$ is less than $k$. We take $c_1$ as an example. In the case of $c_1 < k$, if $c_m \ge k-c_1$ holds, we assign $k-c_1$ groups in \agroupmix to \agroupone (line 6); otherwise, we assign all groups in \agroupmix to \agroupone (line 8). For the groups in \agrouptwo with attribute $a_2$, we also use $c_m$ to expand $c_2$ as we expand $c_1$ (lines 9-13). Finally, we can easily derive that $ED(u)= \mathop{\min}{\{c_1, c_2\}}$.

Based on the enhanced colorful degree, we define the enhanced colorful $k$-core in the following.

\begin{defn}
	\label{def:enhancedcolorcore}
	(\kw{Enhanced} \kw{colorful} \kw{k}-\kw{core}) Given an attributed graph ${G}= (V, E, A)$ with $A_{val} = \{a_1, a_2\}$ and an integer $k$, a subgraph $H = (V_H, E_H, A)$ of $G$ is an enhanced colorful $k$-core if: (1) for each $u \in V_H$, $ED(u) \ge k$; (2) there is no subgraph $H' \subseteq G$ that satisfies (1) and $H \subset H'$.
\end{defn}

By Definition~\ref{def:enhancedcolorcore}, we hold the following lemma, that is, any relative fair clique must be contained in the enhanced colorful $(k-1)$-core. Due to the space limitation, we omit the proof of \lamref{lem:enhancedcolorcoreksfc} as it is similar to that of \lamref{lem:kccorekwfc} and \lamref{lem:faircoreksfc}. 

\begin{algorithm}[t]
	\scriptsize
	\caption{\enhancedcolordeg}
	\label{alg:enhancedcolorcal}
	{\bf Procedure} $\enhancedcolordeg(u, \group, k)$\\
	{
		$c_1 \leftarrow 0$; $c_2 \leftarrow 0$; $c_m \leftarrow 0$\;
		Calculate $c_1, c_2, c_m$ as lines 19-22 of Algorithm \ref{alg:fairnessreduction}\;
		\comment{
		\For{each color $cr$}
		{
			{\bf if} $\group(u, cr, a_1) \ge 1$ and $\group(u, cr, a_2)=0$ {\bf then } $c_1\leftarrow c_1+1$\;
			{\bf if} $\group(u, cr, a_2) \ge 1$ and $\group(u, cr, a_1)=0$ {\bf then } $c_2\leftarrow c_2+1$\;
			{\bf if} $\group(u, cr, a_1) \ge 1$ and $\group(u, cr, a_1)\ge 1$ {\bf then } $c_m\leftarrow c_m+1$\;
		}
		}	
		\If{$c_1 < k$}
		{
			\If{$c_m \ge (k-c_1)$}
			{	
				$c_1 \leftarrow k$; $c_m \leftarrow c_m-(k - c_1)$\;
			}
			\Else{
				$c_1 \leftarrow c_1+c_m$; $c_m \leftarrow 0$\;
			}
		}
		\If{$c_2 < k$}
		{
			\If{$c_m \ge (k-c_2)$}
			{	
				$c_2 \leftarrow k$; $c_m \leftarrow c_m-(k - c_2)$\;
			}
			\Else{
				$c_2 \leftarrow c_2+c_m$; $c_m \leftarrow 0$\;
			}
		}
		$ED(u) \leftarrow \mathop{\min}{\{c_1, c_2\}}$\;
		{\bf return } $ED(u)$\;
	}
\end{algorithm}

\begin{Lemma}
	\label{lem:enhancedcolorcoreksfc}
	Given an attributed graph $G = (V, E, A)$ with $A_{val} = \{a_1, a_2\}$ and a parameter $k$, any relative fair clique must be contained in the  enhanced colorful $(k-1)$-core of $G$.
\end{Lemma}

We also derive a peeling algorithm, i.e., \enhancedcolor, to compute the enhanced colorful $k$-core. The pseudo-code of \enhancedcolor is similar to that of \colorful (Algorithm \ref{alg:fairnessreduction}) and we only need to make slightly modifying as follows. Specifically, in line 6, we perform the procedure \enhancedcolordeg (Algorithm \ref{alg:enhancedcolorcal}) instead of \fairdeg to calculate the enhanced colorful degrees of all vertices. In line 7, we modify the condition to be $ED(u)<k$ to add the vertices with initial enhanced colorful degrees less than $k$ to the queue $Q$. Then, we iteratively remove the vertices with the enhanced colorful degrees less than $k$, and maintain the enhanced colorful degrees for their neighbors and the queue $Q$ (line 14). Due to the space limitation, we omit the pseudo-code of \enhancedcolor.

\begin{example}
	Reconsider the attributed graph in \figref{fig:expgraphnew}. Suppose that we search all relative fair cliques with $k=4$. We need to calculate $3$-colorful core or $3$-enhanced colorful core first. Take vertex $v_9$ as an example. $v_9$ has four neighbors with attribute $a$, i.e., $v_{10}, v_{11}, v_{13}$ and $v_{14}$, and three neighbors with attribute $b$, i.e., $v_5, v_6$ and $v_{12}$. Based on Definition \ref{def:colorfuldeg}, we have $D_a(v_9) = 4$ and $D_b(v_9) = 3$, and further $D_{\min}(v_9) = D_b(v_9) = 3$. Due to $D_{\min}(v_9) = 3$, $v_9$ cannot be removed according to the colorful core pruning technique (Definition \ref{def:colorfulcore}). However, $v_9$ is not contained in any $4$-relative fair clique. This is because $v_6$ with attribute $b$ and $v_{10}$ with attribute $a$ have the same color (green), that is, there are no edge between them, thus $v_6$ and $v_{10}$ cannot coexist in a clique. Analogously, the neighbors colored yellow, i.e., $v_{12}$ with attribute $b$ and $v_{14}$ with attribute $a$, also cannot form a clique. While considering the enhanced colorful degree, we have $ED(u_9)=2$. Clearly, $ED(u_9)=2<3$, thus \enhancedcolor can safely remove $v_9$ from $G$. Hence, the enhanced colorful degree has a stronger pruning effect than the colorful degree. \enhancedcolor repeatedly removes vertices until all the remaining vertices satisfying $ED(*) \ge 3$. Finally, we can obtain an enhanced colorful $3$-core induced by $\{v_1, v_2, v_3, v_4, v_5, v_6, v_7, v_8\}$. \eop	
\end{example}

\begin{theo}
\label{the:enhancedcolor}
	The \enhancedcolor algorithm consumes $O((E+V)\times {\kw{color}})$ time using $O(V\times \kw{color})$ space.
\end{theo}

As aforementioned, the \enhancedcolor algorithm is devised by slightly modifying the \colorful (Algorithm \ref{alg:fairnessreduction}), thus the proof of the complexity analysis for \enhancedcolor is similar to that of \colorful. Here, we omit the proof details due to the limited space.

\stitle{Remark.} Note that the \enhancedcolor pruning technique is more efficient than \colorful, because the enhanced colorful degree provides a tighter upper bound on the minimum number of neighbors of $u$ for arbitrary attributes benefitting from the property of graph coloring. In addition, the \enhancedcolor can work on all proposed fairness-aware clique models in the case of 2D attributes. Specifically, in weak fair clique enumeration, we can use \enhancedcolor instead of \colorful to achieve a stronger pruning effect in line 2 of Algorithm \ref{alg:weakenumeration}. In Algorithm \ref{alg:fairnessreduction} for strong fair clique enumeration, we can also apply \enhancedcolor in line 1. For the relative fair clique search, we will introduce the enumeration algorithms in the following subsections which are also equipped with the \enhancedcolor pruning technique.

\begin{algorithm}[t]
	\scriptsize
	\caption{\relativeweak}
	\label{alg:relativeweakenumeration}
	\KwIn{$G = (V, E, A)$, two integers $k$ and $\delta$}
	\KwOut{The set of relative fair cliques $Res$}
	$Res \leftarrow \emptyset$; $C \leftarrow \emptyset$;\\
	{\bf {if}} $\delta=0$ {\bf then} $Res \leftarrow \strong(G, k)$; {\bf return} $Res$\;
	{\bf {if}} $|A_{val}|=2$ {\bf then} $\hat{G}=(\hat{V}, \hat{E}) \leftarrow \enhancedcolor(G, k-1)$\;
	{\bf {else}} $\hat{G}=(\hat{V}, \hat{E}) \leftarrow \colorful(G, k-1)$\;
	$C \leftarrow \weak(\hat{G}, k)$\;
	\For{$C_i \in C$}
	{
		\For{$a_i \in A_{val}$}{
			$V(a_i) \leftarrow \emptyset$; $Cnt({a_i}) \leftarrow 0$\;	
		}
		\For{$u \in C_i$}{
			$Cnt({u.val})$++; $V(u.val) \leftarrow V(u.val) \cup \{u\}$\;
		}
		$a_{min} \leftarrow \mathop{\arg\min}\limits_{a_i \in A_{val}}{Cnt({a_i})}$; $a_{max} \leftarrow a_{min}+\delta$\;
		$L_A \leftarrow \{a_i \in A _{val}|Cnt({a_i})>a_{max}\}$\;
		\If{$L_A = \emptyset$}{
			$Res \leftarrow C_i$; {\bf continue}\;
		}
		${\cal V}({L_A}) \leftarrow \emptyset$; 
		${\cal V}({L_A}) \leftarrow {\cal V}({L_A}) \cup \{V(a_i)|a_i \in L_A\}$\;
		${\cal V}({C_P}) \leftarrow \emptyset$; 
		${\cal V}({C_P}) \leftarrow {\cal V}({C_P}) \cup \{V(a_i)|a_i \notin L_A\}$\;
		Let $a_s$ be the first attribute element in $L_A$\;
		$\rweakenum({\cal V}({C_P}), {\cal V}({L_A}), L_A, a_{max}, Res, a_s, 0)$\;
	}
	{\bf return} $Res$\;
	
	\vspace*{0.1cm}
	{\bf Procedure} $\rweakenum(C, {\cal V}({L_A}), L_A, a_{max}, Res, a_c, cnt_c)$\\
	{	
		\If{$a_c$ is the last attribute element in $L_A$}{
			$Res \leftarrow Res \cup C$; {\bf return}\;
		}
		\For{$u \in V(a_c)$}{
			$C \leftarrow C \cup \{u\}$; $V(a_c) \leftarrow V(a_c)-\{u\}$\;
			\If{$cnt_c+1<a_{max}$}{
				$\rweakenum(C, {\cal V}({L_A}), L_A, a_{max}, Res, a_c, cnt_c+1)$\;
			}
			\Else{
				$a_{nc} \leftarrow$ the next attribute element of $a_c$ in $L_A$\;
				$\rweakenum(C, {\cal V}({L_A}), L_A, a_{max}, Res, a_{nc}, 0)$\;
			}
			$C \leftarrow C-\{u\}$; $V(a_c) \leftarrow V(a_c) \cup \{u\}$\;
		}
	}
\end{algorithm}

\subsection{The \relativeweak algorithm} \label{sec:onlineupdate}
Reviewing Definition \ref{def:relativefairclique}, a relative fair clique must be contained in weak fair cliques. Therefore, a feasible idea is to find all the weak fair cliques, and then enumerate the relative fair cliques contained in them. Following this idea, we propose an algorithm, called \relativeweak, which is shown as Algorithm \ref{alg:relativeweakenumeration}. 

The \relativeweak algorithm works as follows. If $\delta=0$, it performs \strong to find all relative fair cliques since the relative fair clique model is equivalent to the strong fair clique model in this case (line 2); otherwise, the \relativeweak performs \enhancedcolor or \colorful to prune the original graph for 2D or high-dimensional attributes (lines 3-4). Then it invokes \weak to find all weak fair cliques and refines relative fair cliques contained in them (lines 5-18). For each weak fair clique $C_i$, the \relativeweak computes the number of vertices $Cnt(a_i)$ for each attribute $a_i$, and identifies the minimum $Cnt(a_i)$ as $a_{min}$. Based on $a_{min}$ and $\delta$, at most how many vertices of each attribute has in a relative fairness clique is determined, which we denoted by $a_{max}$ (lines 7-11). The algorithm then collects those attributes with the number of vertices greater than $a_{max}$ into $L_A$, which we call the lacking attribute set (line 12). Clearly, if $L_A$ is empty, the current weak fair clique is a $(k, \delta)$-relative fair clique, and we add it into the result set $Res$ (line 14). In the negative case, \relativeweak refines the vertices with lacking attributes and non-lack attributes into ${\cal V}(L_A)$ and ${\cal V}(C_P)$, respectively (lines 15-16). It then selects a lacking attribute $a_s \in L_A$ and performs the \rweakenum procedure to expand the partial clique induced by ${\cal V}(C_P)$  to search all relative fair cliques (lines 17-18).

\begin{algorithm}[t]
	\scriptsize
	\caption{\relativestrong}
	\label{alg:relativestrongenumeration}
	\KwIn{$G = (V, E, A)$, two integers $k$ and $\delta$}
	\KwOut{The set of relative fair cliques $Res$}
	$Res \leftarrow \emptyset$; $R \leftarrow \emptyset$; $X \leftarrow \emptyset$; $C \leftarrow \emptyset$\;
	{\bf {if}} $\delta=0$ {\bf then} $Res \leftarrow \strong(G, k)$; {\bf return} $Res$\;
	{\bf {if}} $|A_{val}|=2$ {\bf then} $\hat{G}=(\hat{V}, \hat{E}) \leftarrow \enhancedcolor(G, k-1)$\;
	{\bf {else}} $\hat{G}=(\hat{V}, \hat{E}) \leftarrow \colorful(G, k-1)$\;
	Initialize an array $B$ with $B(i) = false, 1 \le i \le |\hat{V}|$\;
	\For{$u \in {\hat V}$}
	{
		\If{$B(u) = false$}
		{	
			$C \leftarrow \connectedcpn(u, B)$\;
			${\mathcal{O}} \leftarrow \text{\wforder}(C)$\;
			$R \leftarrow \emptyset$; $X \leftarrow \emptyset$\; 
			$\rstrongenum(R, C, X, {\mathcal {O}}, a_0, -1)$;
		}
	}
	{\bf return} $Res$\;
\end{algorithm}

In the \rweakenum procedure, there are two important parameters: $a_c$ and $cnt_c$. The parameter $a_c$ indicates that the current round needs to pick a vertex with the lacking attribute $a_c$ into the partial clique $C$. And $cnt_c$ is used to record the number of vertices with the lacking attribute $a_c$. In each recursion of \rweakenum, it tries to add each vertex $u$ with attribute $a_c$ to $C$ to perform a deeper search for relative fair clique enumeration (lines 23-30). If $cnt_c+1 < a_{max}$, that means the number of vertices with $a_c$ in $C$ has not yet reached $a_{max}$, thus we perform the \rweakenum with the parameters: $a_c$ and $cnt_c+1$ (line 26). On the other hand, once the number of vertices with $a_c$ in $C$ is up to $a_{max}$, we invoke the \rweakenum to select vertices for the next lacking attribute $a_{nc}$ with $cnt_c$ equals 0 (lines 28-29). When all lacking attributes in $L_A$ are processed, a relative fair clique $C$ is found and the \rweakenum adds it into the result set $Res$ (lines 20-21). 

\subsection{The \relativestrong algorithm} \label{sec:onlineupdate}
The \relativeweak algorithm is not very efficient for relative fair clique enumeration because a relative fair clique may be contained in many weak fair cliques, which causes a lot of repeated enumeration calculation in \relativeweak. To solve this issue, we propose the \relativestrong algorithm which applies the attribute-alternative-selection search method in \strong to find all relative fair cliques.

The \relativestrong algorithm is outlined in Algorithm \ref{alg:relativestrongenumeration}. Similar to \weak and \strong, $R$ is the currently-found clique and $C$ is the candidate set that can be used to extend $R$. All the relative fair cliques are stored in the set $Res$. To avoid the repeated enumeration, we still use the set $X$ to maintain the vertices that can be used to expand the current clique $R$ but have already been visited in previous search paths. The \relativestrong algorithm performs \strong directly to find all relative fair cliques for $\delta=0$ like the \relativeweak (line 2). In other cases, it first removes the vertices that are definitely not contained in any relative fair clique with the pruning techniques. For the graph $G$ with two types of attributes, that is, $|A_{val}| = 2$, \relativestrong performs \enhancedcolor to prune the original graph (line 3), and \colorful is called for high-dimensional attributes (line 4). Then, the \relativestrong alternatively selects a vertex of a specific attribute in each backtracking round to enumerate all relative fair cliques, i.e., the \rstrongenum procedure (lines 6-11). 

\begin{algorithm}[t]
	\scriptsize
	\caption{\rstrongenum}
	\label{alg:relativestrongbacktrack}
	{\bf Procedure} $\rstrongenum(R, C, X, {\mathcal O}, a_{\phi}, a_{max})$\\
	{\bf for} $u \in C$ {\bf do} $C_A(u.val) \leftarrow C_A(u.val) \cup u$\;
	{\bf for} $u \in R$ {\bf do} $R_A(u.val) \leftarrow R_A(u.val) \cup u$\;
	\If{$|C_A(a_{\phi})|=0$ and $a_{max}=-1$}
	{
		$a_{min} \leftarrow |R_A(a_{\phi})|$;
		$a_{max} \leftarrow a_{min}+\delta$\;
	}
	\For{$a_i \in A_{val}$}
	{
	{\bf {if}} $|R_A(a_i)|=a_{max}$ {\bf then} $C \leftarrow C-C_A(a_i)$; $C_A(a_i) \leftarrow \emptyset$\;
	}
	\If{$C=\emptyset$}
	{
		$isMaximal \leftarrow true$\;
		\If{$X \neq \emptyset$}
		{
			$a_{min} \leftarrow \mathop{\min}_{a_i \in A_{val}}{|R_A(a_i)|}$\;
			\For{$u \in X$}
			{
				\If{$u.val=a_{min}$ or $|R_A(u.val)|+1<=a_{max}$}{
					$isMaximal \leftarrow false$; {\bf break}\; 	
				}
			}	
		}
		{\bf {if}} $isMaximal=true$ {\bf then} $Res \leftarrow Res \cup R$; {\bf return}\;
	}
	\If{$C_A(a_\phi)=\emptyset$}{$\rstrongenum({R}, {C}, {X}, {\mathcal {O}}, a_{\phi+1}, a_{max}$); {\bf return}\;}

		\For{$u \in C_A(a_{\phi})$}
		{
			$\hat{R} \leftarrow R \cup u$; ${\hat C} \leftarrow \emptyset$; $flag \leftarrow false$\;
			\For{$v \in C$}
			{
				\If{$v \in N(u)$ and $\mathcal{O}(v) > \mathcal{O}(u)$}
				{
					${\hat C} \leftarrow$ ${\hat C} \cup v$; ${{\hat C}_{\cnt}}(v.val)$++\;
				}
			}
			{\bf {if}} $|{\hat C}| + |{\hat R}| < k * A_n$ {\bf then continue}\;
			{\bf {for}} $v \in {\hat R}$ {\bf {do}} ${{\hat R}_{\cnt}}(v.val)$++\;
			\For{$a_i \in A_{val}$}
			{
				\If{${{\hat R}_{\cnt}}(a_i) + {{\hat C}_{\cnt}}(a_i) < k$}
				{
					$flag \leftarrow true$; {\bf break};\
				}
			}
			{\bf if} $flag = true$ {\bf then continue}\;

			${\hat X} \leftarrow X \cap N(u)$\;

			$\rstrongenum({\hat R}, {\hat C}, {\hat X}, {\mathcal {O}}, a_{\phi+1}, a_{max}$)\;
			$X \leftarrow X \cup u$\;
		}
\end{algorithm}

The workflow of the \rstrongenum procedure is depicted in Algorithm \ref{alg:relativestrongbacktrack}. The input parameter $a_\phi$ is used to indicate the attribute value of the vertices to be selected in the current iteration. $a_{max}$ is the upper bound of the number of vertices for an arbitrary attribute $a_i$ in the current search space, which is initialized to $-1$ (line 11 in Algorithm \ref{alg:relativestrongenumeration}). In each iteration with attribute $a_{\phi}$, the \rstrongenum procedure first divides the vertices in the candidate set $C$ and the current partial clique $R$ into $A_n$ collections according to their attributes, respectively (lines 2-3). For the specified $a_{\phi}$, if the current candidate set has no vertex with $a_{\phi}$ and $a_{max}$ is equal to the initial $-1$, that means the lower bound of the number of vertices for an arbitrary attribute $a_i$ is determined. And further, $a_{max}$ is also fixed based on the difference threshold $\delta$ (lines 4-5). The \rstrongenum then identifies whether the number of vertices for attribute $a_i \in A_{val}$ in the current clique $R$ has reached $a_{max}$. In the affirmative case, adding any vertex with $a_i$ to $R$ would violate the definition of a relative fair clique, and thus the procedure removes all vertices with $a_i$ from the candidate set $C$ (lines 6-7). Since $a_{\phi}$ is specified for the current round, for each candidate $u$ in $C_A(a_{\phi})$, the \rstrongenum picks one vertex at a time to add to the currently-found clique and call itself to perform a deeper search for the next attribute $a_{\phi+1}$ (lines 18-31). Note that if $C_A(a_{\phi})$ is empty, the \rstrongenum directly invokes a recursion by specifying the attribute $a_{\phi+1}$ (lines 16-17).

\stitle{Maximality checking.} Once the candidate set $C$ is empty, we check the maximality of $R$. As previously mentioned, the vertices in $X$ can expand $R$ but have already been visited in previous search paths. Thus, we check the maximality by adding each vertex in $X$ to $R$ (lines 8-15). A variable $isMaximal$, initialized as true, is used to indicate whether $R$ is a relative fair clique (line 9). Consider a vertex $u \in X$, the maximality checking is discussed in two aspects according to whether the attribute of $u$ is the attribute with the least number of vertices in $R$ (line 13). In the case of $u.val=a_{min}$, adding $u$ can increase $a_{min}$ by 1 to obtain a larger relative fair clique. Therefore, $R$ is not an answer because it does not satisfy maximality, i.e., the condition (3) in Definition \ref{def:relativefairclique}. On the other hand, that is, $u.val \neq a_{min}$, the \rstrongenum procedure identifies whether the number of vertices in $R$ with the attribute $u.val$ is up to $a_{max}$. If no, adding $u$ into $R$ still satisfies the definition of a relative fair clique, thus $R$ is not an answer due to the violation of the maximality. Once there is a vertex $u$ that can make $R$ break the maximality, we set the variable $isMaximal$ to false. After checking all the vertices in the set $X$, the \rstrongenum adds $R$ into the answer set $Res$ if $isMaximal$ equals true. 

\section{Experiments} \label{sec:experiments}

\subsection{Experimental setup} \label{sec:setup}

\begin{table}[t!]
\caption{Datasets}
\label{tab:datasets}
\vspace*{-0.3cm}
\begin{center}
{
    \scriptsize
\begin{tabular}{c|c|c|c|c} \hline
	{\bf Dataset} & $n=|V|$ & $m=|E|$ & $d_{\max}$ &Description\\ \hline
	\slashdot &	82,169 & 504,230 &2,252 & Social network\\
    \themarker & 69,414	& 1,644,843	& 8,930 & Social network\\
    \wiki &	2,394,385	& 5,021,410 & 100,029 & Communication network\\
	\flixster &	2,523,387	&	7,918,801 & 1,474 & Social network\\
	\hline
\end{tabular}
}
\end{center}
\vspace*{-0.4cm}
\end{table}

\begin{figure*}[t!]
	\begin{center}
		\begin{tabular}[t]{c}
			\subfigure[{\scriptsize \slashdot (vary $k$)}]{
				\includegraphics[width=0.4\columnwidth, height=2.5cm]{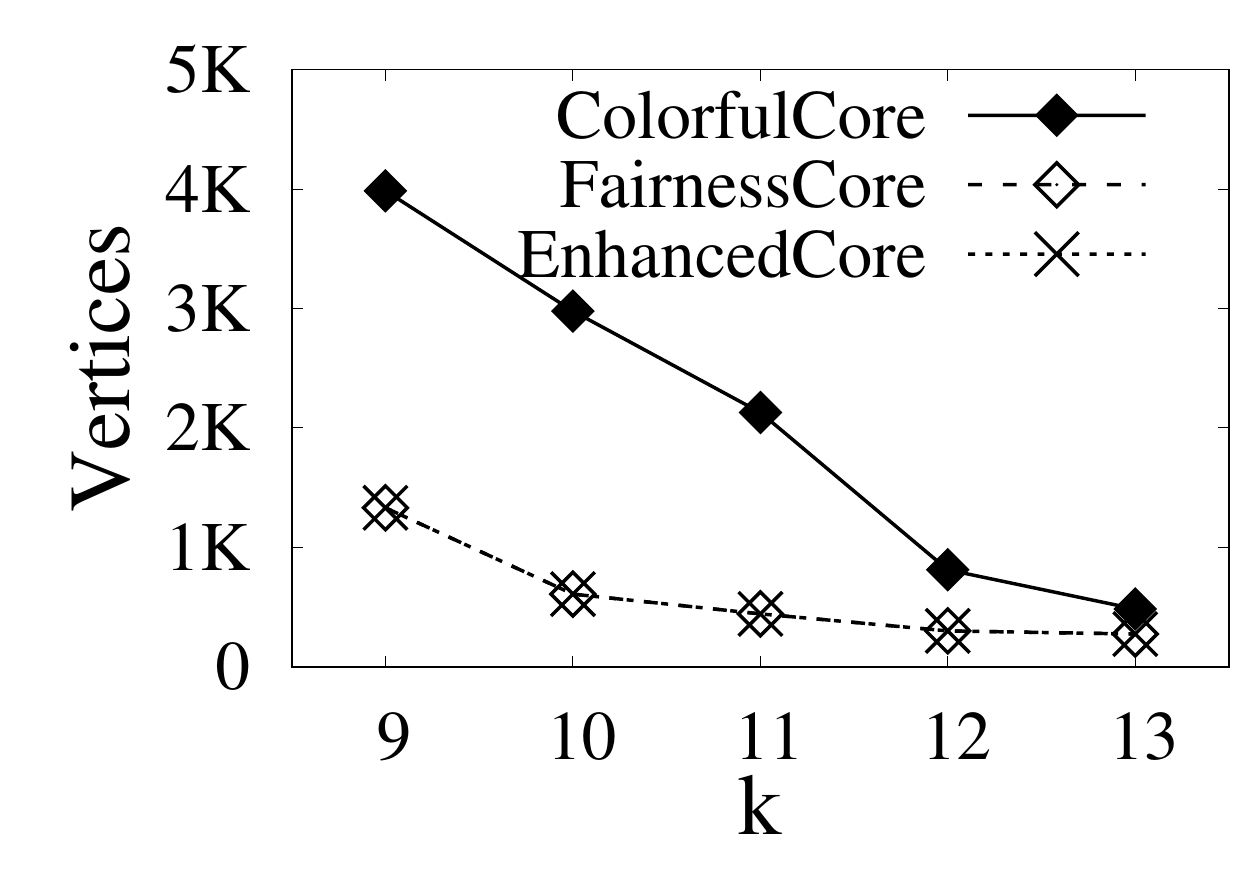}
			}
			\subfigure[{\scriptsize \themarker (vary $k$)}]{
				\includegraphics[width=0.4\columnwidth, height=2.5cm]{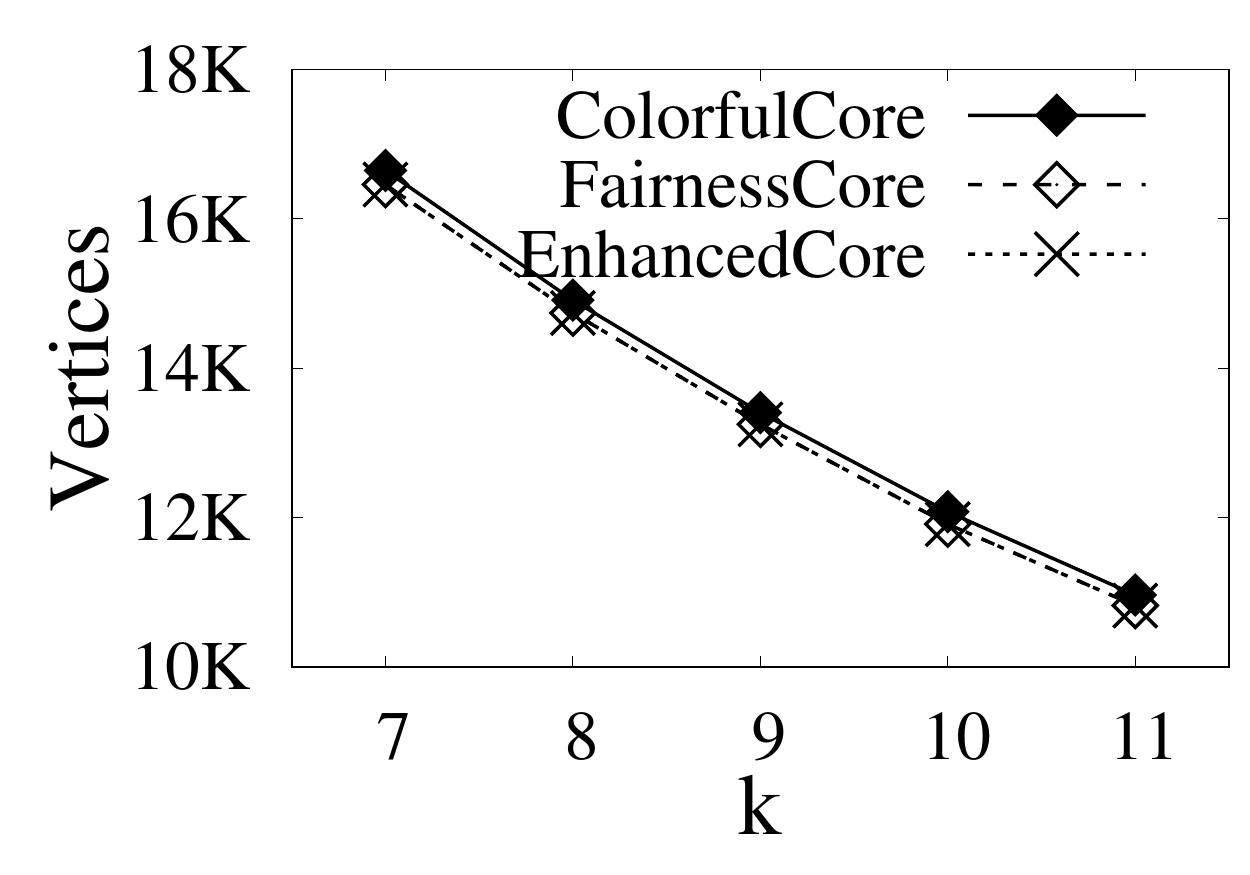}
			}
			\subfigure[{\scriptsize \wiki (vary $k$)}]{
				\includegraphics[width=0.4\columnwidth, height=2.5cm]{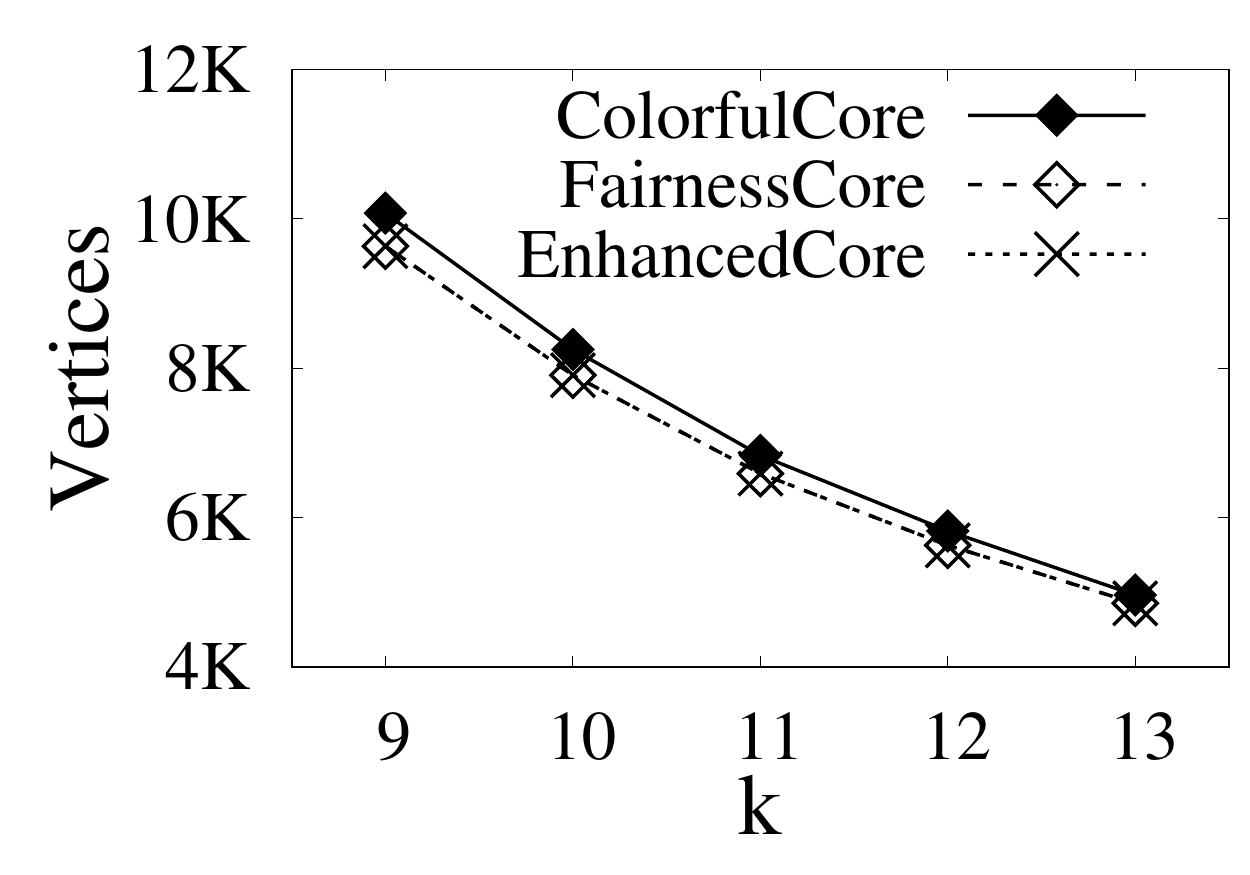}
			}
			\subfigure[{\scriptsize \flixster (vary $k$)}]{
				\includegraphics[width=0.4\columnwidth, height=2.5cm]{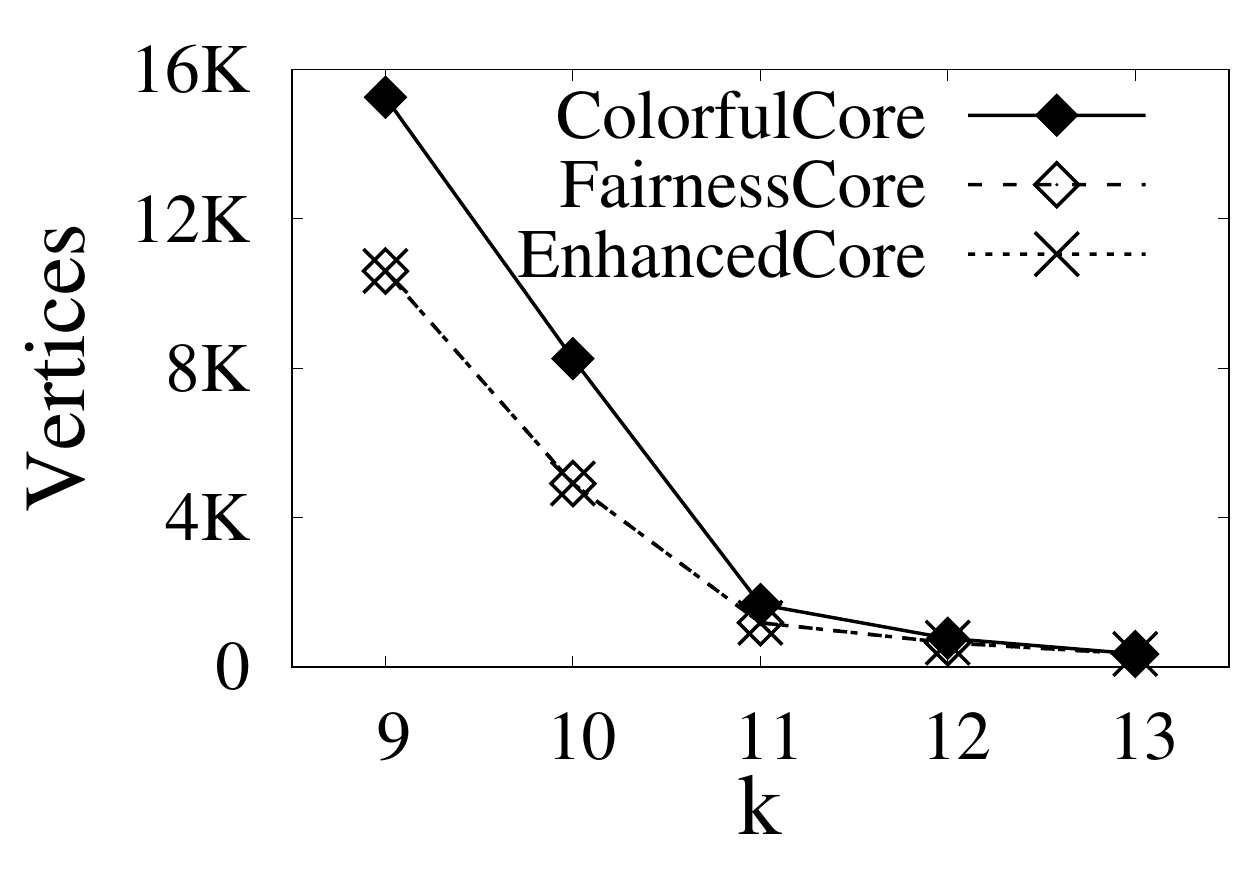}
			}
			\vspace*{-0.2cm} \\
			
			\subfigure[{\scriptsize \slashdot (vary $d$)}]{
				\includegraphics[width=0.4\columnwidth, height=2.5cm]{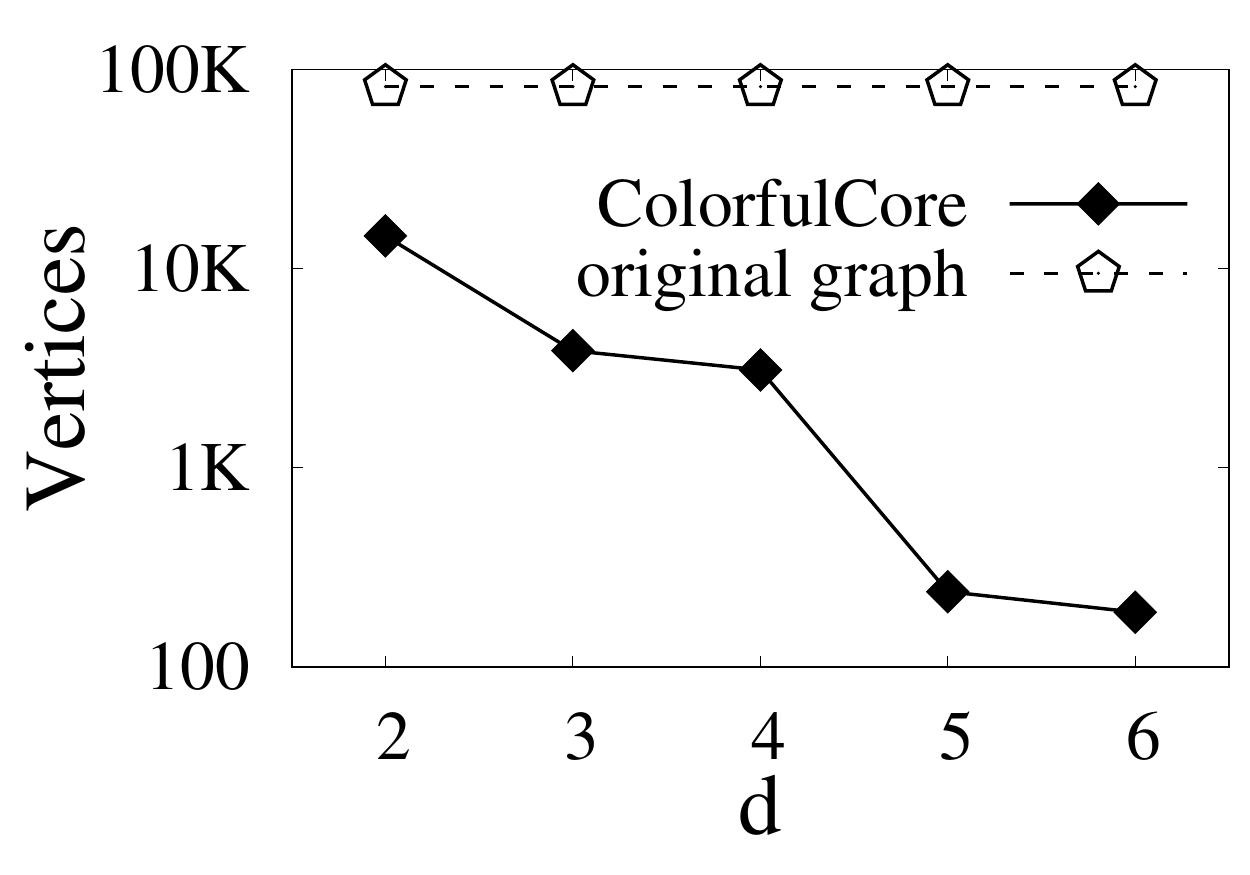}
			}
			\subfigure[{\scriptsize \themarker (vary $d$)}]{
				\includegraphics[width=0.4\columnwidth, height=2.5cm]{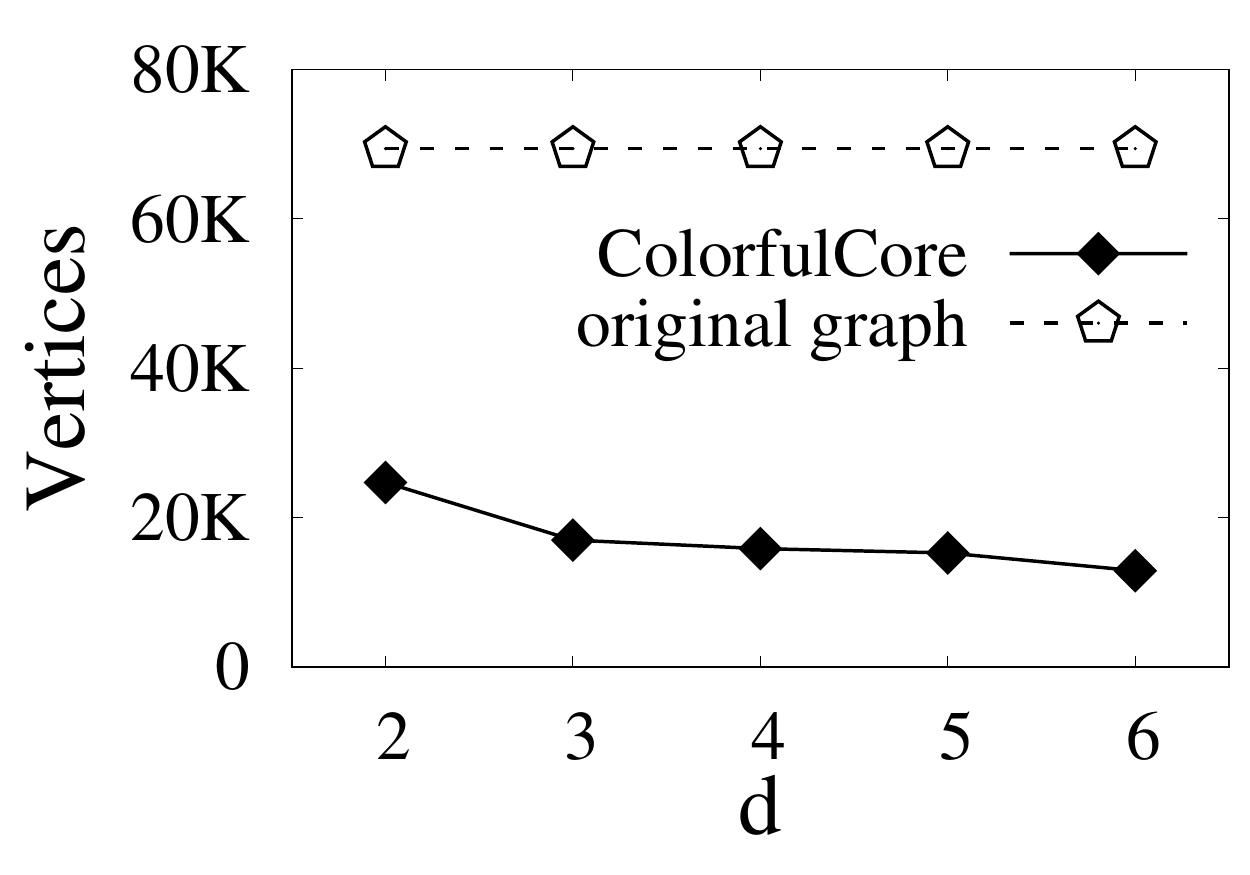}
			}
			\subfigure[{\scriptsize \wiki (vary $d$)}]{
				\includegraphics[width=0.4\columnwidth, height=2.5cm]{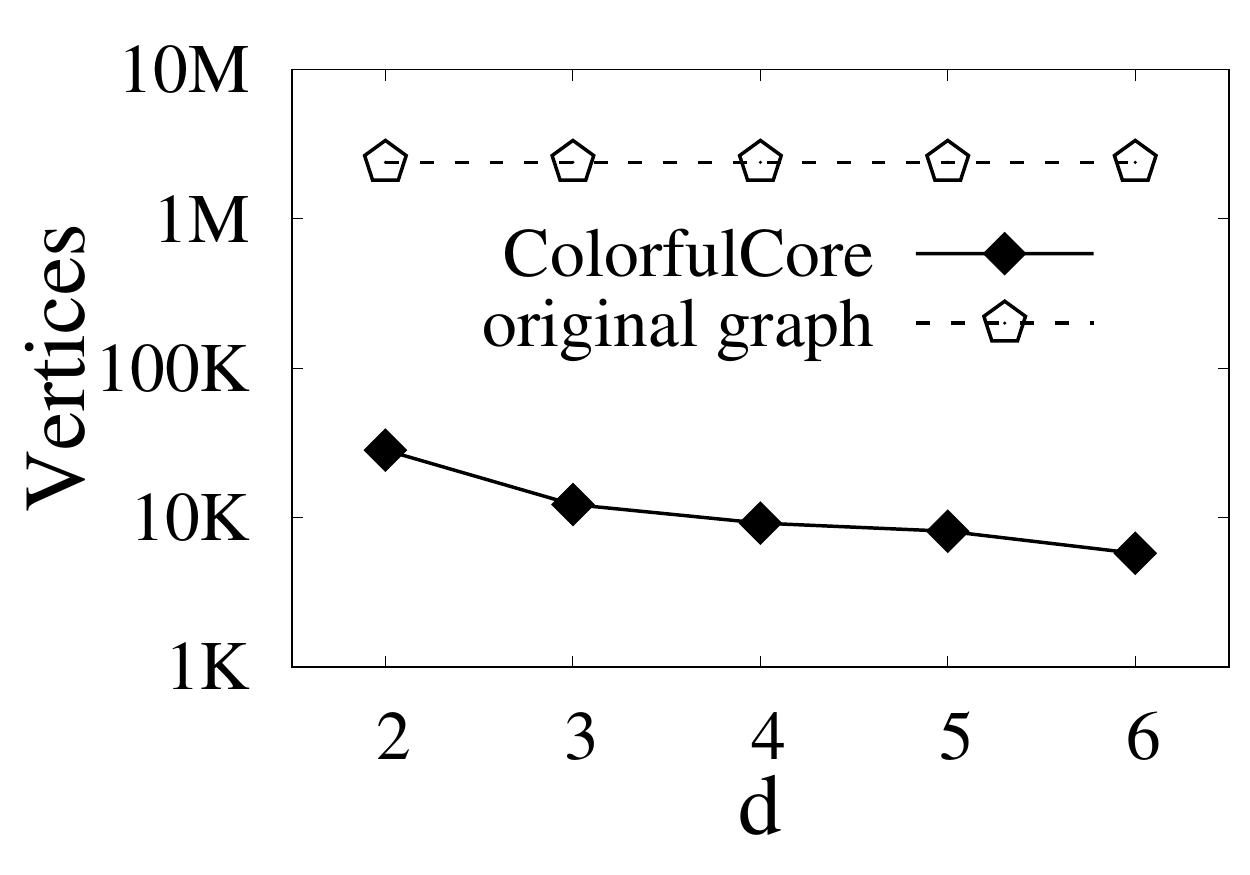}
			}
			\subfigure[{\scriptsize \flixster (vary $d$)}]{
				\includegraphics[width=0.4\columnwidth, height=2.5cm]{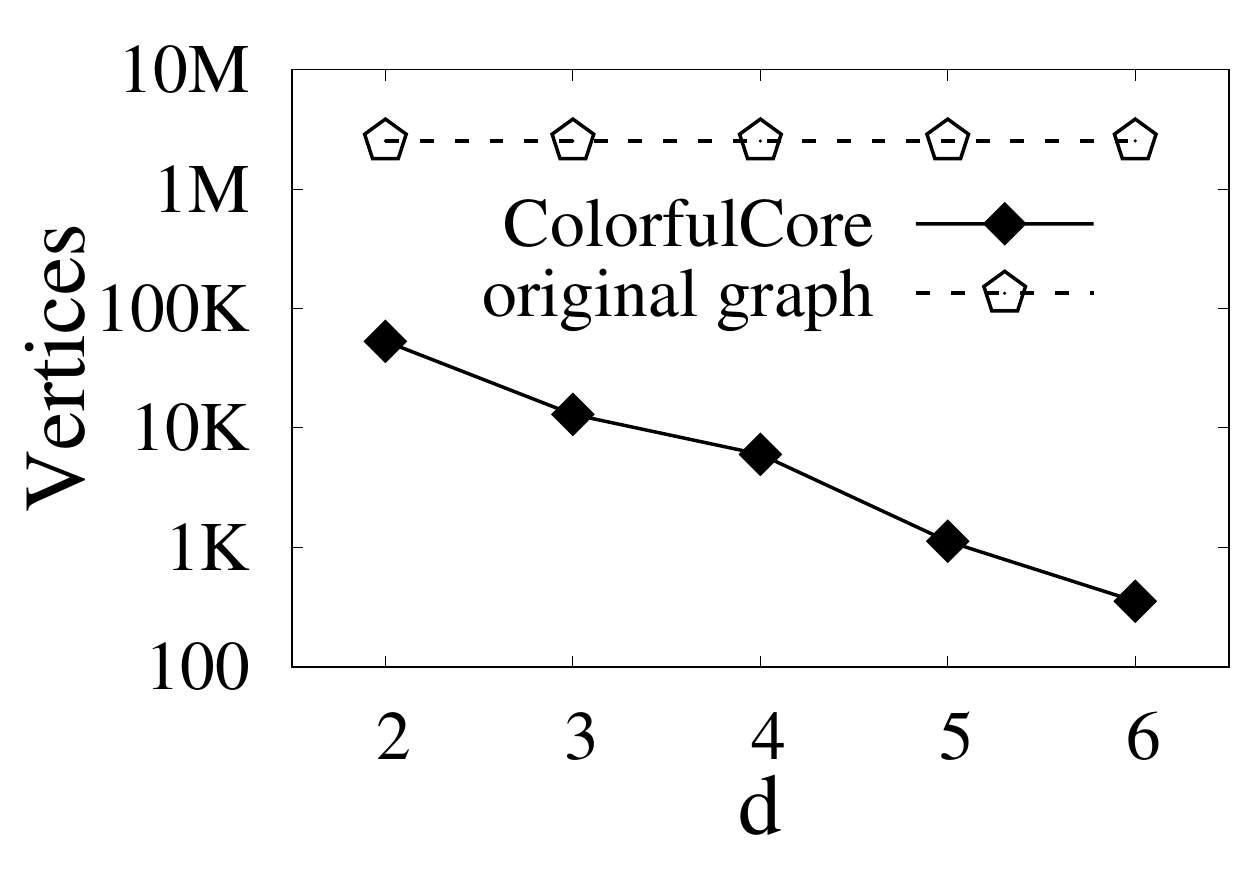}
			}
		\end{tabular}
	\end{center}
	\vspace*{-0.4cm}
	\caption{The number of remaining vertices after performing \colorful, \enhancedcolor and \fairness}
	\vspace*{-0.5cm}
	\label{fig:exp:Core}

\end{figure*}

We implement \weak (Algorithm \ref{alg:weakenumeration}) for weak fair clique enumeration. For strong fair clique enumeration, we implement \strong (Algorithm \ref{alg:strongenumeration}) equipped with 1) the pruning technique \fairness (Algorithm \ref{alg:fairnessreduction}) and the ordering \frorder for the 2D case; and 2) the pruning technique \colorful and the heuristic ordering \heurorder calculated by Algorithm \ref{alg:heuristic} for the high-dimensional case. For relative fair clique enumeration, we implement \relativeweak (Algorithm \ref{alg:relativeweakenumeration}) and \relativestrong (Algorithm \ref{alg:relativestrongenumeration}) equipped with the pruning techniques \enhancedcolor and \colorful for 2D and high-dimensional cases. Since there is no existing algorithm that can be directly used to enumerate fairness-aware cliques, we implement three baseline algorithms, called \baseweak, \basestrong and \baserelative. For the weak (relative) fair clique enumeration, \baseweak (\baserelative) first finds all maximal cliques using the state-of-the-art Bron-Kerbosch algorithm with pivoting technique \cite{bron1973algorithm,tomita2006worst}, and then filters them based on attribute constraint to identify weak (relative) fair cliques. For the strong fair clique enumeration, \basestrong enumerates all cliques with size larger than $k\times A_n$, and then selects the strong fair cliques among them based on the attribute and maximality constraints. In addition, we also introduce two different basic orderings for our fairness-aware clique enumeration algorithms. The first ordering, called \bfsorder, is obtained by performing breadth-first search (BFS) to explore the graph (i.e., the BFS visiting ordering of vertices); and the second ordering, called \idorder, is obtained by sorting the vertices based on the vertices' IDs. We compare the \baseweak (\basestrong) with the \weak (\strong) algorithms equipped with different orderings, i.e., \bfsorder, \idorder and our proposed orderings. All algorithms are implemented in C++. We conduct all experiments on a PC with a 2.10GHz Inter Xeon CPU and 256GB memory. We set the time limit for all algorithms to $3$ hours, and use the symbol ``INF'' to denote that the algorithm cannot terminate within 3 hours.

\stitle{Datasets.} We make use of four real-world graphs to evaluate the efficiency of the proposed algorithms. \tabref{tab:datasets} summarizes the statistics of the datasets in our experiments. \wiki is a communication network. \themarker, \slashdot and \flixster are social networks. All datasets can be downloaded from \url{networkrepository.com/} and \url{snap.stanford.edu}. Note that all these datasets are non-attributed graphs, thus we randomly assign an attribute to each vertex to generate attributed graphs which will be used to evaluate the efficiency of all algorithms. 
	
\stitle{Parameters.} There are two parameters in our weak fair clique enumeration and strong fair clique enumeration algorithms: $k$ and $d=A_n$. The parameter $k$ is the threshold for fair cliques and $d$ is the number of attribute values (i.e., the attribute dimension). For the relative fair clique search algorithms, there is an extra parameter $\delta$ which is the maximum difference in the number of vertices of the attribute in addition to $k$ and $d$. Since different datasets have various scales, the parameter $k$ is set within different integers. For \themarker, $k$ is chosen from the interval $[7, 11]$ with a default value of $k = 4$. For the other datasets, $k$ is chosen from the interval $[9, 13]$ with a default value $k=5$. The parameter $d$ is chosen from the interval $[2, 6]$ with a default value of $d = 2$. The parameter $\delta$ is selected from the interval $[1, 5]$ with a default value of $\delta = 3$. Unless otherwise specified, the values of the other parameters are set to their default values when varying a parameter.

\subsection{Efficiency testing} \label{sec:results}

\begin{figure*}[t!] 
	\begin{center}
		\begin{tabular}[t]{c}
            \subfigure[{\scriptsize \slashdot (vary $k$)}]{
				\includegraphics[width=0.4\columnwidth, height=2.5cm]{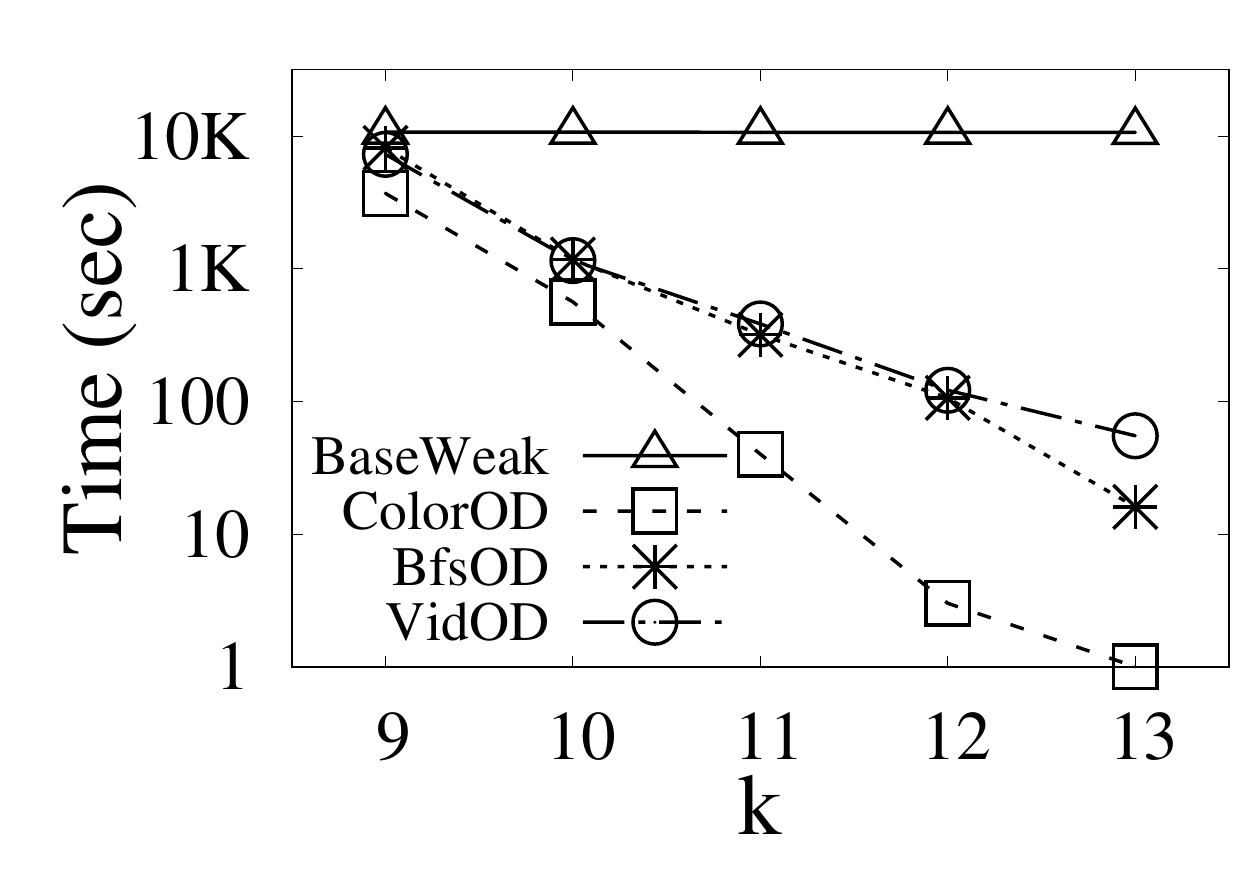}
			}
			\subfigure[{\scriptsize \themarker (vary $k$)}]{
				\includegraphics[width=0.4\columnwidth, height=2.5cm]{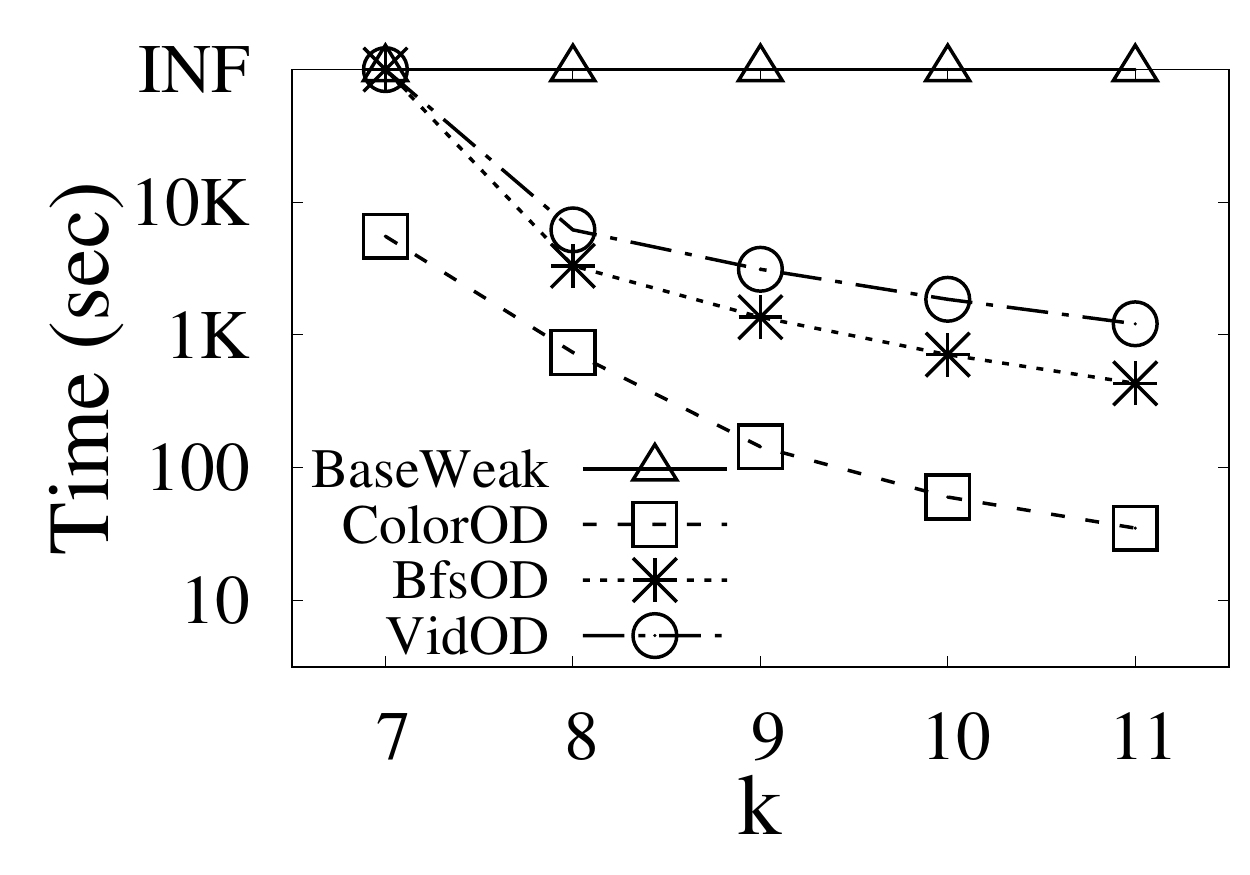}
			}
			\subfigure[{\scriptsize \wiki (vary $k$)}]{
				\includegraphics[width=0.4\columnwidth, height=2.5cm]{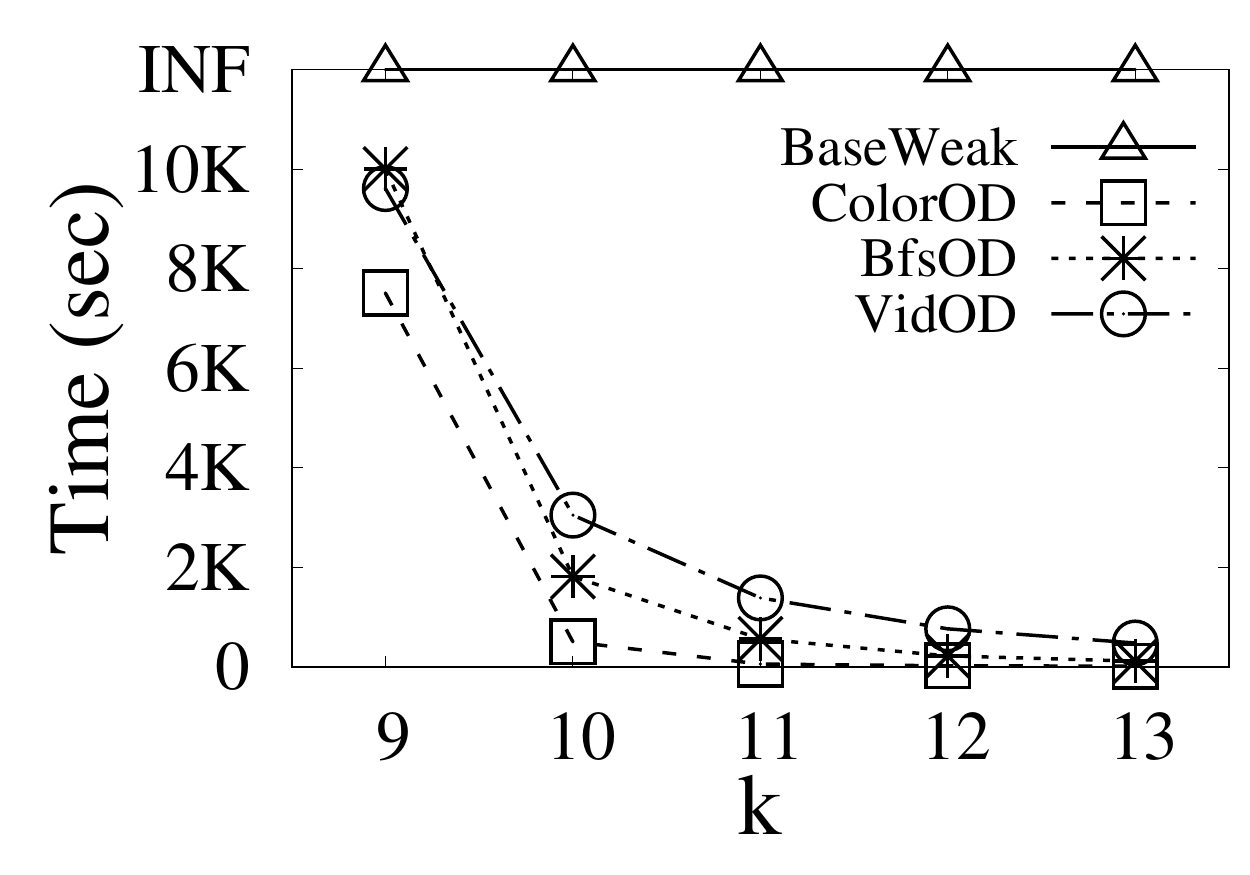}
			}
			\subfigure[{\scriptsize \flixster (vary $k$)}]{
				\includegraphics[width=0.4\columnwidth, height=2.5cm]{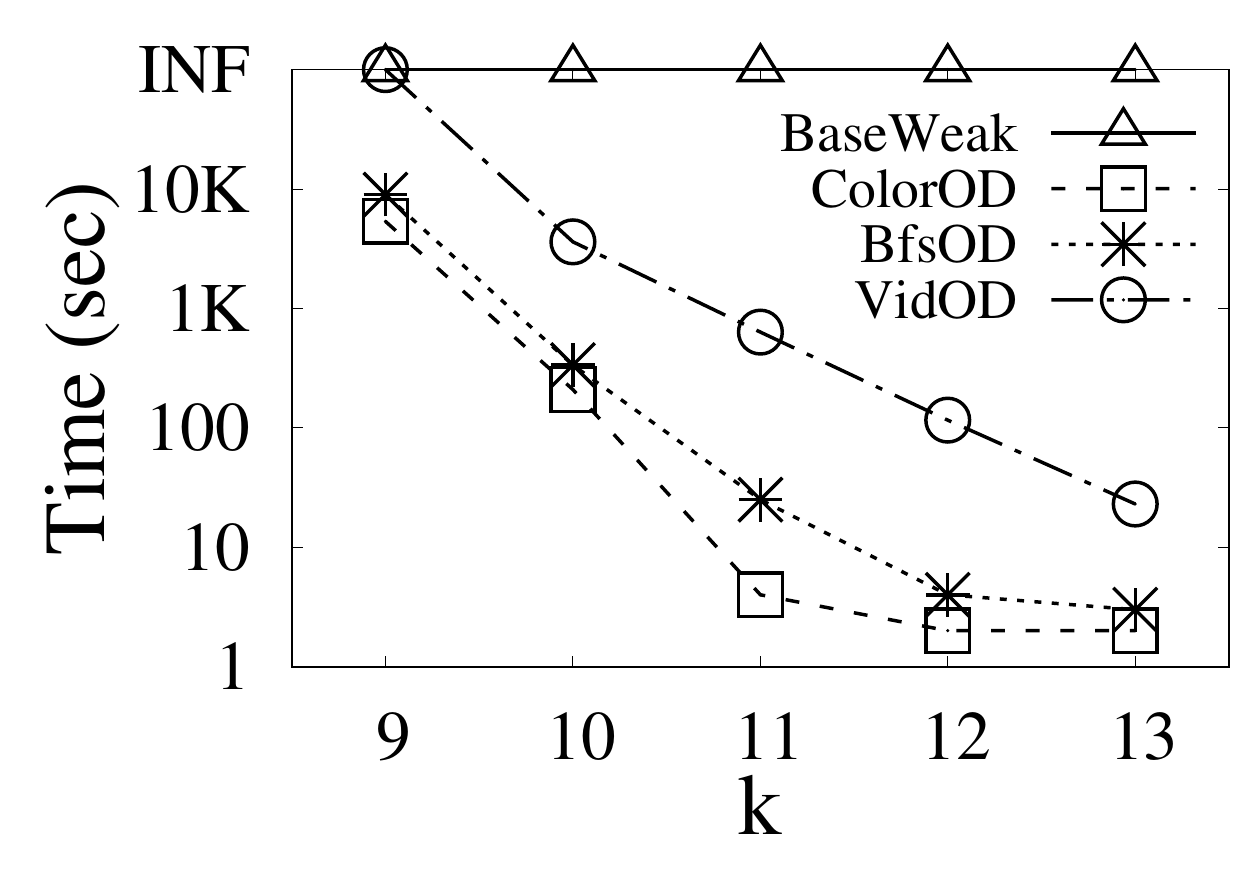}
			}
			\vspace*{-0.2cm} \\
			
            \subfigure[{\scriptsize \slashdot (vary $d$)}]{
				\includegraphics[width=0.4\columnwidth, height=2.5cm]{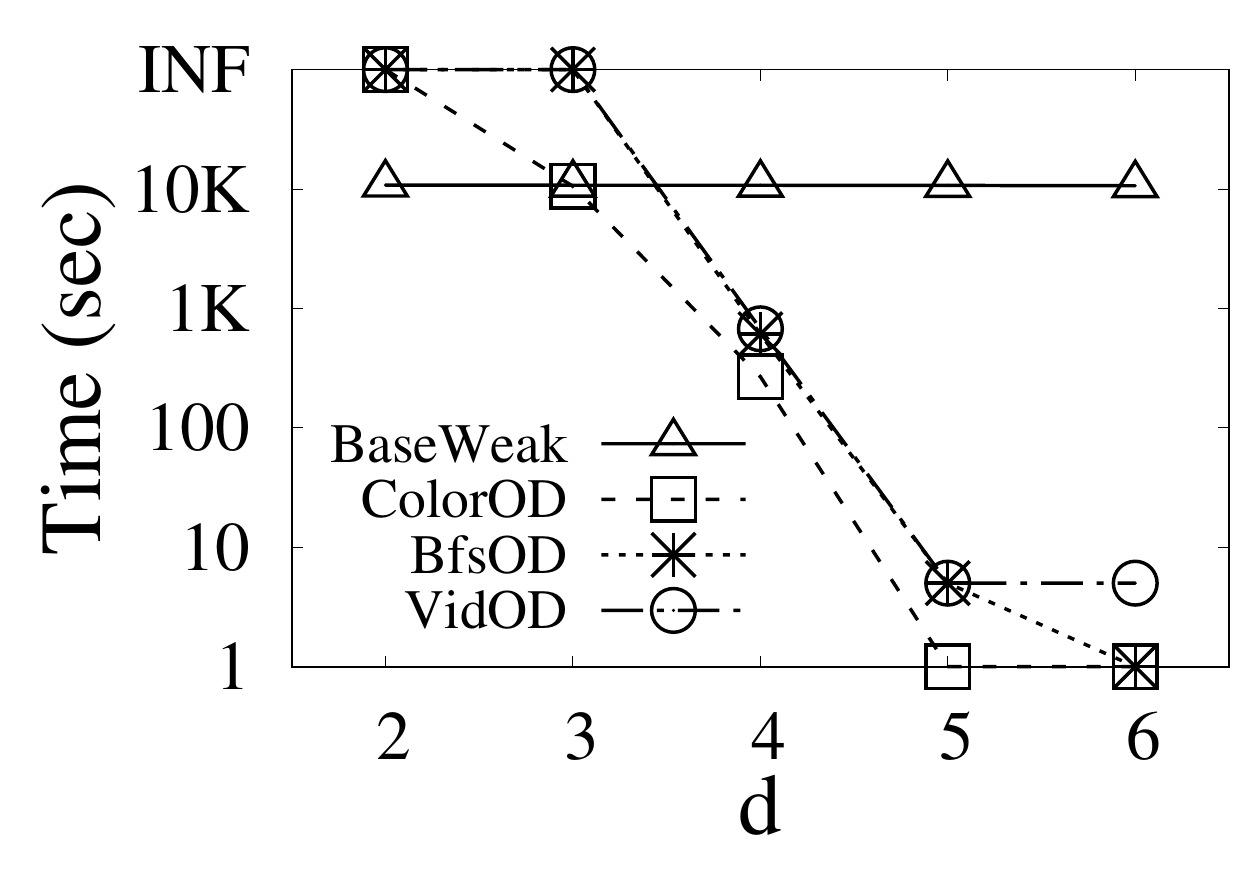}
			}
			\subfigure[{\scriptsize \themarker (vary $d$)}]{
				\includegraphics[width=0.4\columnwidth, height=2.5cm]{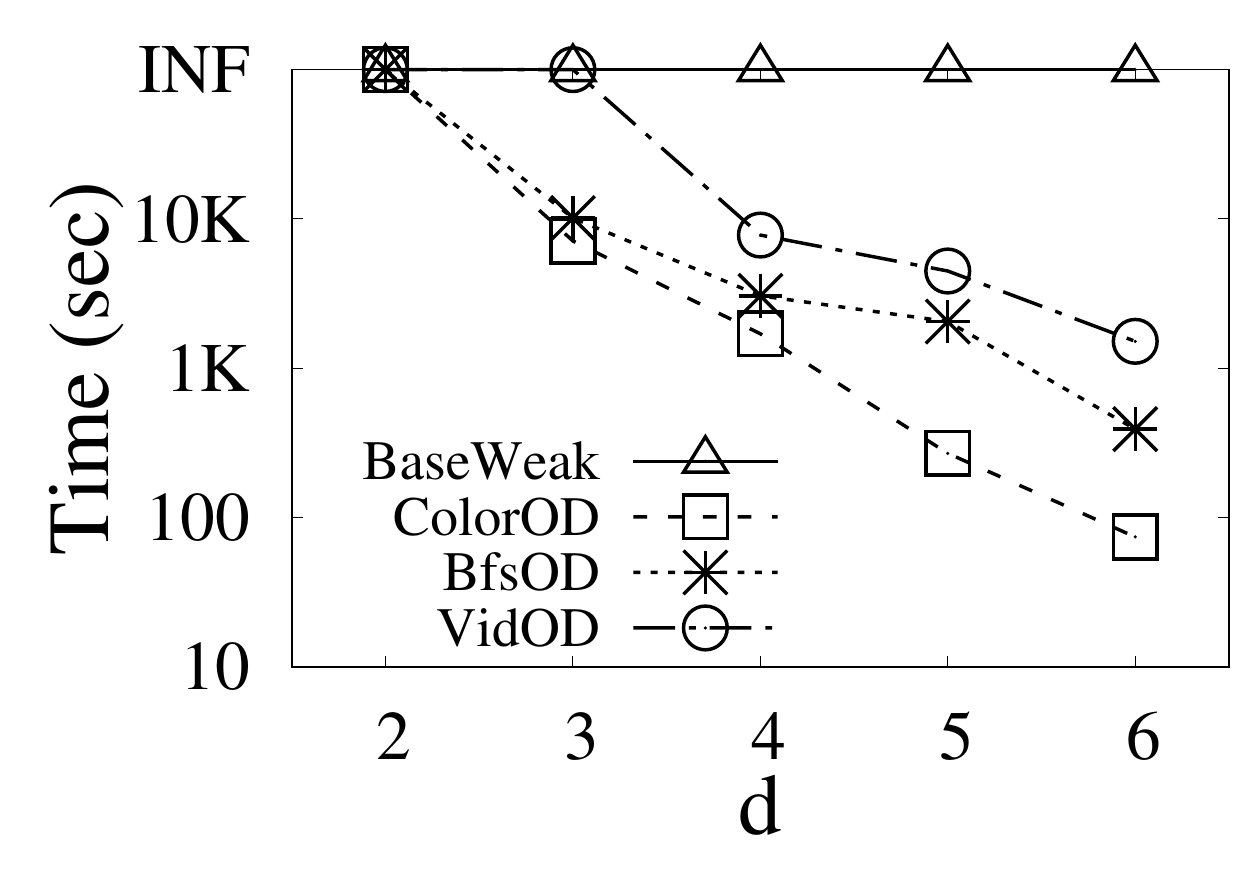}
			}
			\subfigure[{\scriptsize \wiki (vary $d$)}]{
				\includegraphics[width=0.4\columnwidth, height=2.5cm]{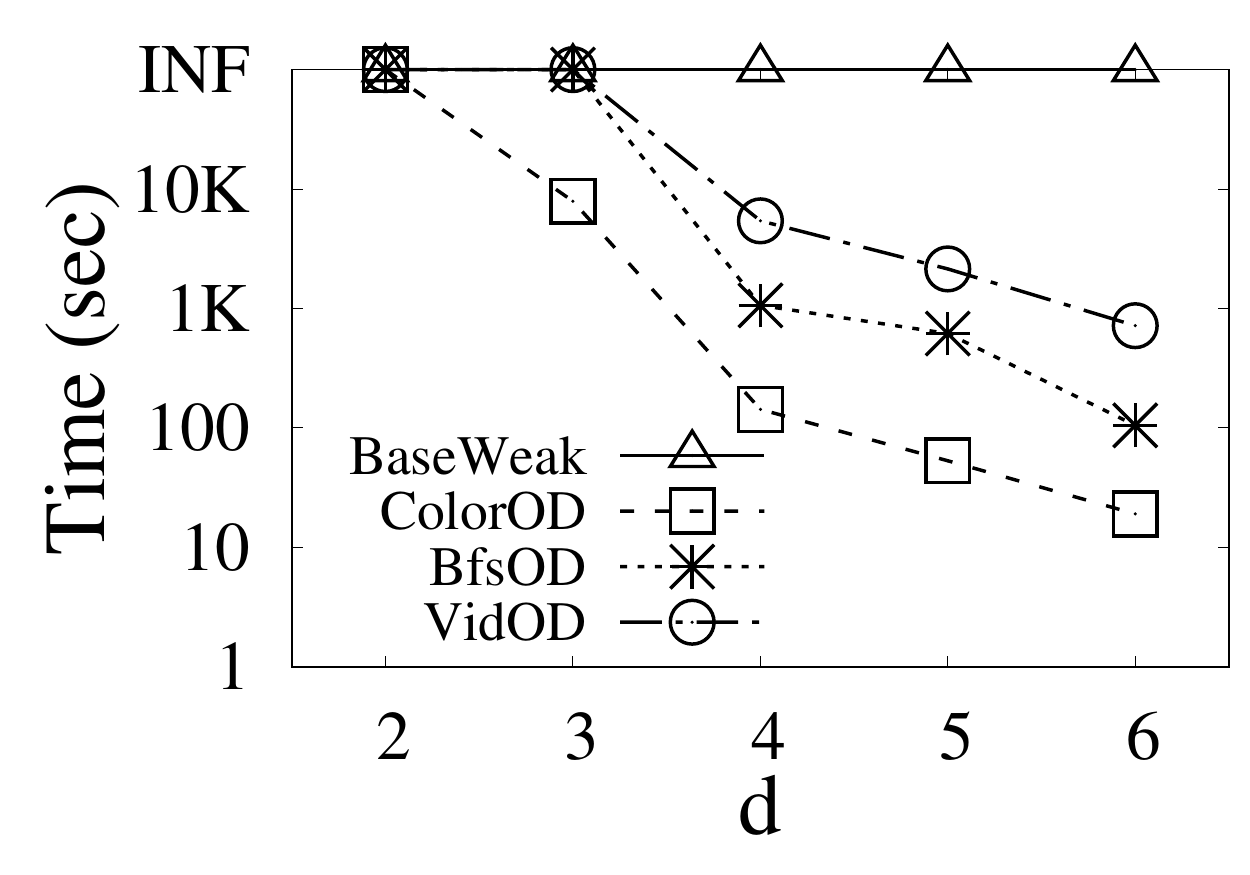}
			}
			\subfigure[{\scriptsize \flixster (vary $d$)}]{
				\includegraphics[width=0.4\columnwidth, height=2.5cm]{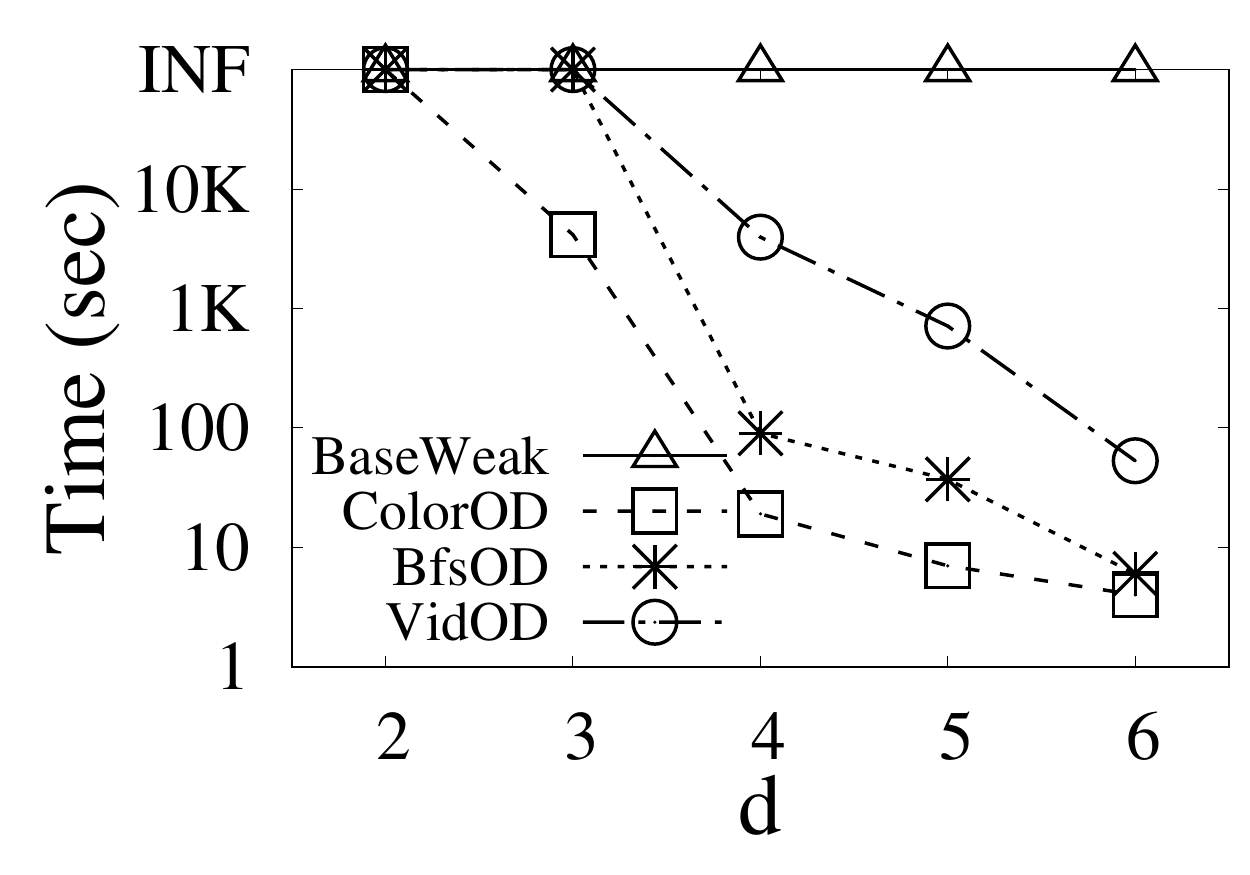}
			}
		\end{tabular}
	\end{center}
	\vspace*{-0.4cm}
	\caption{Running time of the \baseweak algorithm and \weak algorithms with different orderings}
	\vspace*{-0.3cm}
	\label{fig:exp:WFCE}
\end{figure*}

\begin{figure*}[t!] 
	\begin{center}
		\begin{tabular}[t]{c}
			\subfigure[{\scriptsize \slashdot (vary $k$)}]{
				\includegraphics[width=0.4\columnwidth, height=2.5cm]{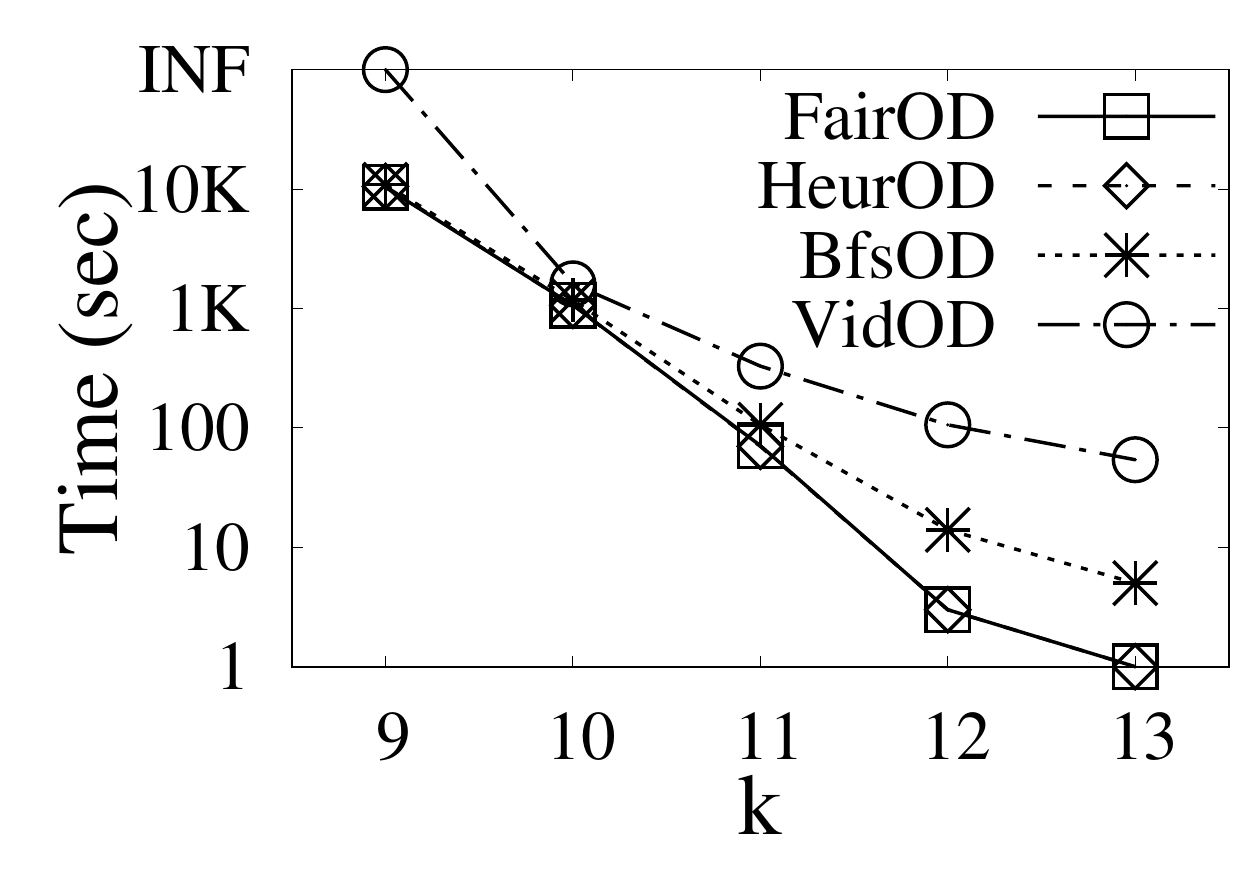}
			}
			\subfigure[{\scriptsize \themarker (vary $k$)}]{
				\includegraphics[width=0.4\columnwidth, height=2.5cm]{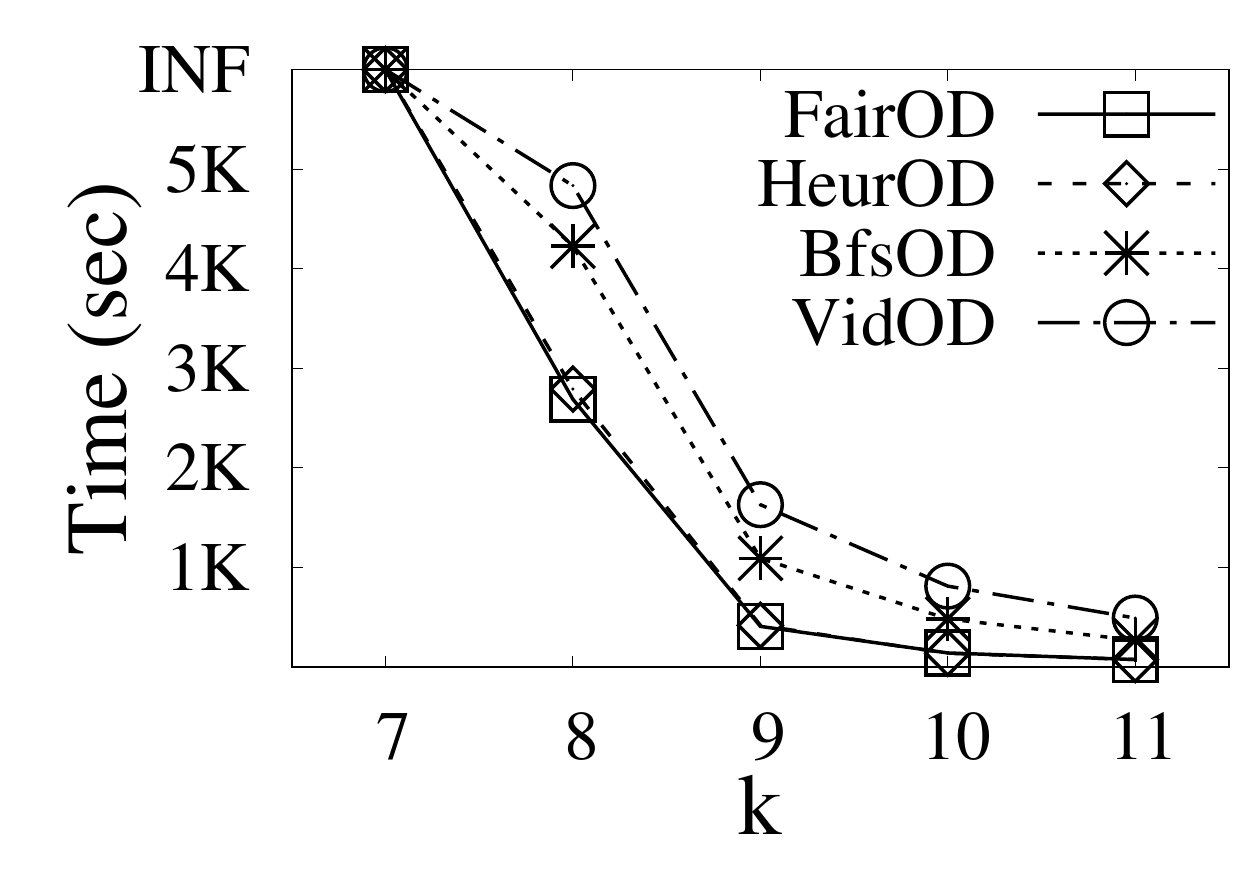}
			}
			\subfigure[{\scriptsize \wiki (vary $k$)}]{
				\includegraphics[width=0.4\columnwidth, height=2.5cm]{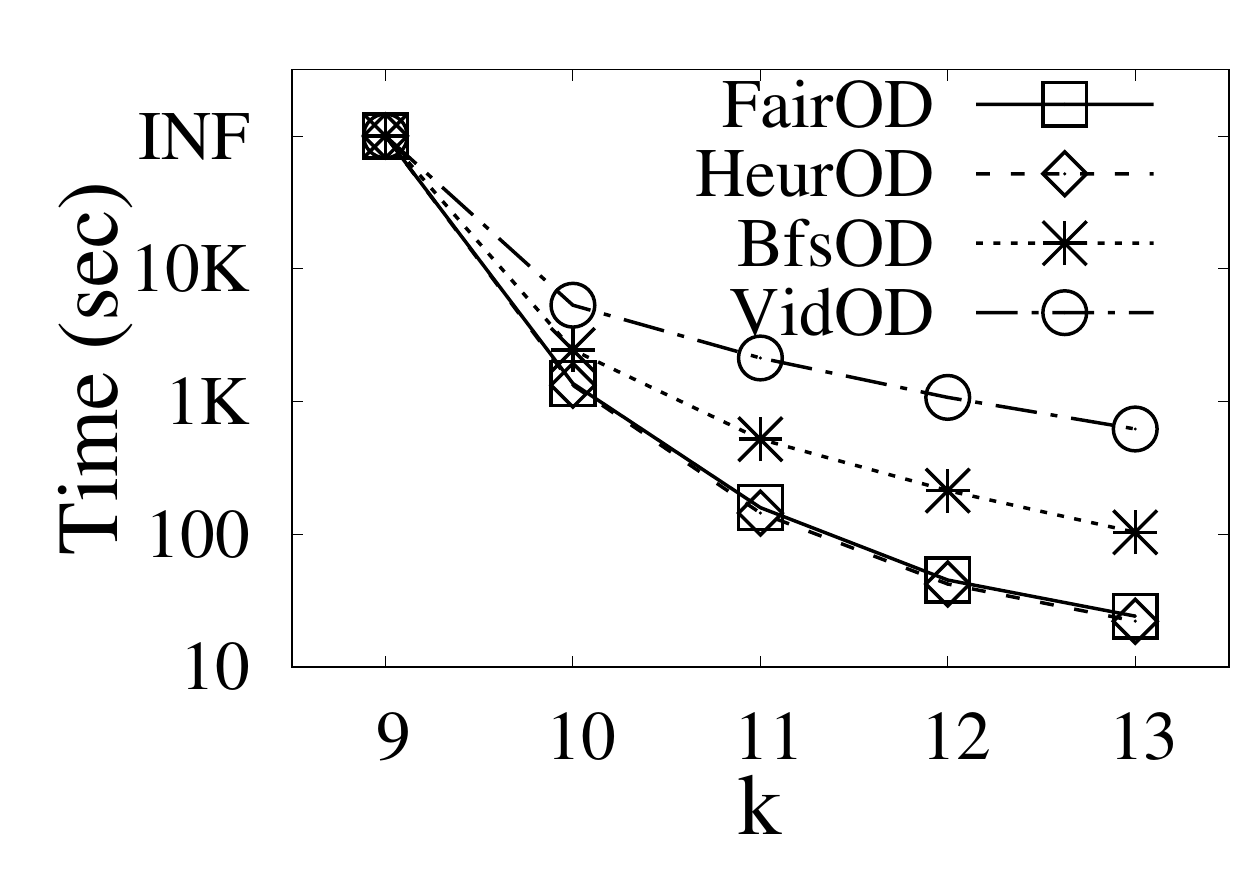}
			}
			\subfigure[{\scriptsize \flixster (vary $k$)}]{
				\includegraphics[width=0.4\columnwidth, height=2.5cm]{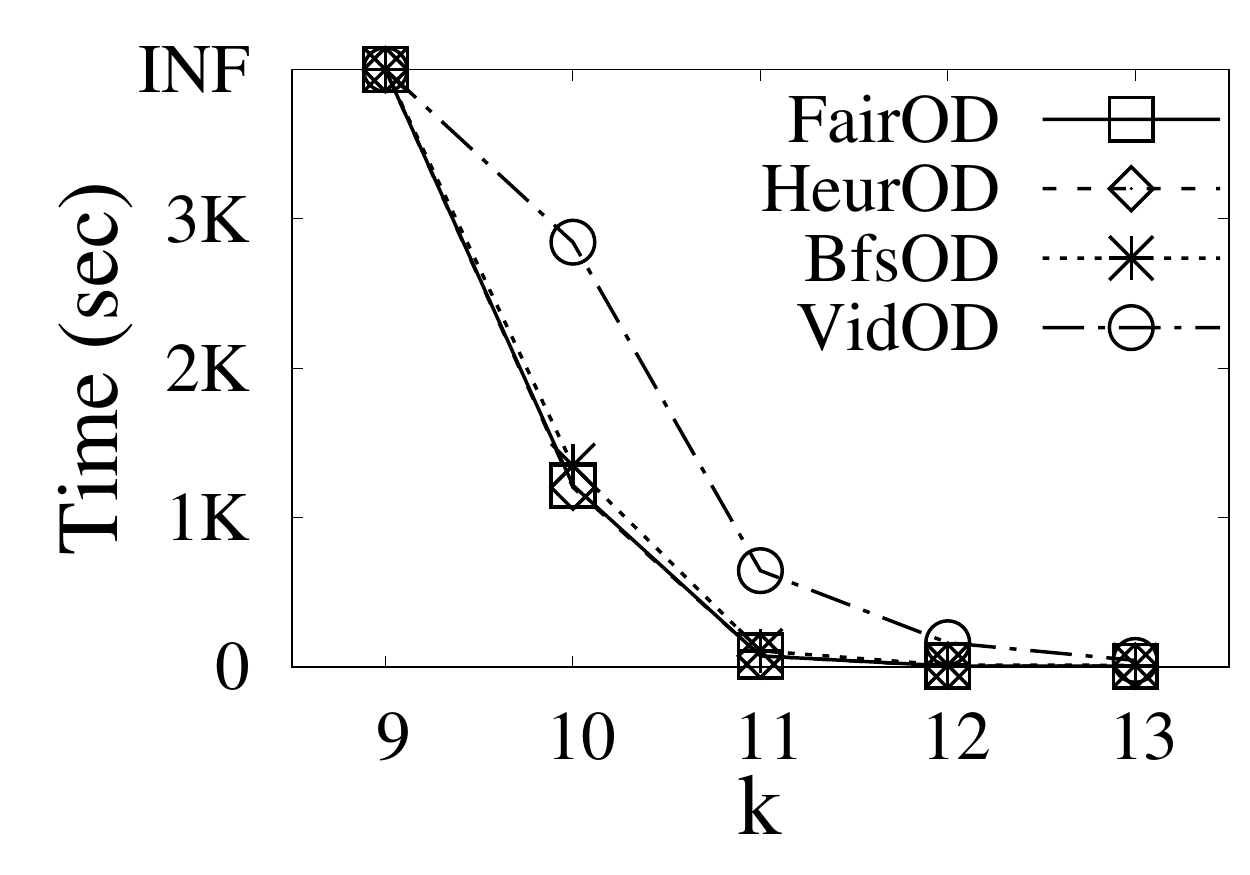}
			}
			\vspace*{-0.2cm} \\
			
			\subfigure[{\scriptsize \slashdot (vary $d$)}]{
				\includegraphics[width=0.4\columnwidth, height=2.5cm]{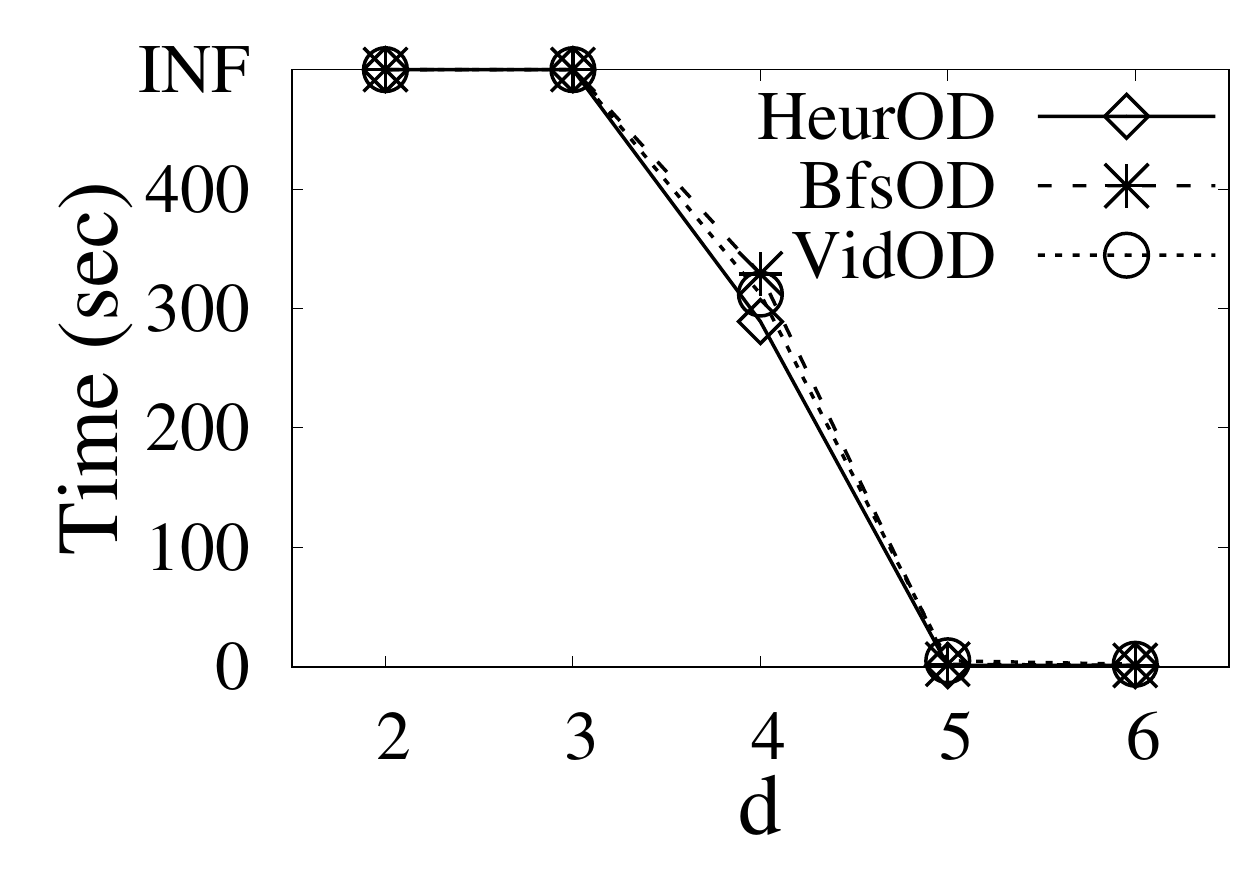}
			}
			\subfigure[{\scriptsize \themarker (vary $d$)}]{
				\includegraphics[width=0.4\columnwidth, height=2.5cm]{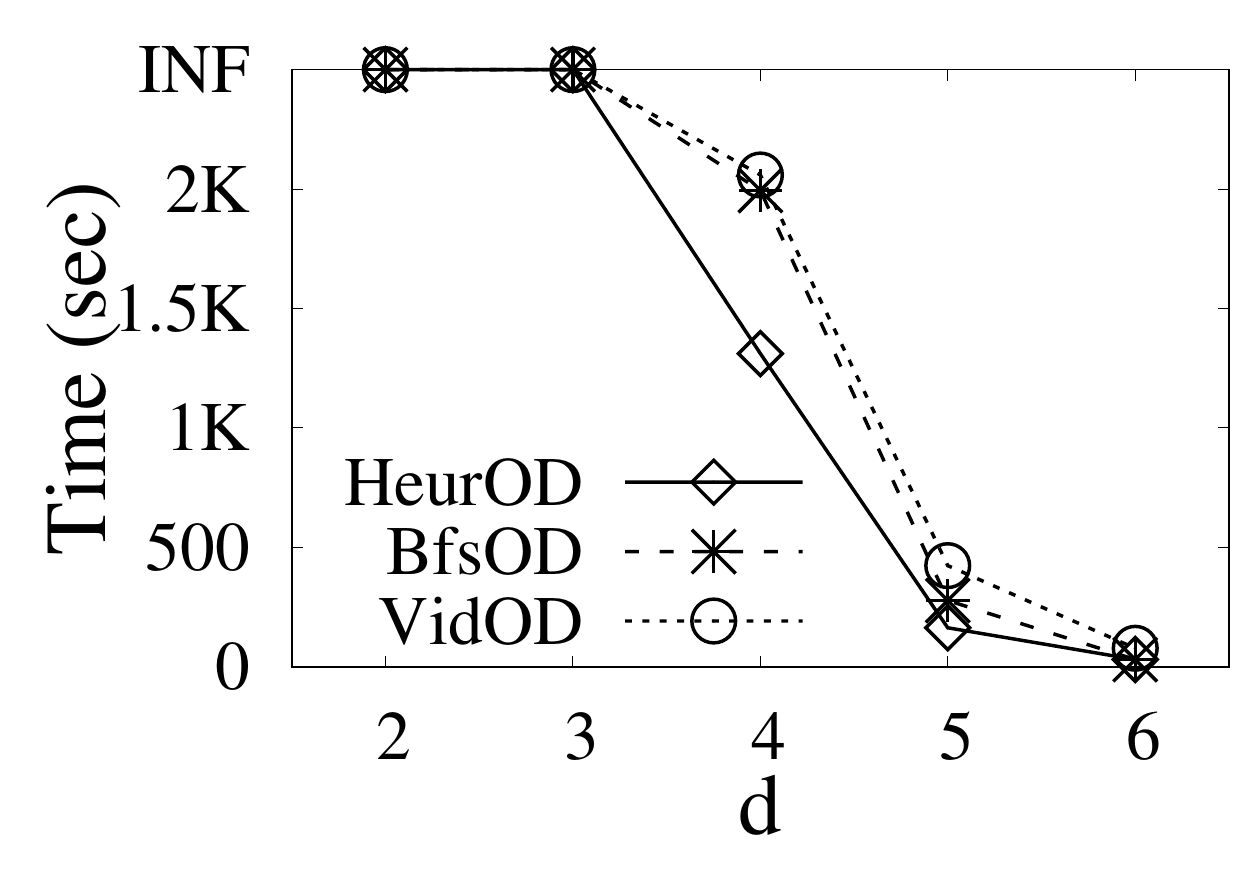}
			}
			\subfigure[{\scriptsize \wiki (vary $d$)}]{
				\includegraphics[width=0.4\columnwidth, height=2.5cm]{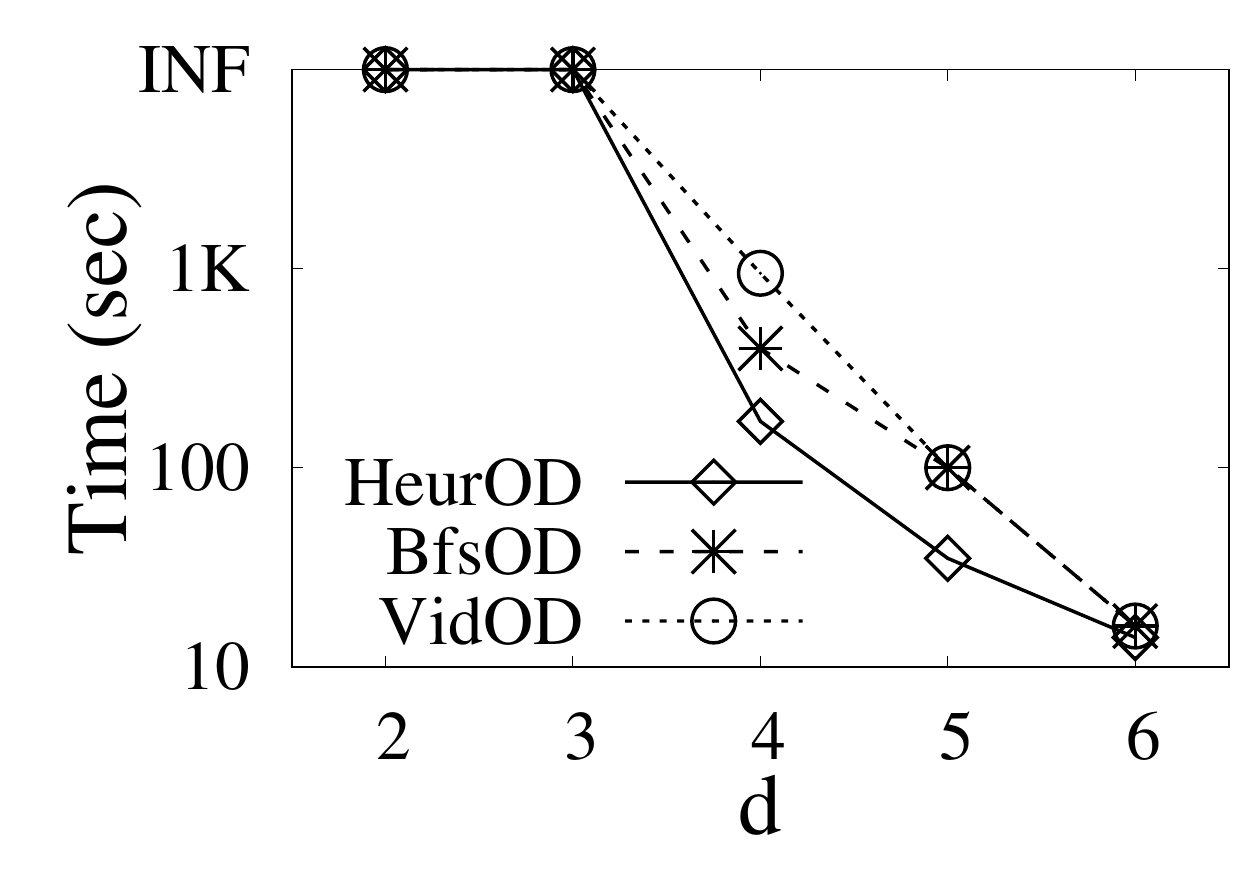}
			}
			\subfigure[{\scriptsize \flixster (vary $d$)}]{
				\includegraphics[width=0.4\columnwidth, height=2.5cm]{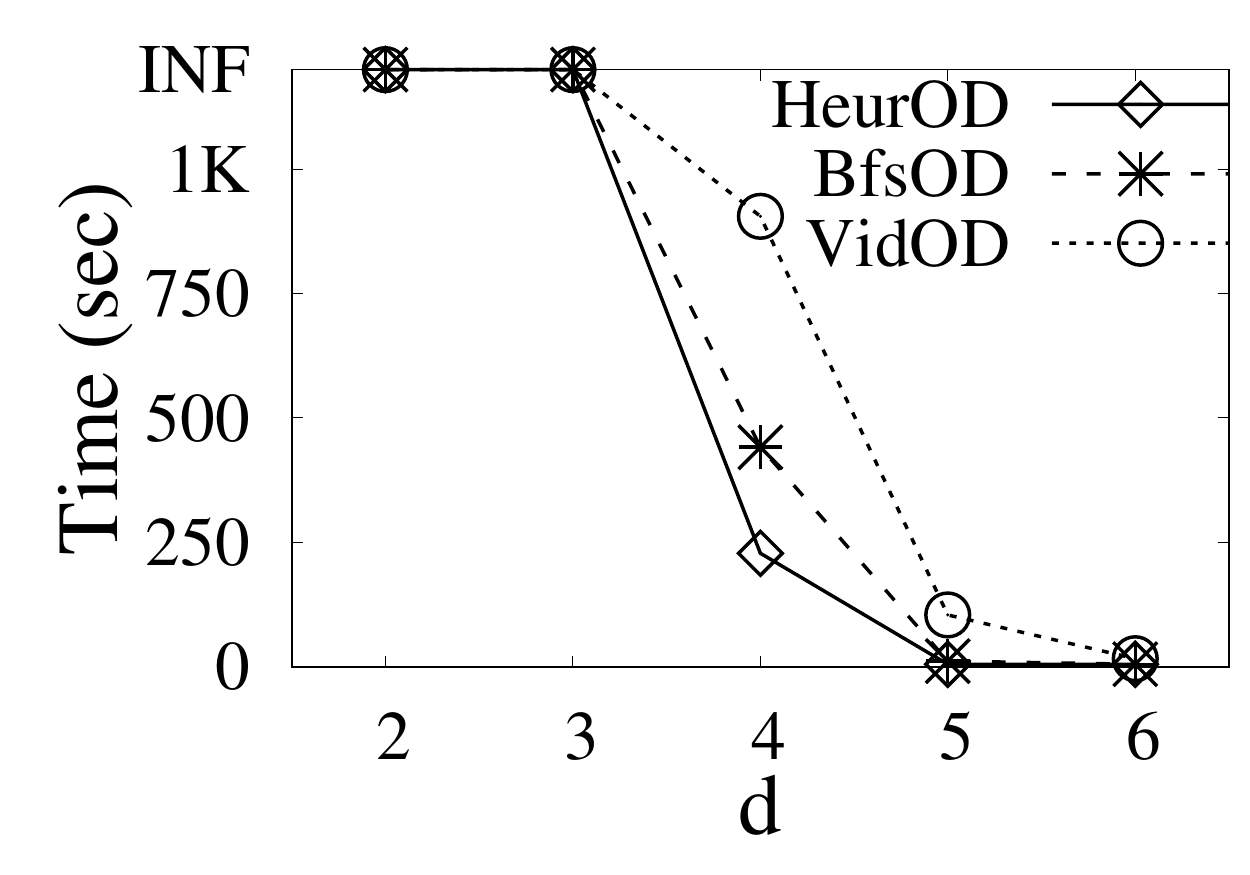}
			}
			\comment{
				\hspace*{-0.5cm}
				\subfigure[{\scriptsize \wiki (vary $d$)}]{
					\includegraphics[width=0.4\columnwidth, height=2.5cm]{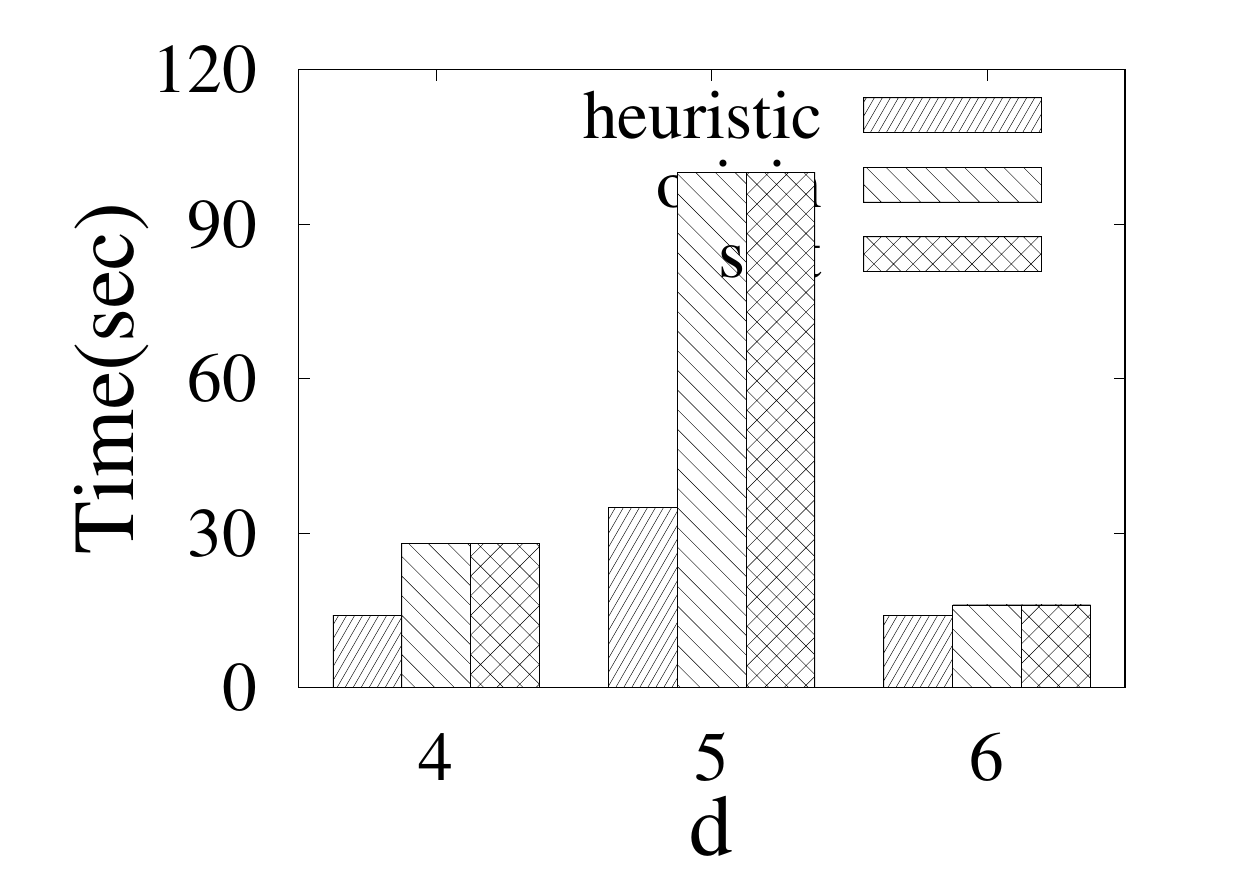}
				}
				\subfigure[{\scriptsize \flixster (vary $d$)}]{
					\includegraphics[width=0.4\columnwidth, height=2.5cm]{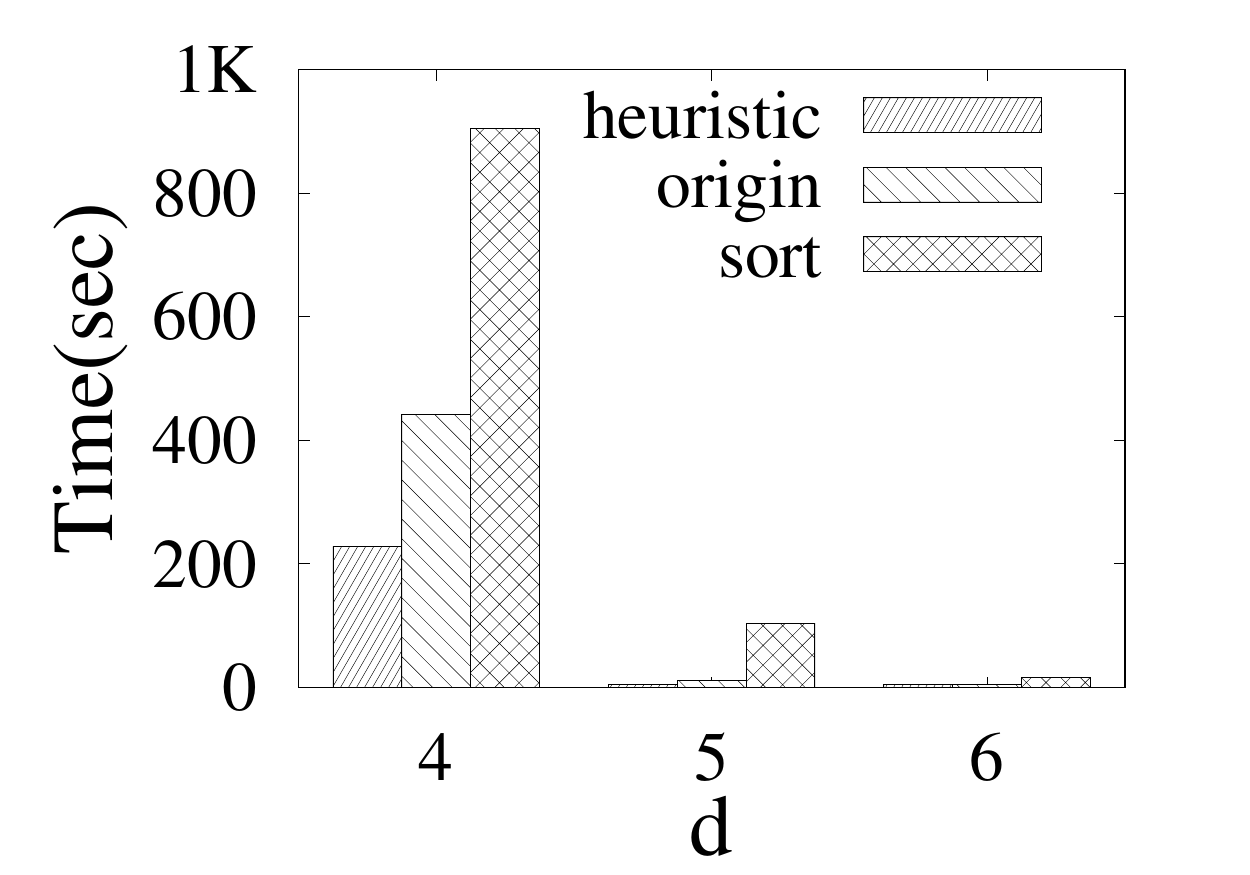}
				}
				\subfigure[{\scriptsize \slashdot (vary $d$)}]{
					\includegraphics[width=0.4\columnwidth, height=2.5cm]{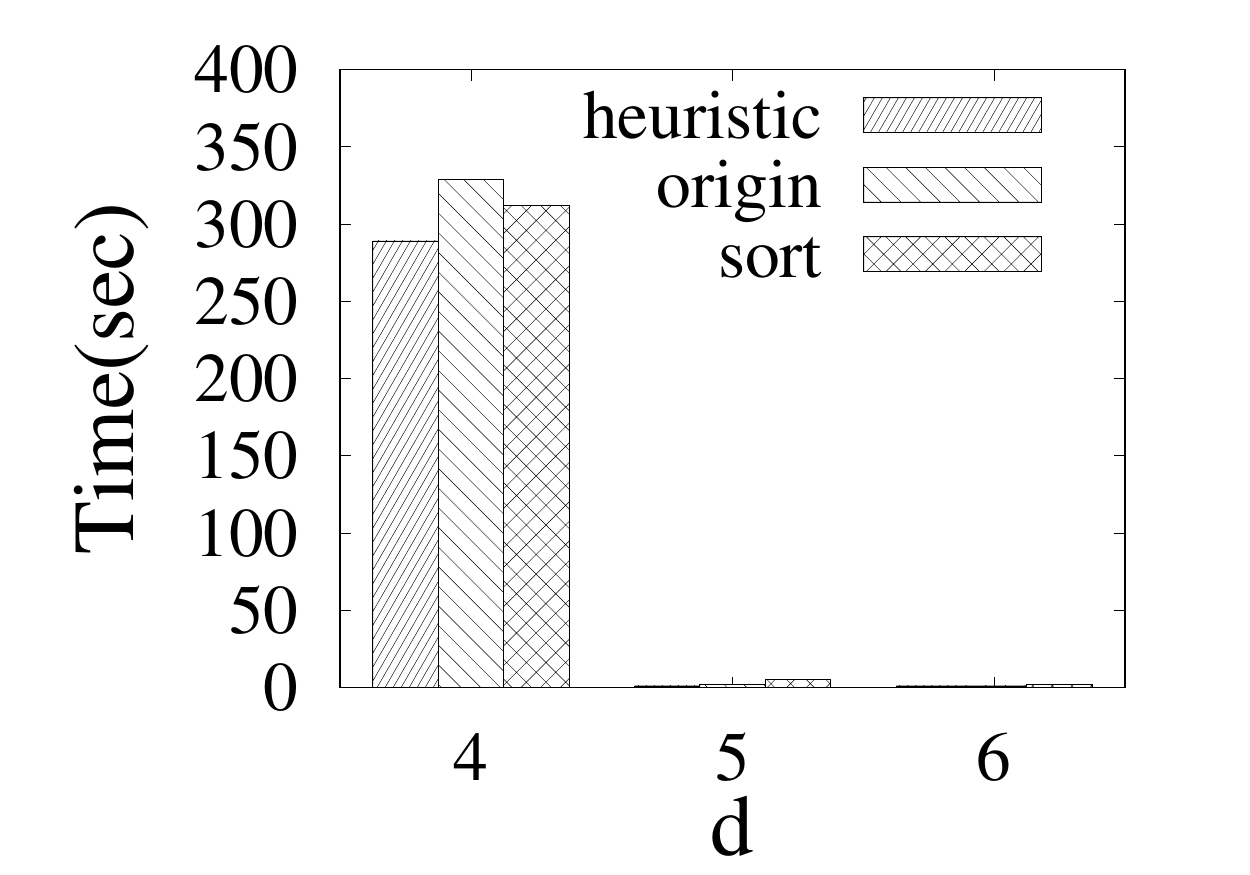}
				}
				\subfigure[{\scriptsize \webwiki (vary $d$)}]{
					\includegraphics[width=0.4\columnwidth, height=2.5cm]{exp/webwiki/clique-num-vary-time-vary-d-strong_hist-eps-converted-to.pdf}
				}
				\subfigure[{\scriptsize \themarker (vary $d$)}]{
					\includegraphics[width=0.4\columnwidth, height=2.5cm]{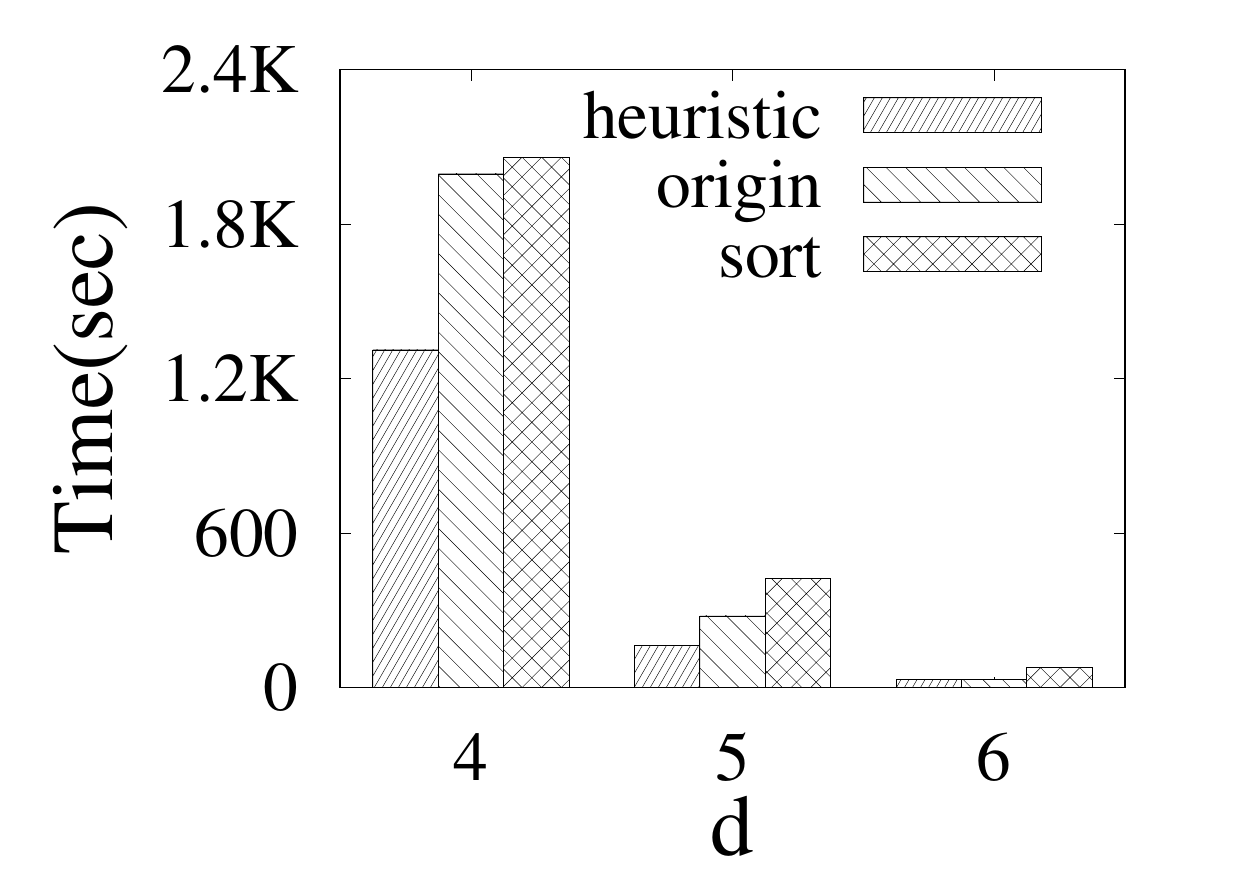}
				}
			}
		\end{tabular}
	\end{center}
	\vspace*{-0.4cm}
	\caption{Running time of the \strong algorithms with different orderings}
	\vspace*{-0.2cm}
	\label{fig:exp:SFCE}
\end{figure*}

\stitle{Evaluation of the pruning techniques.} For the 2D case (i.e., $d=2$), both \colorful and \enhancedcolor can be used to reduce the graph size in \weak, \relativeweak and \relativestrong algorithms. And \colorful and \fairness can be used to reduce the graph size in the \strong algorithm. In this experiment, we evaluate these pruning techniques by comparing the number of remaining vertices after pruning with varying $k$. The results are depicted in \figref{fig:exp:Core} (a)-(d). 

As can be seen from \figref{fig:exp:Core}, in \weak, \relativeweak and \relativestrong algorithms, both \colorful and \enhancedcolor can significantly reduce the number of vertices compared to the original graph as expected. Moreover, the number of remaining vertices decreases as $k$ increases. For example, in \slashdot with $k=9$, \colorful reduces the number of vertices from 82,169 to 3,985; and \enhancedcolor further reduces the number of vertices to 1,330. In general, \enhancedcolor consistently outperforms \colorful in terms of the pruning performance, especially for relatively small $k$ values. When $k$ goes larger, the pruning effect of \colorful is slightly worse than that of \enhancedcolor. This is because \colorful can also prune a large number of vertices for a large $k$; for the \strong algorithm, we can find that \fairness substantially reduces the number of vertices compared to \colorful and the original graph. For instance, in \flixster with $k=9$, the number of remaining vertices after applying \colorful and \enhancedcolor is 15,258 and 10,602 respectively, while there are 2,523,387 nodes in the original graph. Generally, the pruning performance of \fairness is better than that of \colorful with all parameter settings, especially for relatively small $k$ values. For a larger $k$, the pruning effect of \colorful is slightly worse than that of \fairness. This is because \fairness first invokes \colorful to prune unpromising vertices. Since \colorful is already able to prune a large number of vertices when $k$ is large, \fairness cannot further prune too many vertices after invoking \colorful. These results confirm that our pruning techniques are indeed very effective in reducing the graph size.

Note that for the high-dimensional case (i.e., $d \ge 3$), only the \colorful algorithm can be used to prune the unpromising vertices in \weak, \strong, \relativeweak and \relativestrong algorithms. Therefore, we further study how the dimension $d$ affects the pruning performance of \colorful. \figref{fig:exp:Core} (e)-(h) show the number of remaining vertices after invoking \colorful with varying $d$. As can be seen, \colorful can substantially reduce the number of vertices with different $d$ values overall datasets, which is consistent with our previous findings. In general, the number of remaining vertices decreases as $d$ increases. This is because with a larger $d$, the constraints of \colorful become stricter, thus more vertices can be pruned. These results further confirm the effectiveness of the proposed pruning techniques.


\stitle{Evaluation of \weak.} Here we compare the \baseweak and the \weak algorithms equipped with \bfsorder, \idorder and \colororder by varying $k$ and $d$. The results are depicted in \figref{fig:exp:WFCE}. As can be seen,  \baseweak can only output the results on \slashdot and cannot terminate within the time limit on the other datasets. Our \weak algorithm, however, can work well on most datasets. The running time of \baseweak is insensitive w.r.t.\ $k$ and $d$, but the runtime of our \weak algorithm decreases as $k$ or $d$ increases as expected. Moreover, we can see that the runtime of \weak is several orders of magnitude lower than that of \baseweak for a large $k$ or $d$. For example, on \slashdot with $k=11$, \weak takes 268 seconds to enumerate all weak fair cliques, while \baseweak consumes 10,665 seconds. This is because \baseweak needs to enumerate all maximal cliques, which is the main bottleneck of the algorithm. For a large $k$, \weak can prune many vertices by the colorful $k$-core based pruning technique and the search space can also be reduced during the backtracking procedure. For a large $d$, the number of weak fair cliques decreases with an increasing $d$, thus reducing time overheads. These results confirm that the proposed \weak algorithm is much more efficient than \baseweak to find all weak fair cliques on large graphs.

In addition, we can also see that \weak with \colororder is much faster than \weak with \bfsorder and \idorder. For instance, when $k = 11$, \weak with \colororder consumes 4 seconds to output all results on \flixster, while \weak with \bfsorder and \idorder takes 25 and 633 seconds, respectively. On the \themarker dataset, when $k=7$, the running time of \weak with \colororder is 5,550 seconds, while the two baseline algorithms cannot finish within 3 hours. These results indicate that the proposed algorithm is very efficient to enumerate all weak fair cliques in large real-life graphs. Also, the results confirm the effectiveness of the proposed ordering technique \colororder.


\begin{figure*}[t!] 
	\begin{center}
		\begin{tabular}[t]{c}
			\subfigure[{\scriptsize \slashdot (vary $k$)}]{
				\includegraphics[width=0.4\columnwidth, height=2.5cm]{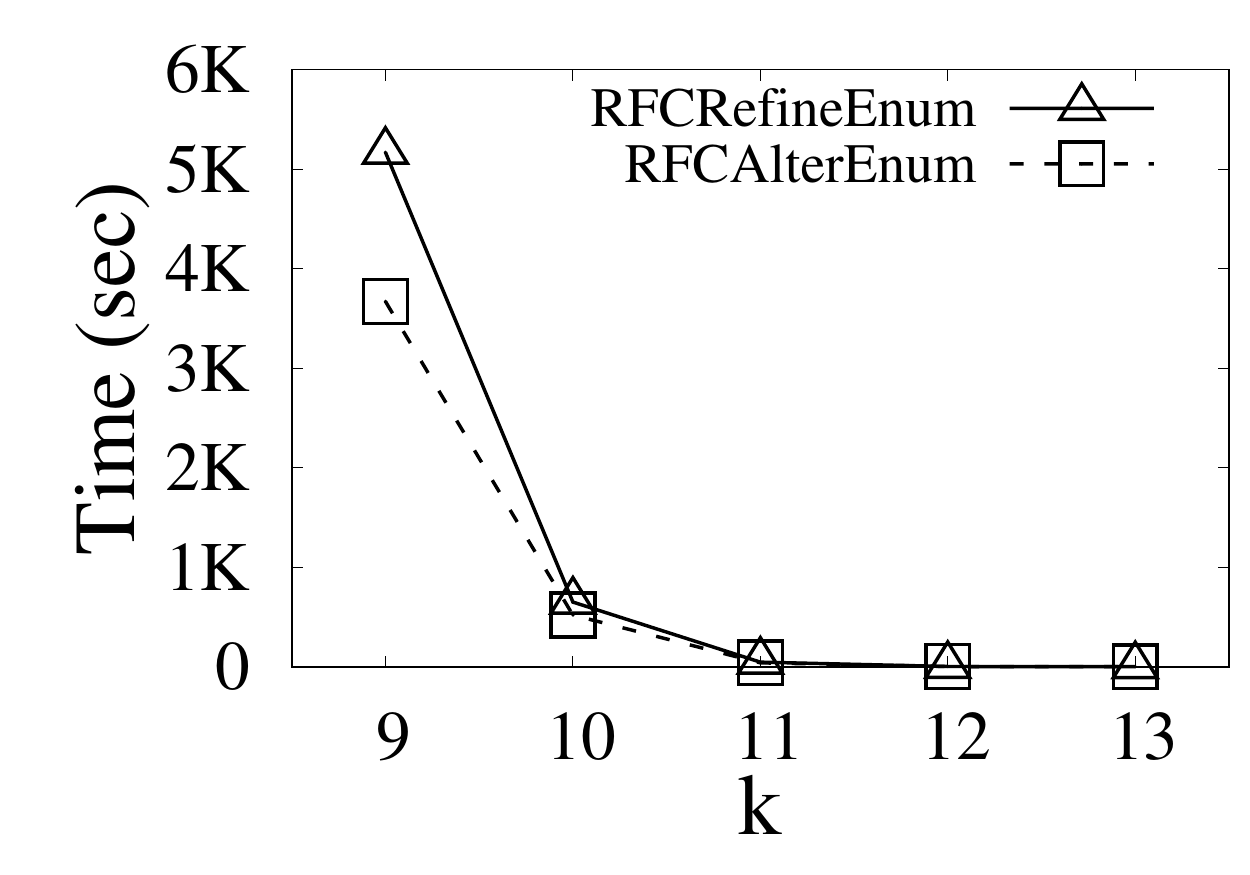}
			}
			\subfigure[{\scriptsize \themarker (vary $k$)}]{
				\includegraphics[width=0.4\columnwidth, height=2.5cm]{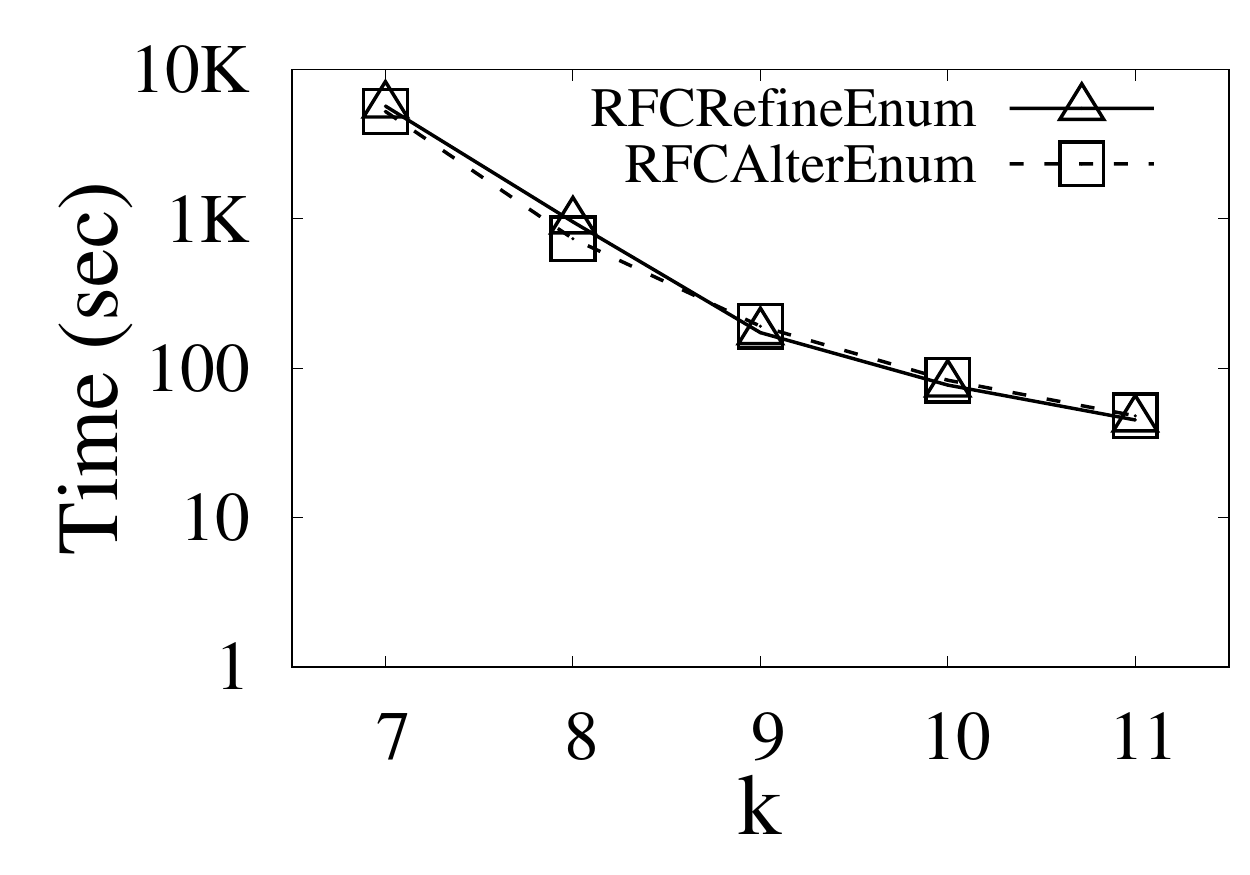}
			}
			\subfigure[{\scriptsize \wiki (vary $k$)}]{
				\includegraphics[width=0.4\columnwidth, height=2.5cm]{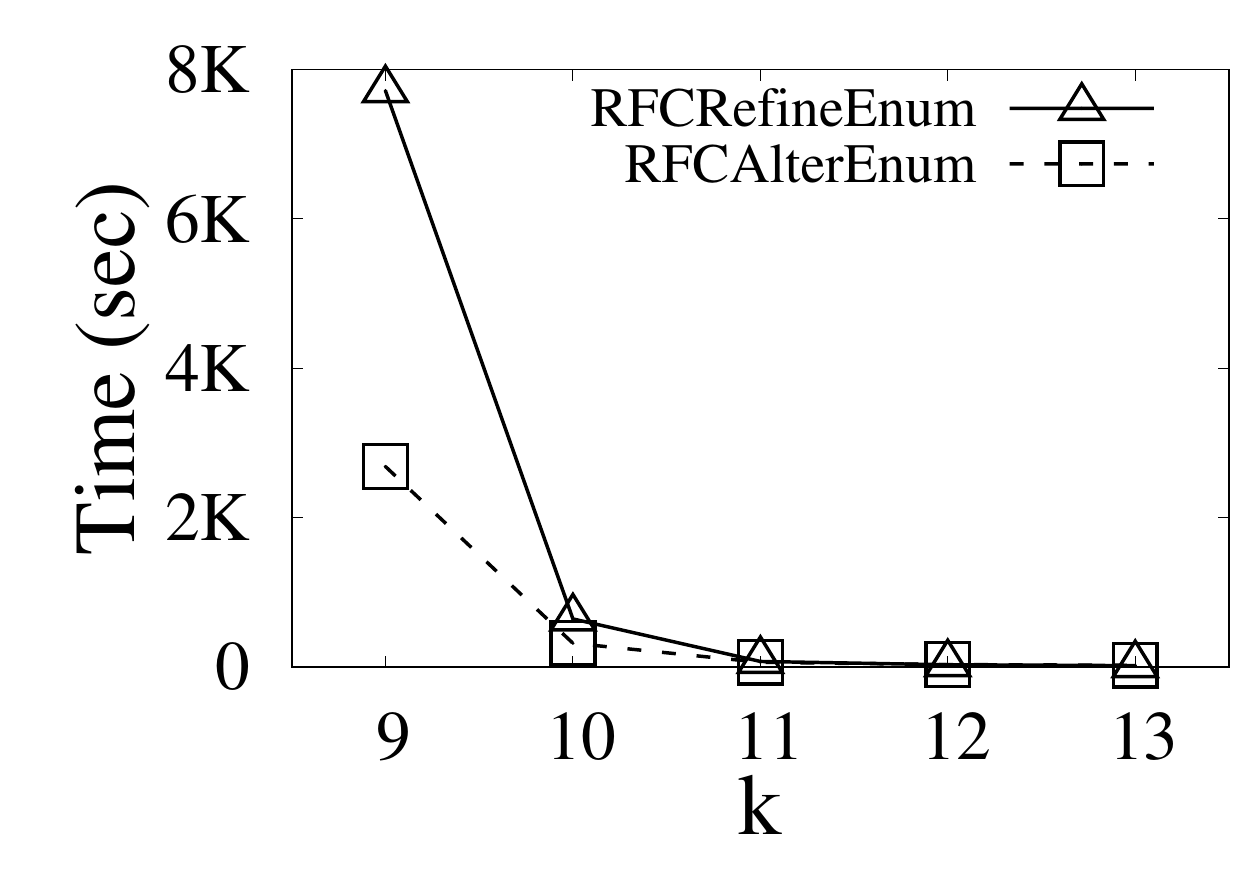}
			}
			\subfigure[{\scriptsize \flixster (vary $k$)}]{
				\includegraphics[width=0.4\columnwidth, height=2.5cm]{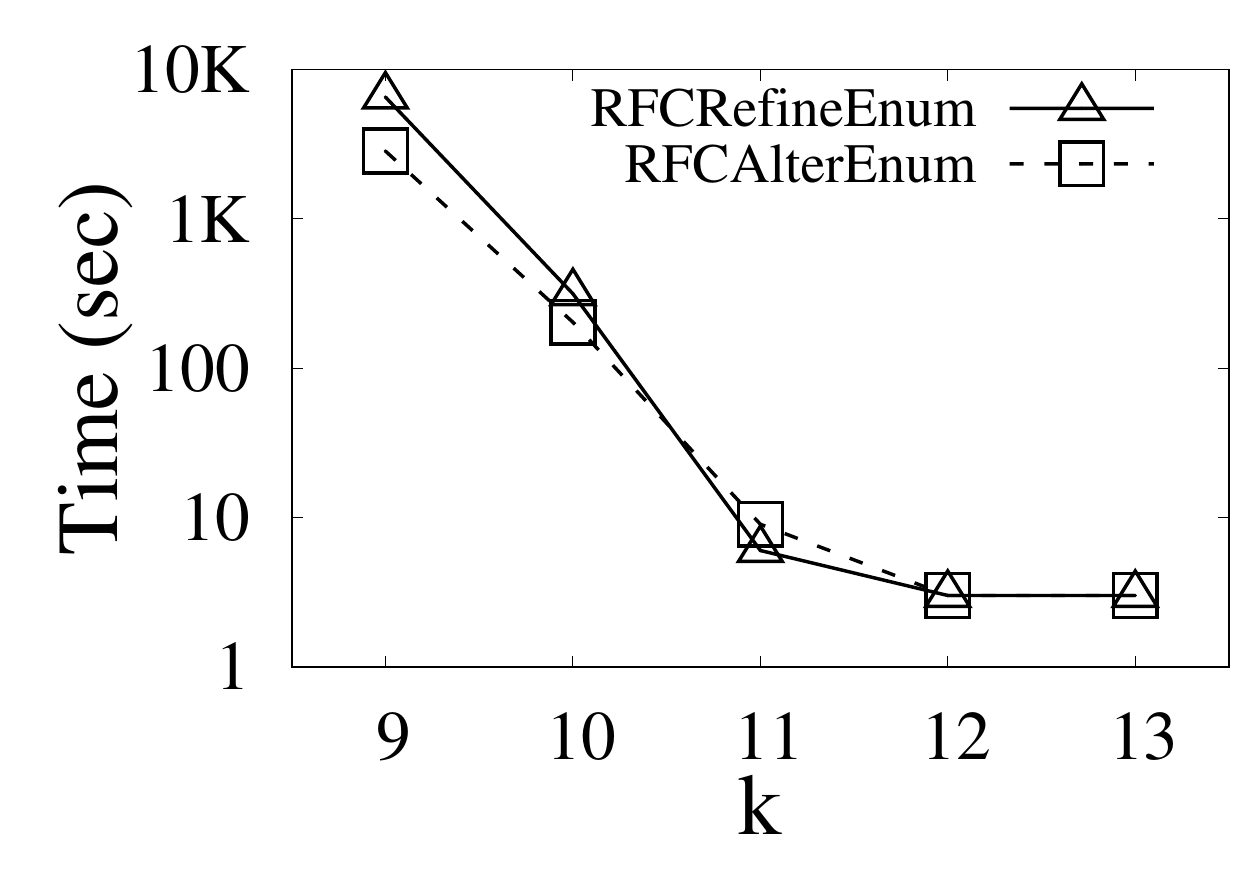}
			}
			\vspace*{-0.2cm} \\
			
			\subfigure[{\scriptsize \slashdot (vary $d$)}]{
				\includegraphics[width=0.4\columnwidth, height=2.5cm]{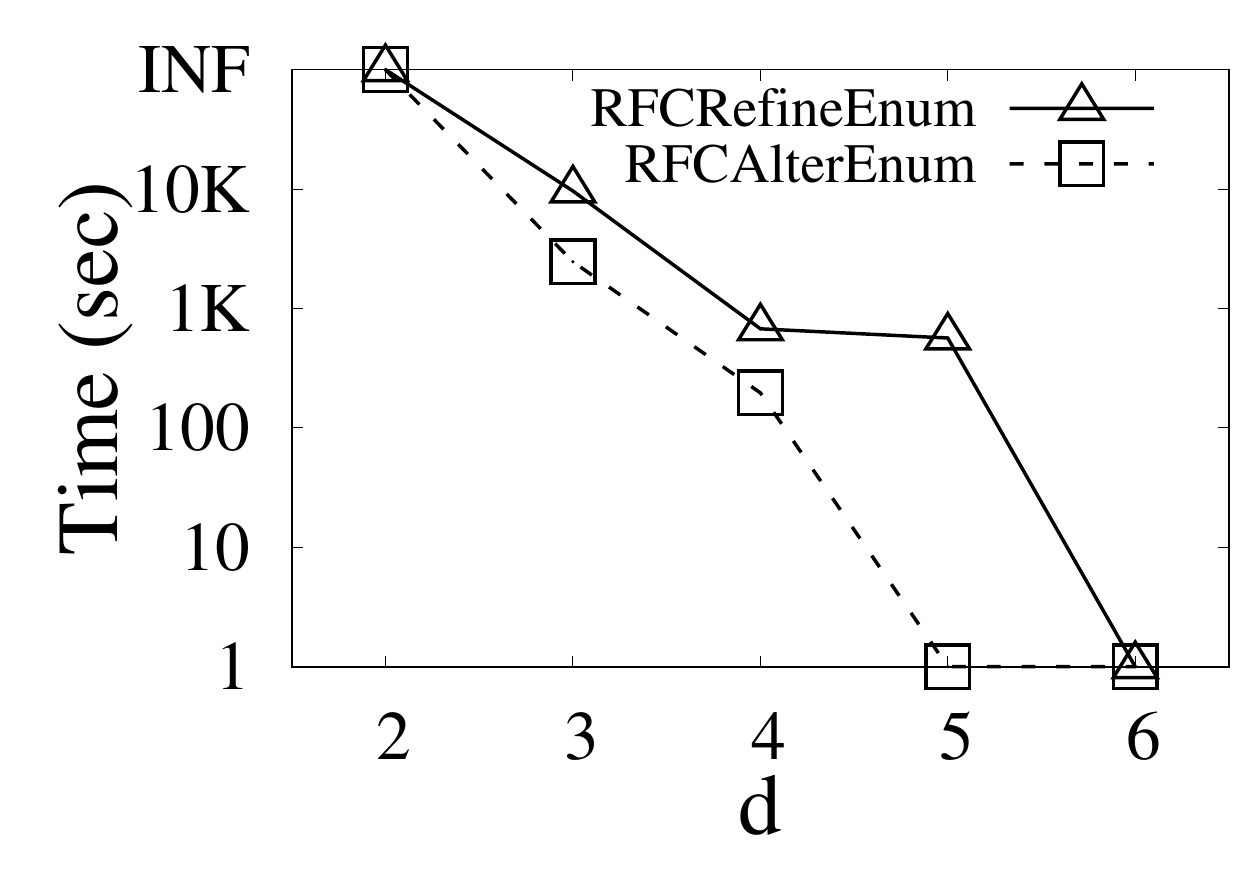}
			}
			\subfigure[{\scriptsize \themarker (vary $d$)}]{
				\includegraphics[width=0.4\columnwidth, height=2.5cm]{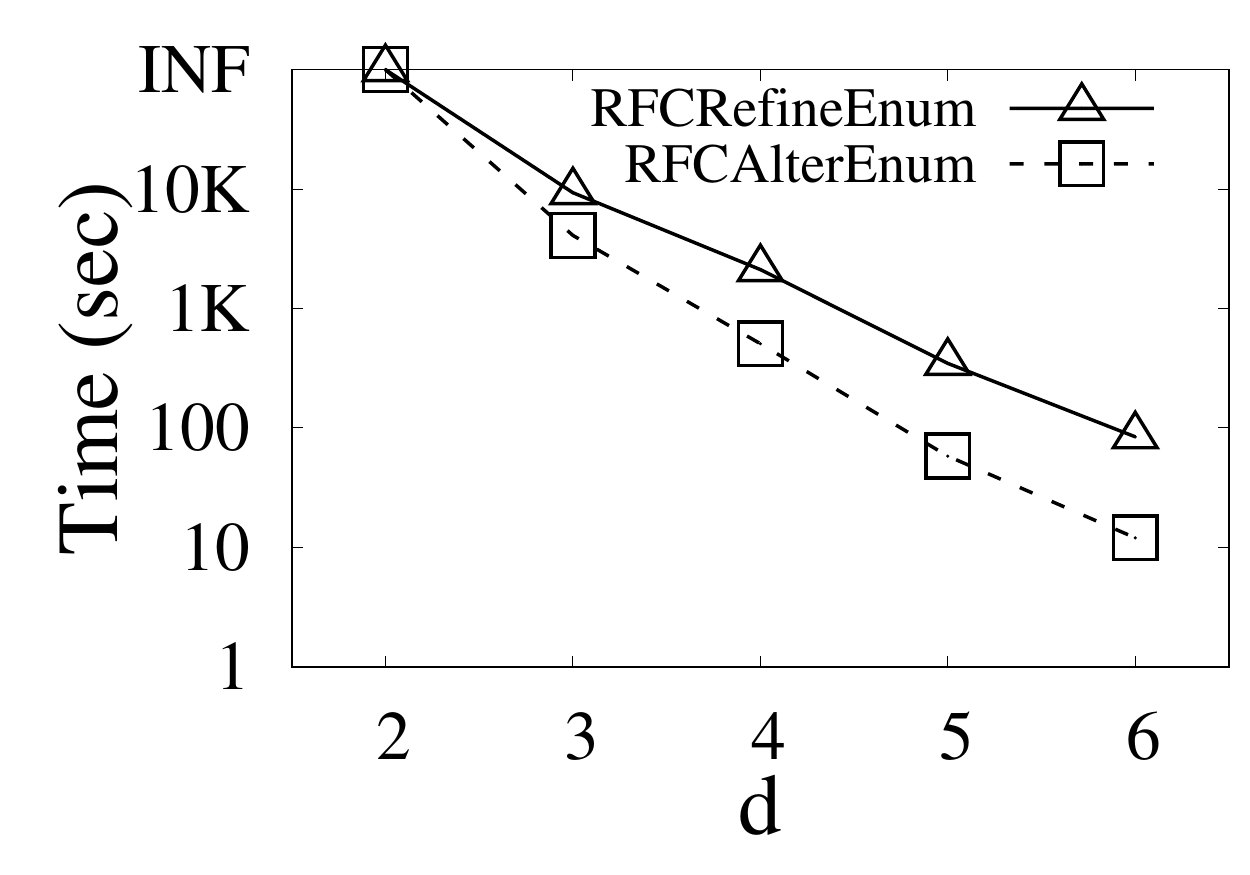}
			}
			\subfigure[{\scriptsize \wiki (vary $d$)}]{
				\includegraphics[width=0.4\columnwidth, height=2.5cm]{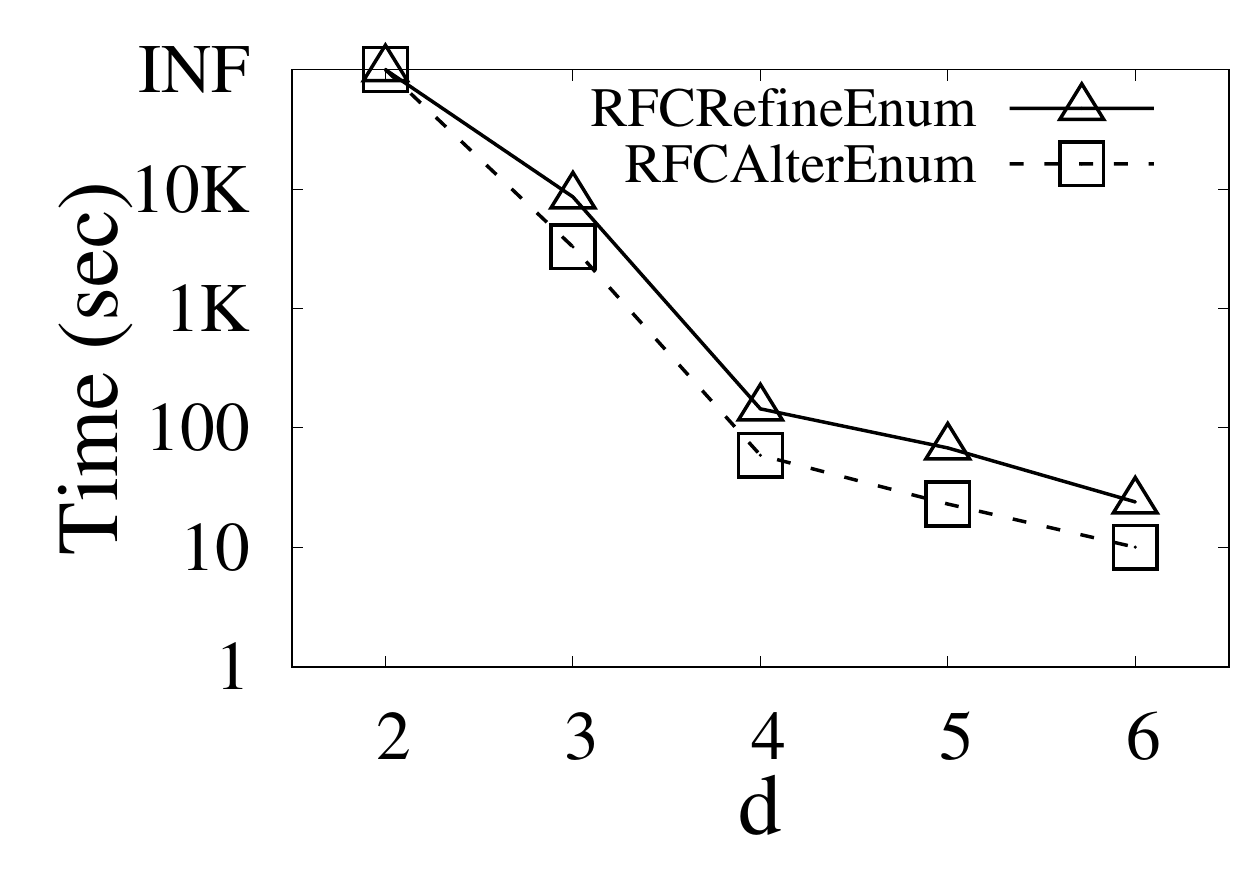}
			}
			\subfigure[{\scriptsize \flixster (vary $d$)}]{
				\includegraphics[width=0.4\columnwidth, height=2.5cm]{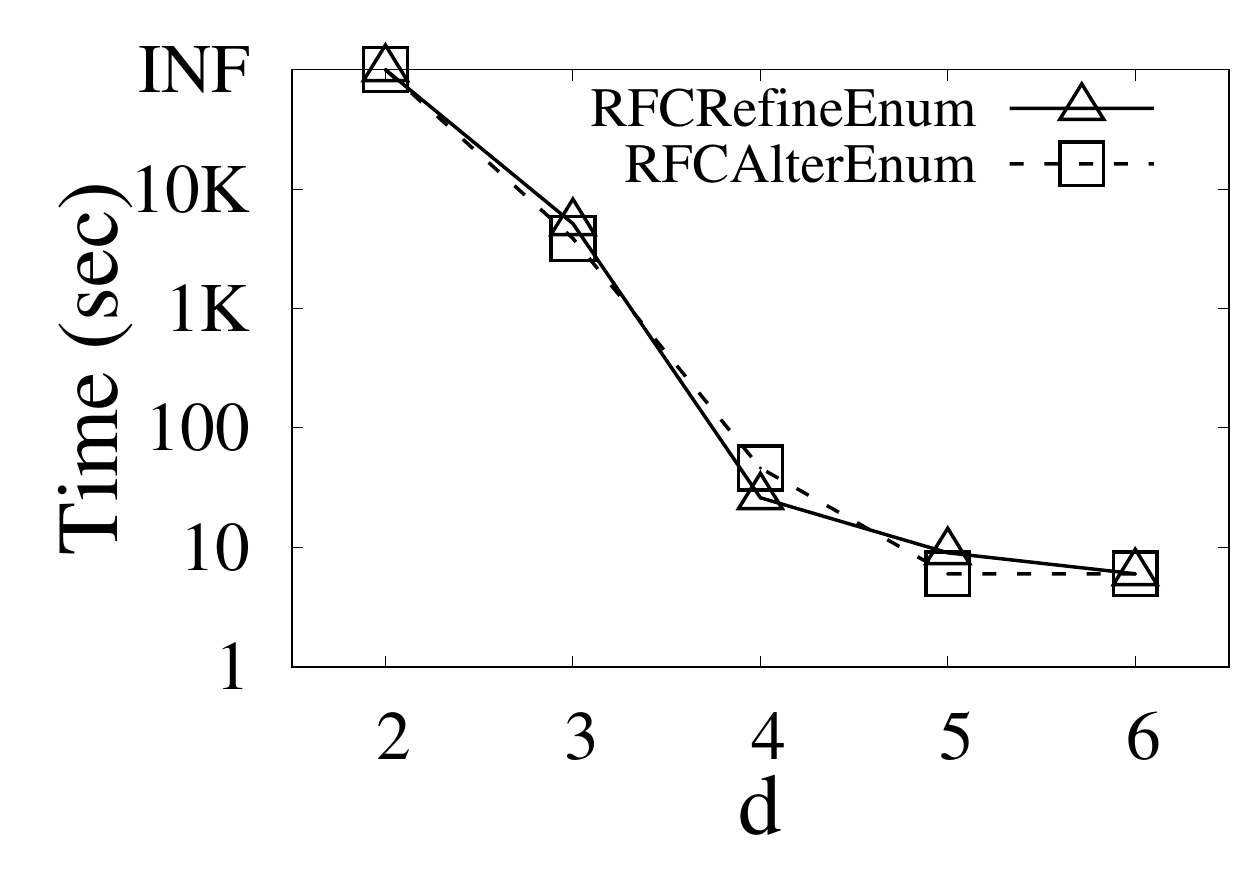}
			}
			\vspace*{-0.2cm} \\
			
			\subfigure[{\scriptsize \slashdot (vary $\delta$)}]{
				\includegraphics[width=0.4\columnwidth, height=2.5cm]{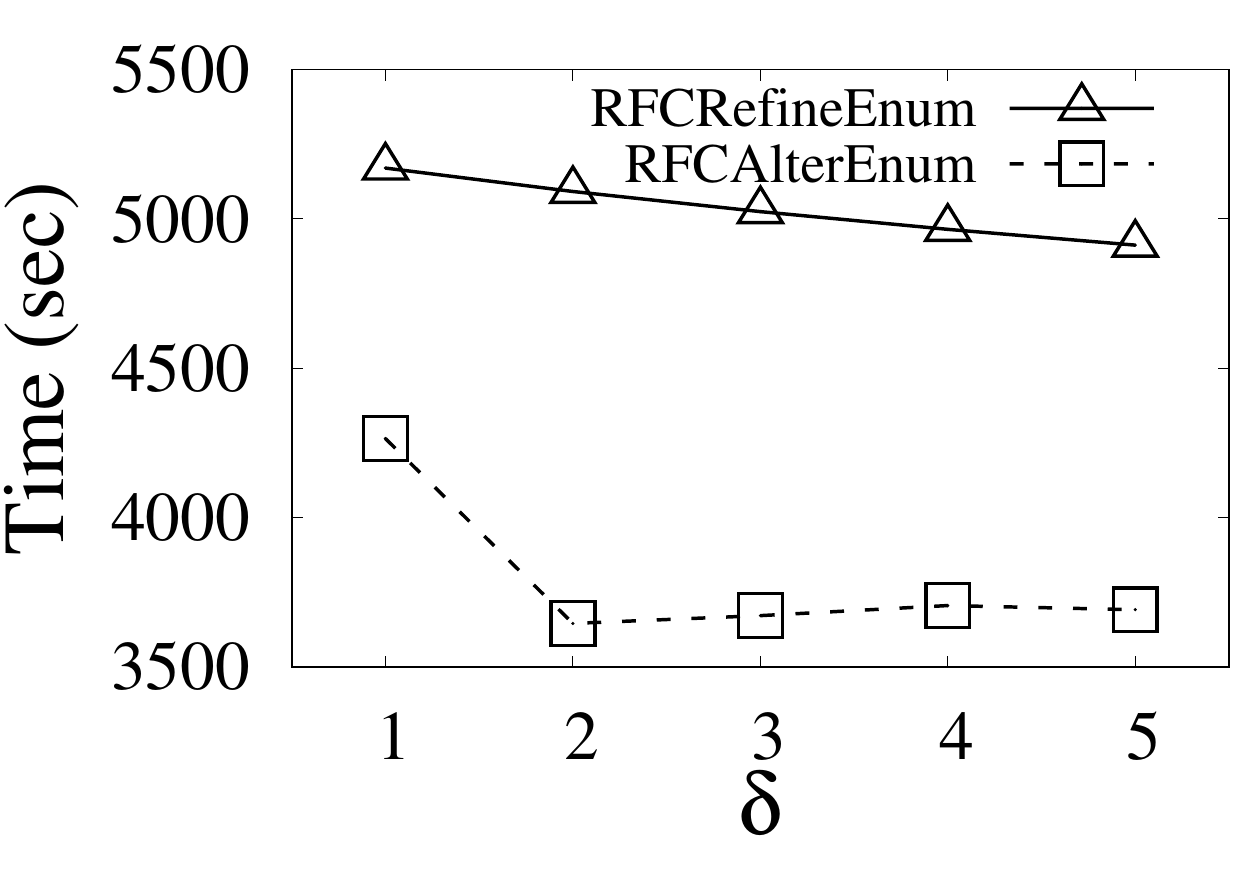}
			}
			\subfigure[{\scriptsize \themarker (vary $\delta$)}]{
				\includegraphics[width=0.4\columnwidth, height=2.5cm]{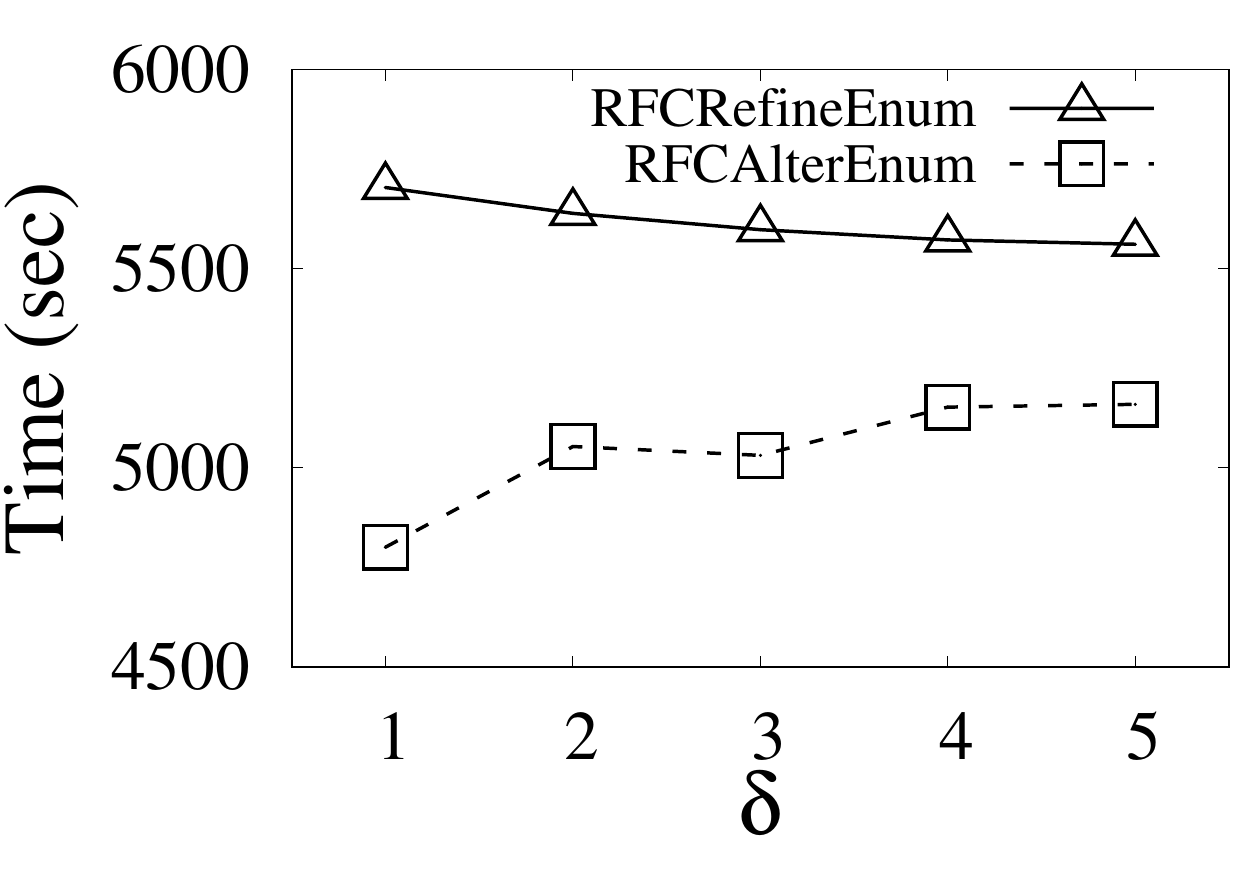}
			}
			\subfigure[{\scriptsize \wiki (vary $\delta$)}]{
				\includegraphics[width=0.4\columnwidth, height=2.5cm]{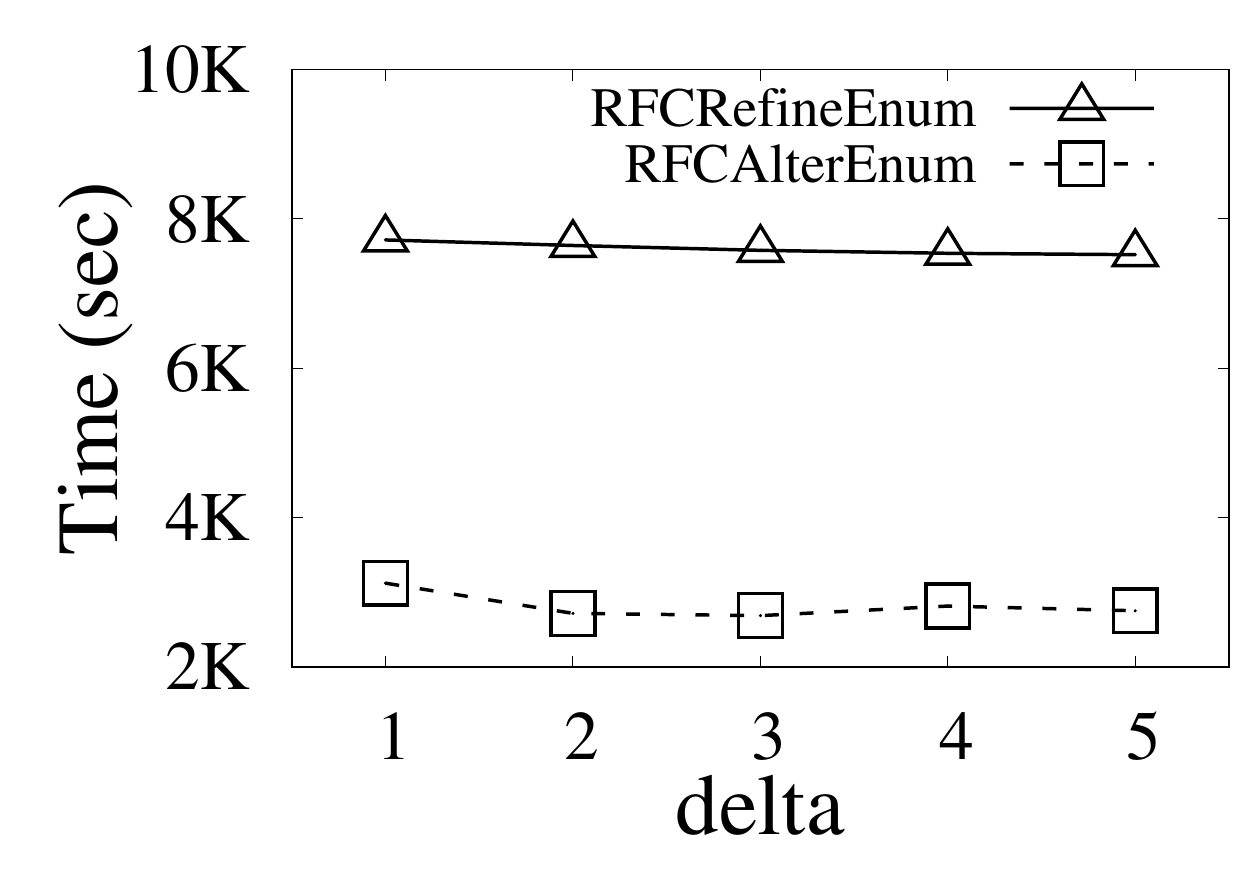}
			}
			\subfigure[{\scriptsize \flixster (vary $\delta$)}]{
				\includegraphics[width=0.4\columnwidth, height=2.5cm]{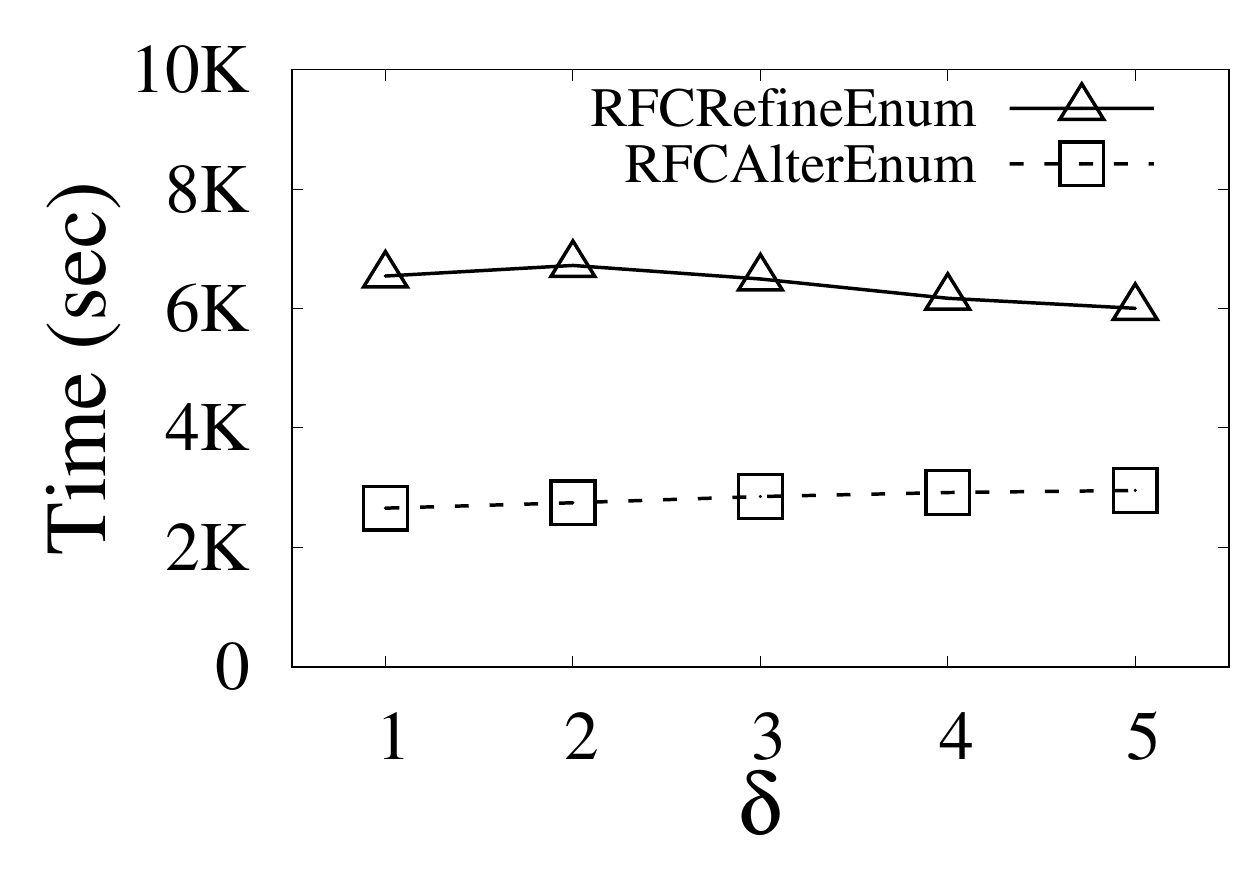}
			}
		\end{tabular}
	\end{center}
	\vspace*{-0.4cm}
	\caption{Running time of the \relativeweak and \relativestrong algorithms}
	\vspace*{-0.3cm}
	\label{fig:exp:RFCE}
\end{figure*}

\stitle{Evaluation of \strong.} We evaluate the runtime of \strong with varying $k$ and $d$. Since the proposed \frorder is tailored for $d = 2$, we only evaluate \strong with \frorder by varying $k$. The experimental results of \strong are illustrated in \figref{fig:exp:SFCE}. In general, the runtime of \strong decreases as $k$ or $d$ increases. This is because for a larger $k$ or $d$, there are fewer cliques satisfying the definition of strong fair clique, thus the runtime for enumerating all strong fair cliques decreases. Additionally, we can see that the \strong algorithms with \frorder and \heurorder are faster than those with \bfsorder and \idorder. For example, for $k = 8$ on \themarker, the \strong algorithms equipped with \frorder and \heurorder consume 2,686 seconds and 2,789 seconds respectively, while the \strong algorithms with \bfsorder and \idorder take 4,225 and 4,834 seconds to output all strong fair cliques respectively. These results confirm the effectiveness of the proposed ordering techniques.

Additionally, by comparing \basestrong and \strong, we find that the running time of \basestrong on all datasets exceeds the time limit, thus we do not show them in \figref{fig:exp:SFCE}. The proposed \strong algorithms, however, work well on most datasets. As aforementioned, to enumerate strong fair cliques, \basestrong needs to find all cliques with size larger than $k\times A_n$ first. The number of such cliques is often extremely large, thus the running time of \basestrong is significantly higher than \strong.


\begin{figure*}[t!]
	\begin{center}
		\begin{tabular}[t]{c}
			\subfigure[{\scriptsize \slashdot (vary $k$)}]{
				\includegraphics[width=0.4\columnwidth, height=2.5cm]{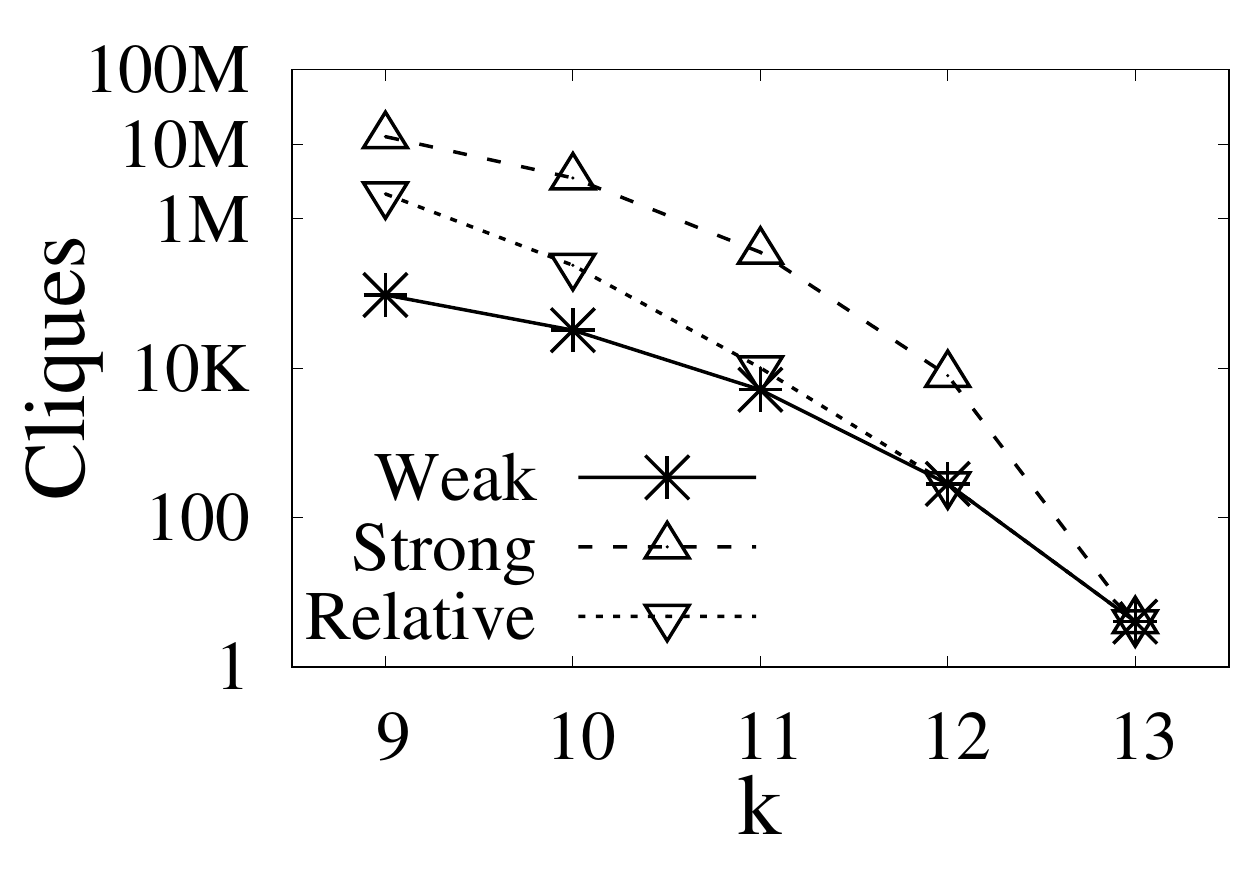}
			}
			\subfigure[{\scriptsize \themarker (vary $k$)}]{
				\includegraphics[width=0.4\columnwidth, height=2.5cm]{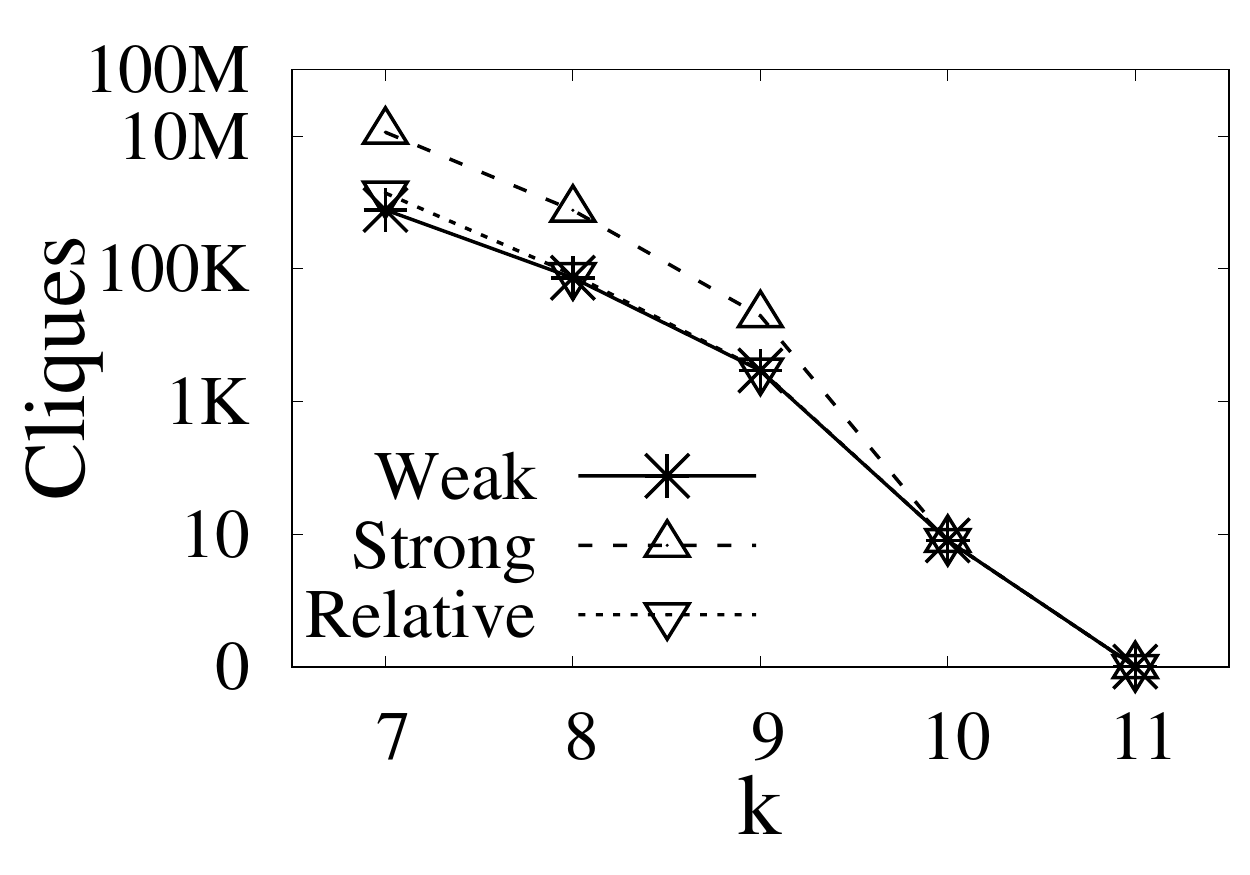}
			}
			\subfigure[{\scriptsize \wiki (vary $k$)}]{
				\includegraphics[width=0.4\columnwidth, height=2.5cm]{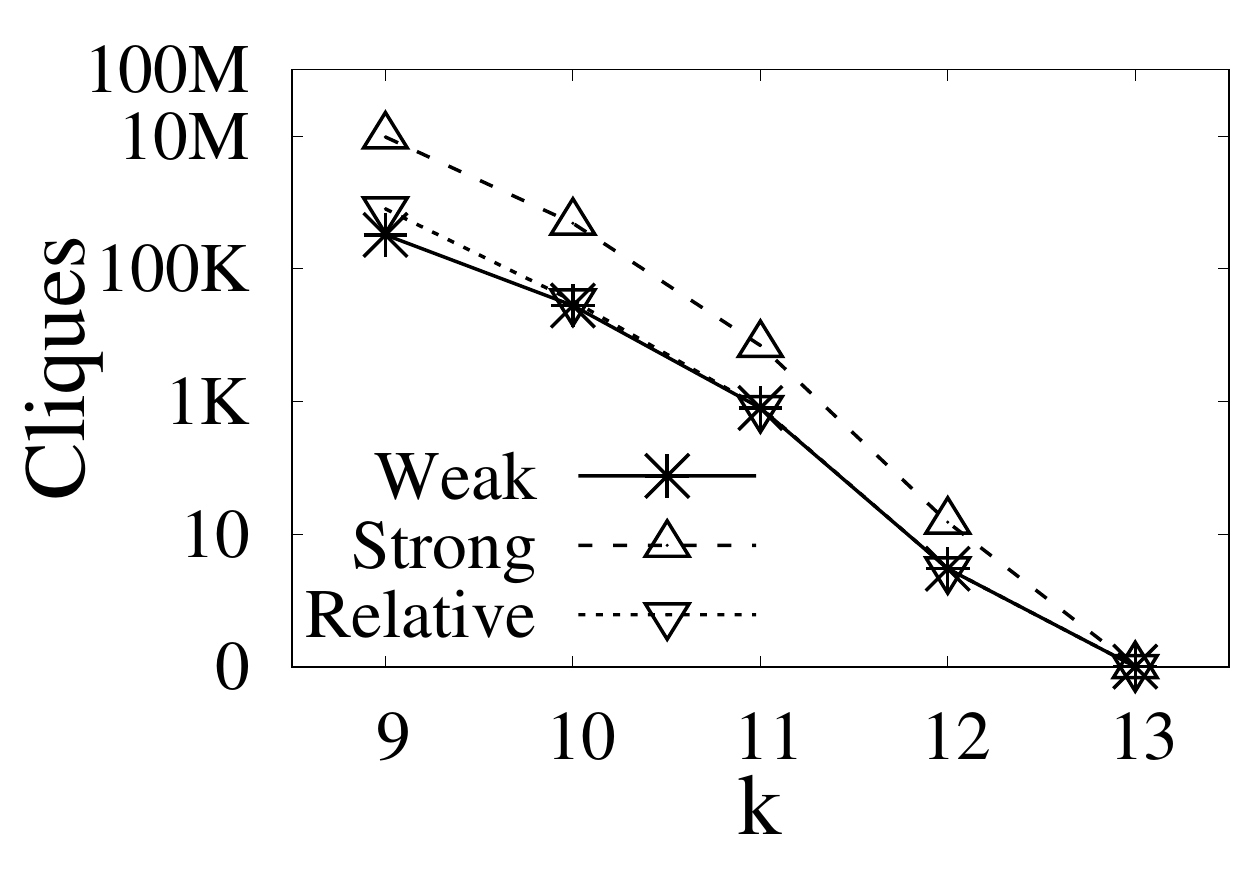}
			}
			\subfigure[{\scriptsize \flixster (vary $k$)}]{
				\includegraphics[width=0.4\columnwidth, height=2.5cm]{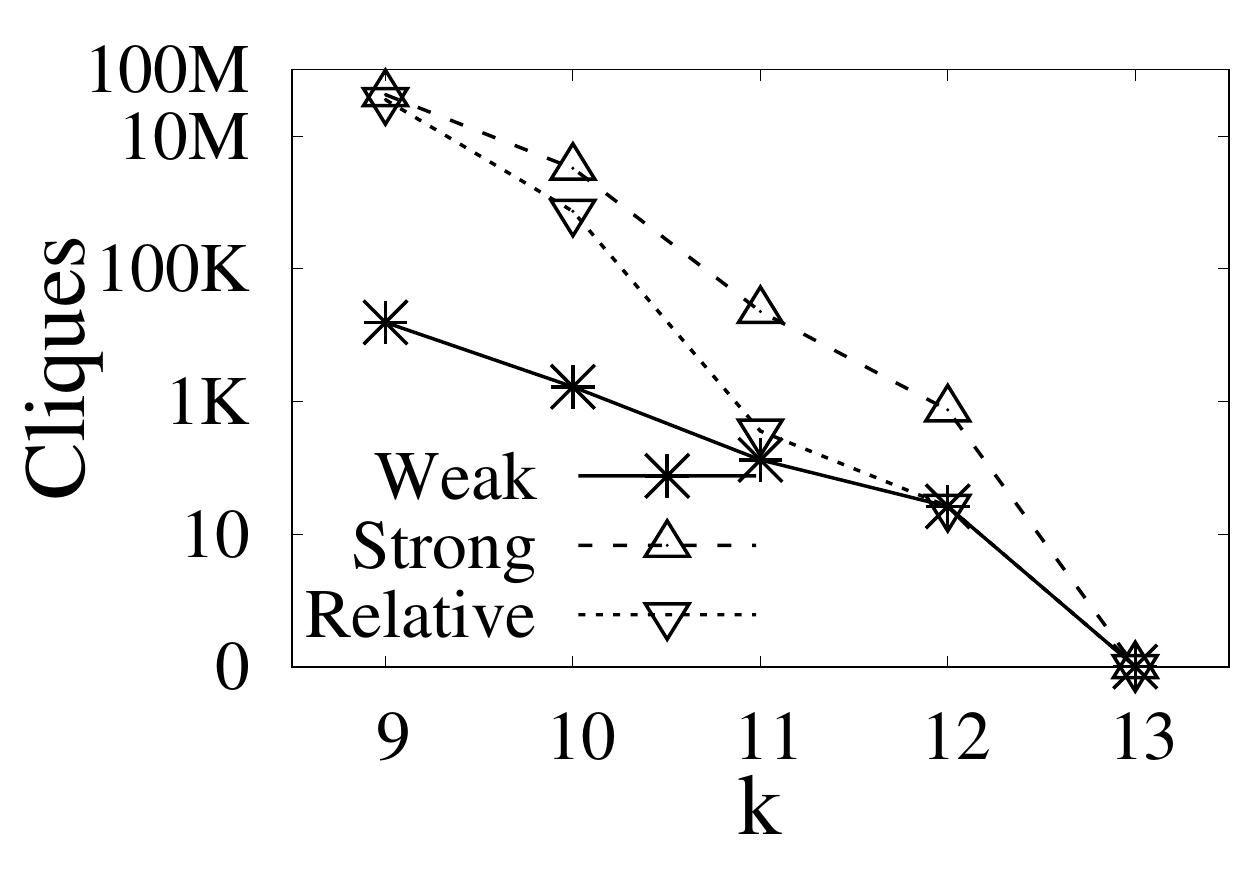}
			}
			\vspace*{-0.2cm} \\
			
			\subfigure[{\scriptsize \slashdot (vary $d$)}]{
				\includegraphics[width=0.4\columnwidth, height=2.5cm]{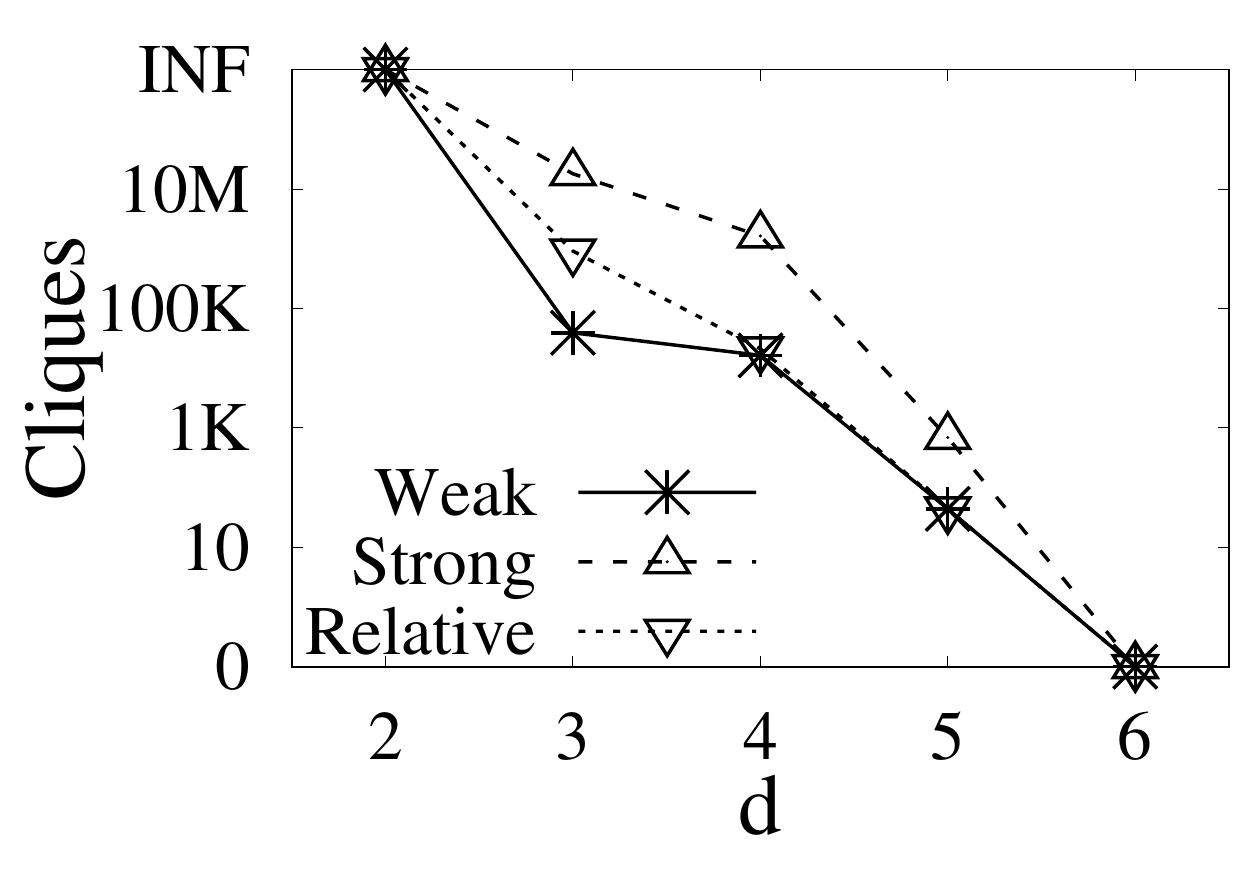}
			}
			\subfigure[{\scriptsize \themarker (vary $d$)}]{
				\includegraphics[width=0.4\columnwidth, height=2.5cm]{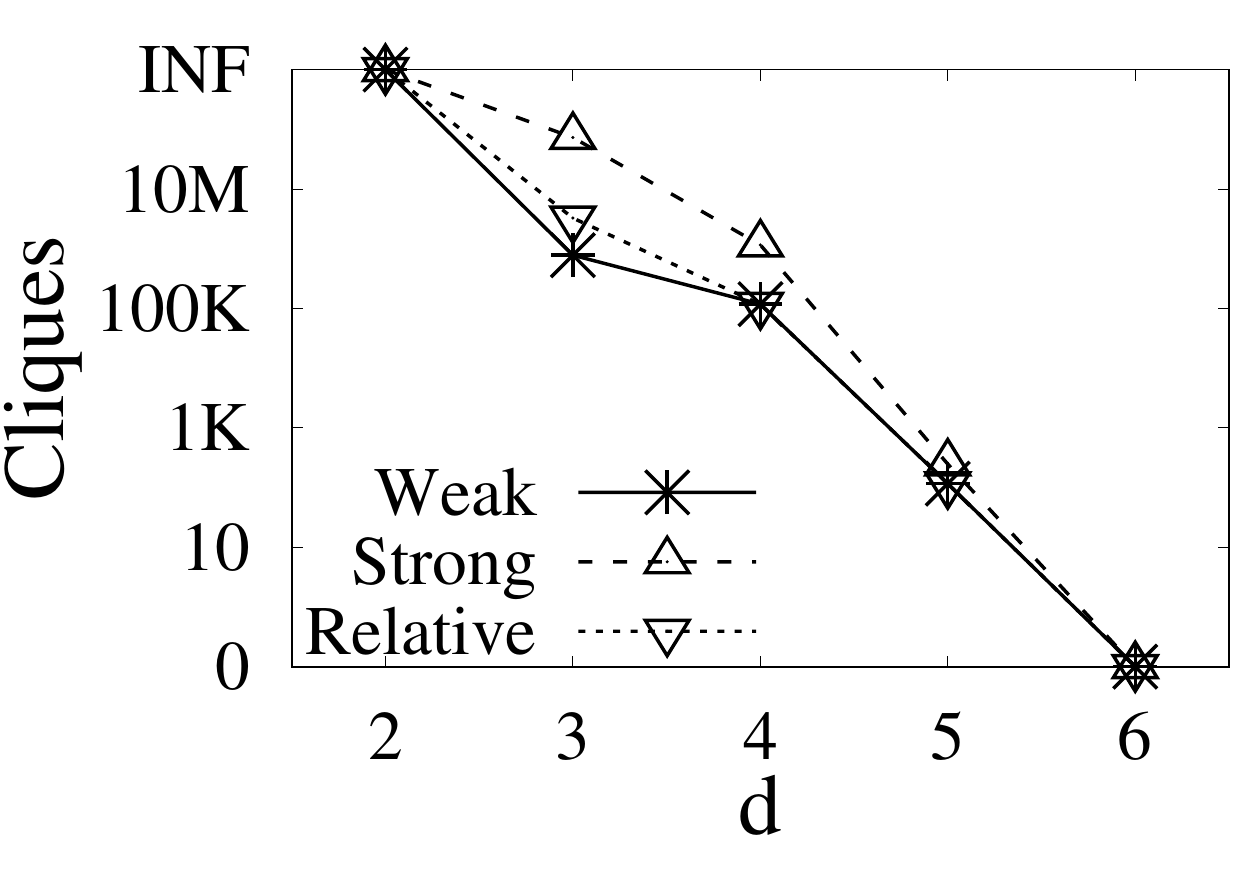}
			}
			\subfigure[{\scriptsize \wiki (vary $d$)}]{
				\includegraphics[width=0.4\columnwidth, height=2.5cm]{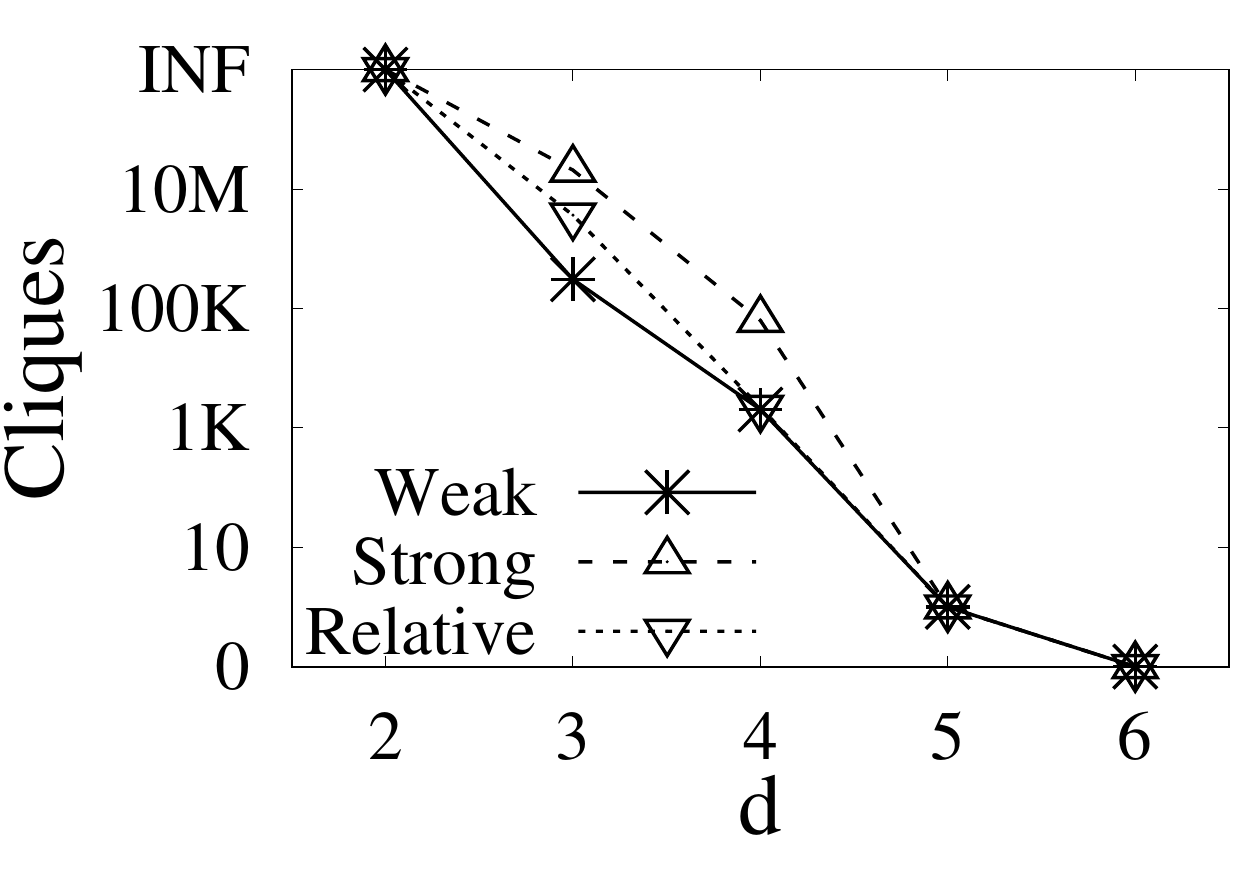}
			}
			\subfigure[{\scriptsize \flixster (vary $d$)}]{
				\includegraphics[width=0.4\columnwidth, height=2.5cm]{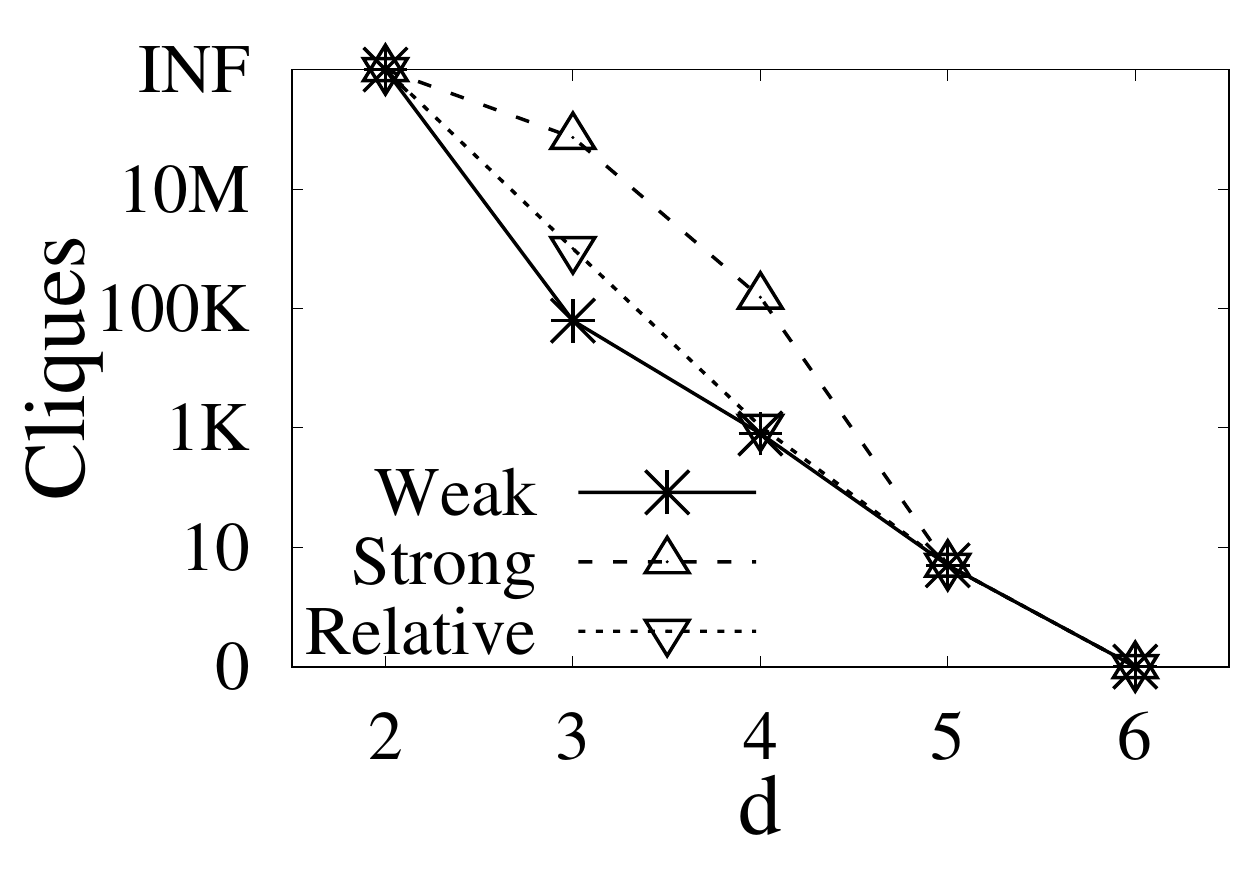}
			}
			\vspace*{-0.2cm} \\
			
			\subfigure[{\scriptsize \slashdot (vary $\delta$)}]{
				\includegraphics[width=0.4\columnwidth, height=2.5cm]{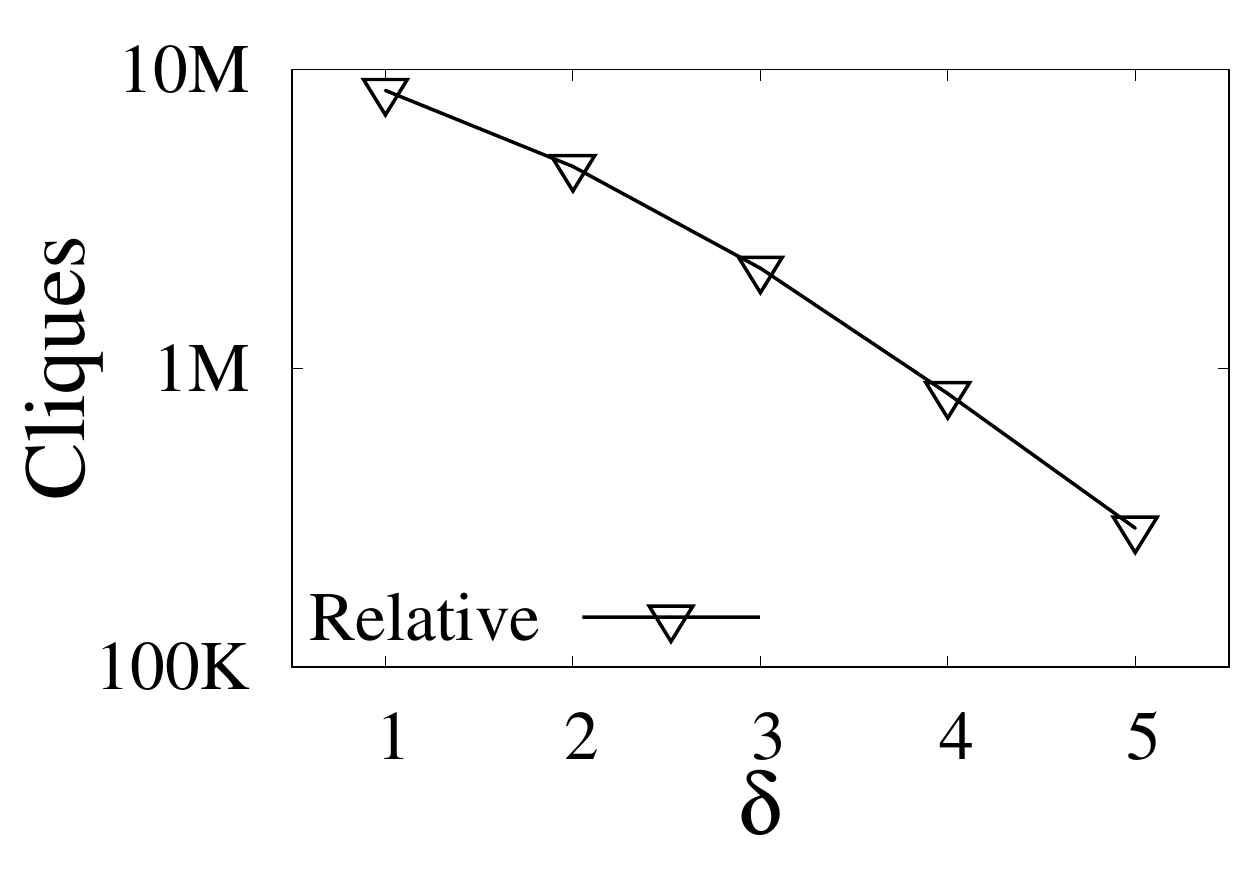}
			}
			\subfigure[{\scriptsize \themarker (vary $\delta$)}]{
				\includegraphics[width=0.4\columnwidth, height=2.5cm]{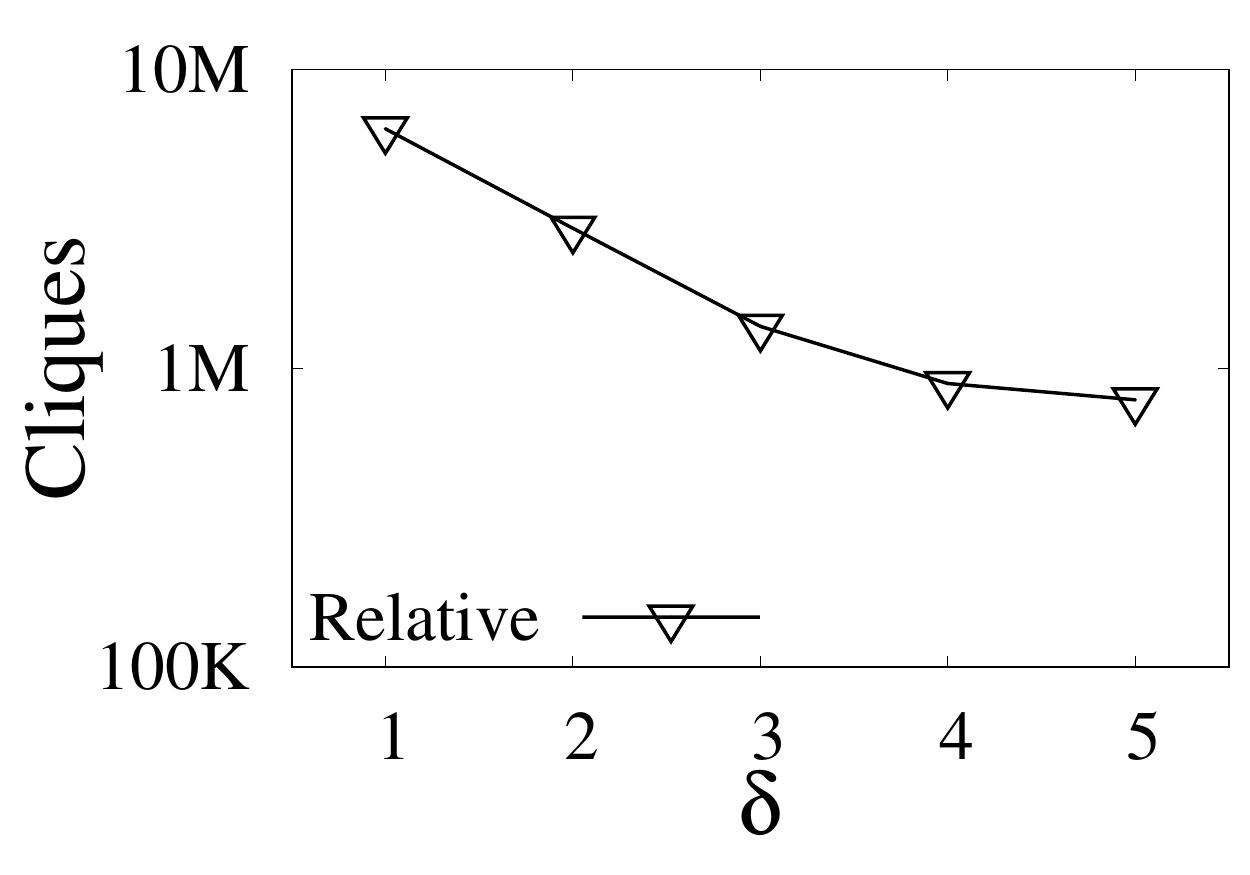}
			}
			\subfigure[{\scriptsize \wiki (vary $\delta$)}]{
				\includegraphics[width=0.4\columnwidth, height=2.5cm]{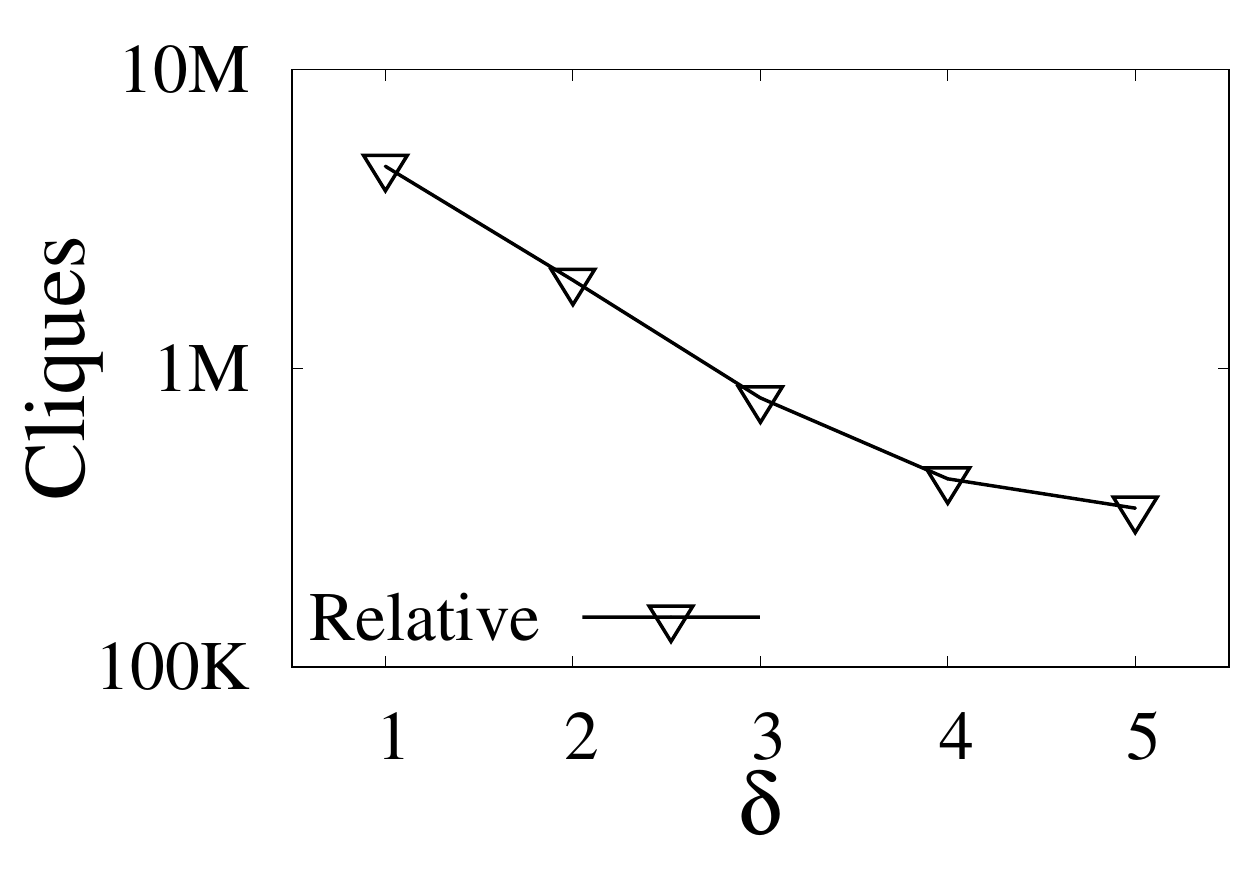}
			}
			\subfigure[{\scriptsize \flixster (vary $\delta$)}]{
				\includegraphics[width=0.4\columnwidth, height=2.5cm]{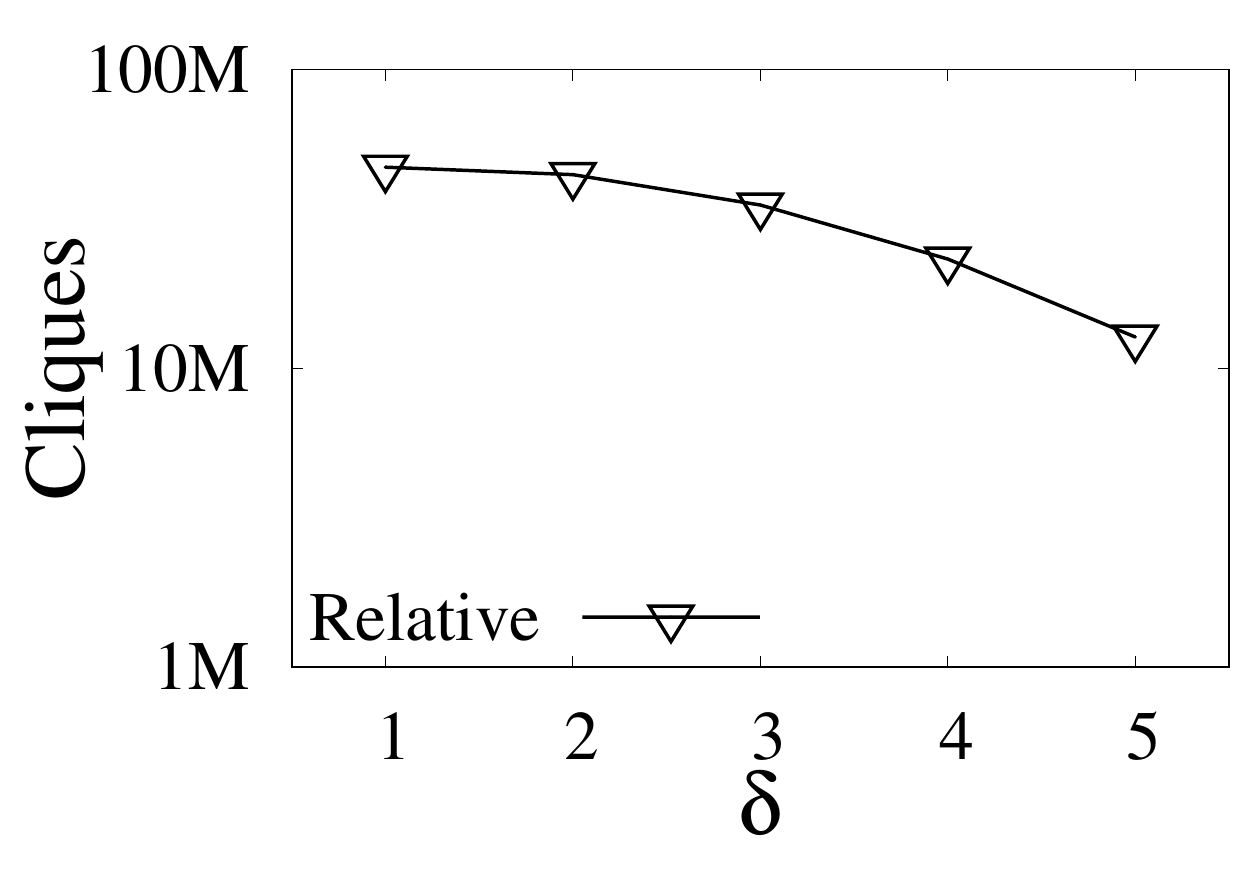}
			}
		\end{tabular}
	\end{center}
	\vspace*{-0.4cm}
	\caption{The number of weak fair cliques, strong fair cliques and relative fair cliques on various datasets}
	\vspace*{-0.4cm}
	\label{fig:exp:cliquenum}
\end{figure*}

\stitle{Evaluation of \relativeweak and \relativestrong.} Here, we evaluate the proposed relative fair clique enumeration algorithms, i.e., \relativeweak and \relativestrong, with varying $k$, $d$ and $\delta$. The experimental results are illustrated in \figref{fig:exp:RFCE}. In general, the runtime of \relativeweak and \relativestrong decreases as $k$ or $d$ increases as expected. This is because for a larger $k$ or $d$, fewer cliques satisfying the definition of a relative fair clique, thus decreasing the runtime for enumerating all relative fair cliques. These results are consistent with the previous findings. For the parameter $\delta$, the runtime of \relativeweak decreases with increasing $\delta$, while the \relativestrong achieves the maximum runtime at $\delta=1$, and then its runtime changes very smoothly with increasing $\delta$. This is because the \relativeweak algorithm performs \weak to find all weak fair cliques and then enumerates relative fair cliques contained in them. A larger $\delta$ implies that a relative fair clique approaches a weak fair clique, thus decreasing the enumeration depth and reducing the time cost. For the \relativestrong, it adopts attribute-alternatively-selection strategy to enumerate relative fair cliques, thus the runtime is insensitive to the difference threshold $\delta$. In particular, when $\delta$ equals $1$ and the attribute with the minimum number of nodes is $a_{\phi}$, the numbers of nodes with attributes $a_0, a_1, ..., a_{\phi-1}$ reach the maximum. Thus, the \relativestrong needs to update candidates sets to be empty for $a_0, a_1, ..., a_{\phi-1}$ which causes a little bit of increase in running time. 

From \figref{fig:exp:RFCE}, we can also see that the \relativestrong algorithm is faster than \relativeweak within all parameter settings over all datasets. For example, in the case of $k=9$, the runtime of \relativeweak and \relativestrong algorithms on \wiki is 7,718 seconds and 2,683 seconds, respectively. Clearly, the former is around 2.877 times slower than the latter. While for $d=3$ on \slashdot, the \relativeweak algorithm consumes 9,628 seconds, while the \relativestrong takes 2,461 seconds to output all relative fair cliques which is roughly 3.912 times faster than that of \relativeweak. For $\delta = 3$ on \flixster, the \relativeweak and \relativestrong algorithms take 6,492 seconds and 2,849 seconds to output the results. The runtime of \relativestrong is roughly 2.279 times faster than that of \relativeweak. In addition, we also evaluate the proposed algorithms by comparing them with the \baserelative algorithm. The running time of \baserelative on all datasets exceeds the time limit, thus we do not show them in \figref{fig:exp:RFCE}. From \figref{fig:exp:RFCE}, the proposed \relativeweak and \relativestrong algorithms work well on most datasets. To search relative fair cliques, the \baserelative algorithm needs to find all maximal cliques first, thus the running time is significantly higher than our proposed algorithms. These results confirm the efficiency of the proposed \relativeweak and \relativestrong algorithms.

\begin{figure}[t!]
	\begin{center}
		\begin{tabular}[t]{c}
			\subfigure[{\scriptsize \flixster, \weak (vary $m$)}]{
				\label{fig:exp-scala-vary-m-weak-flixster}
				\includegraphics[width=0.4\columnwidth, height=2.5cm]{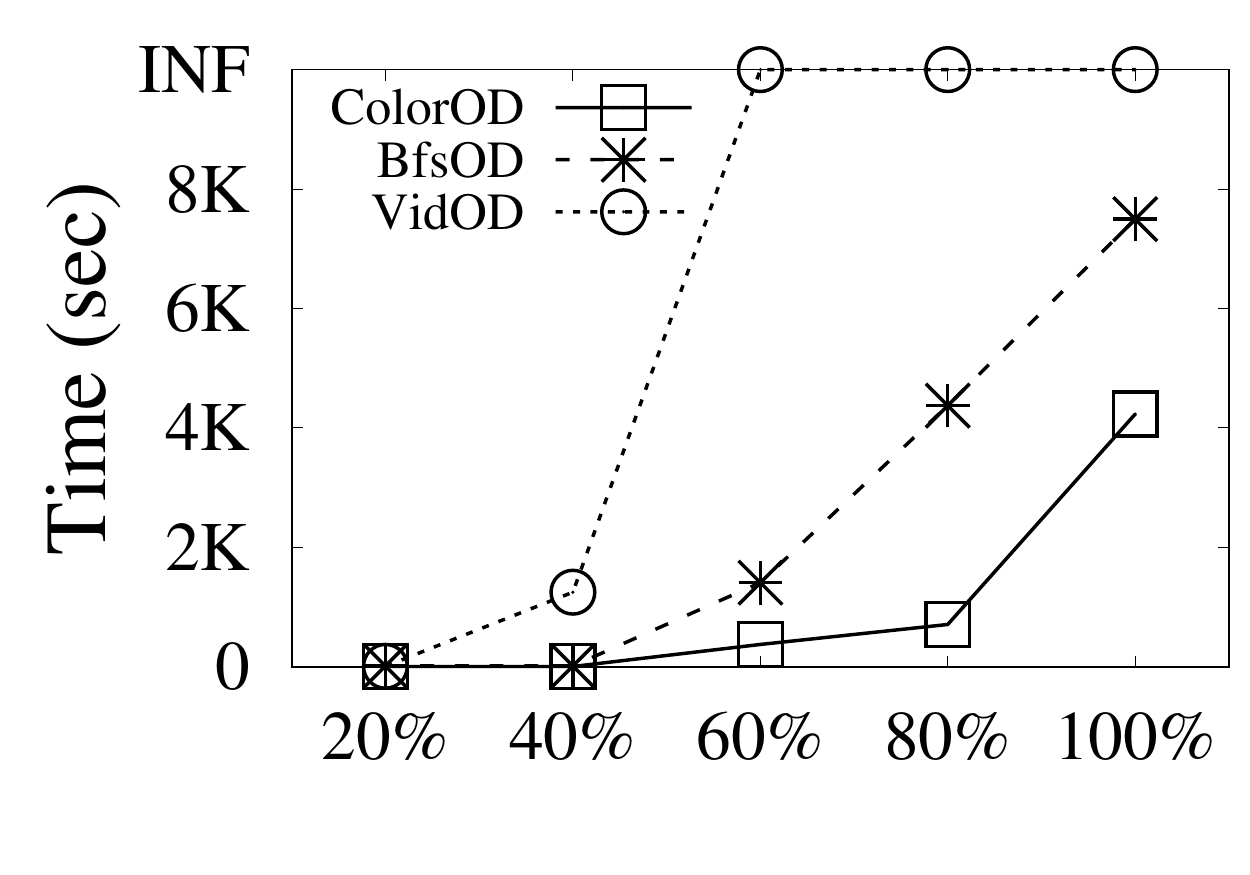}
			}
			\subfigure[{\scriptsize \flixster, \strong (vary $m$)}]{
				\label{fig:exp-scala-vary-m-strong-flixster}
				\includegraphics[width=0.4\columnwidth, height=2.5cm]{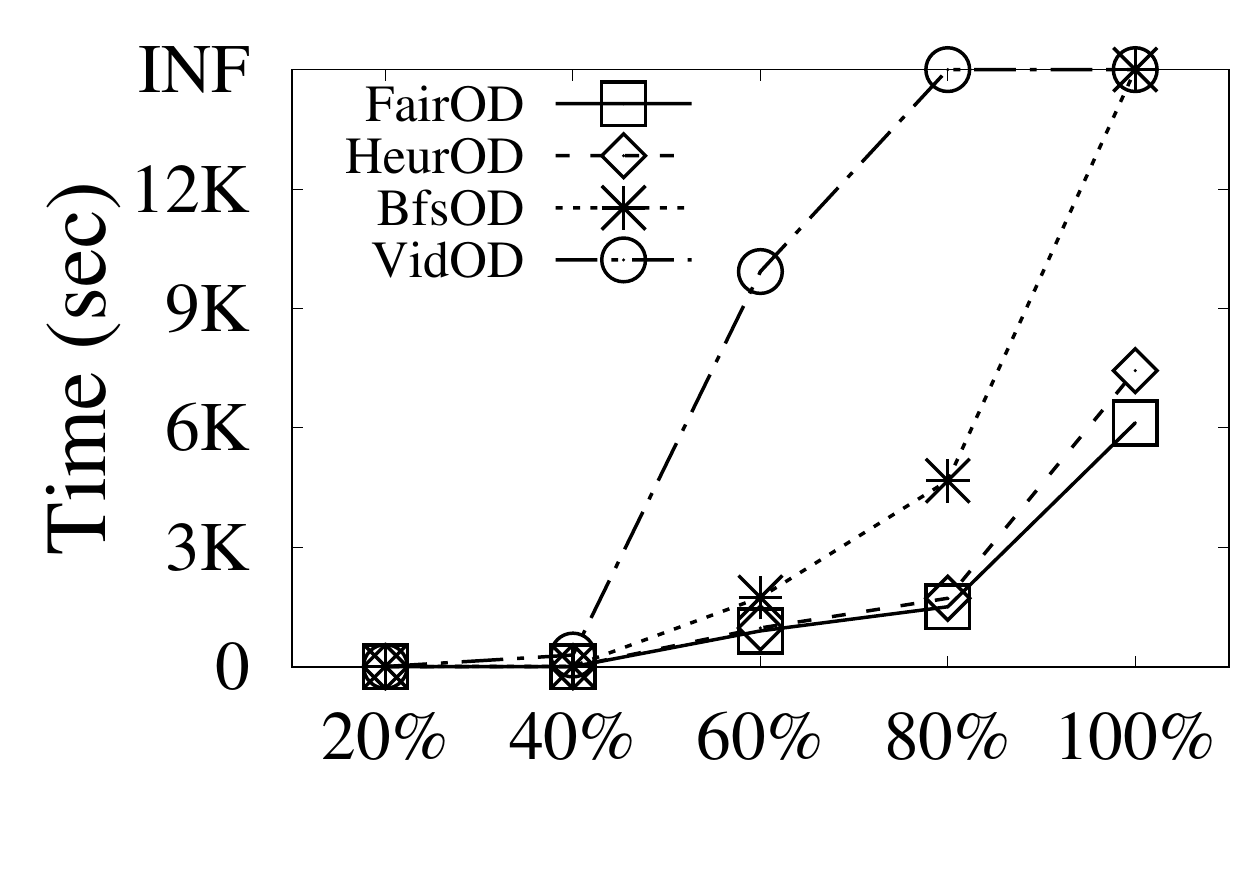}
			}\\
			
			\subfigure[{\scriptsize \flixster, \relativeweak and \relativestrong (vary $m$)}]{
				\label{fig:exp-scala-vary-m-relative-flixster}
				\includegraphics[width=0.4\columnwidth, height=2.5cm]{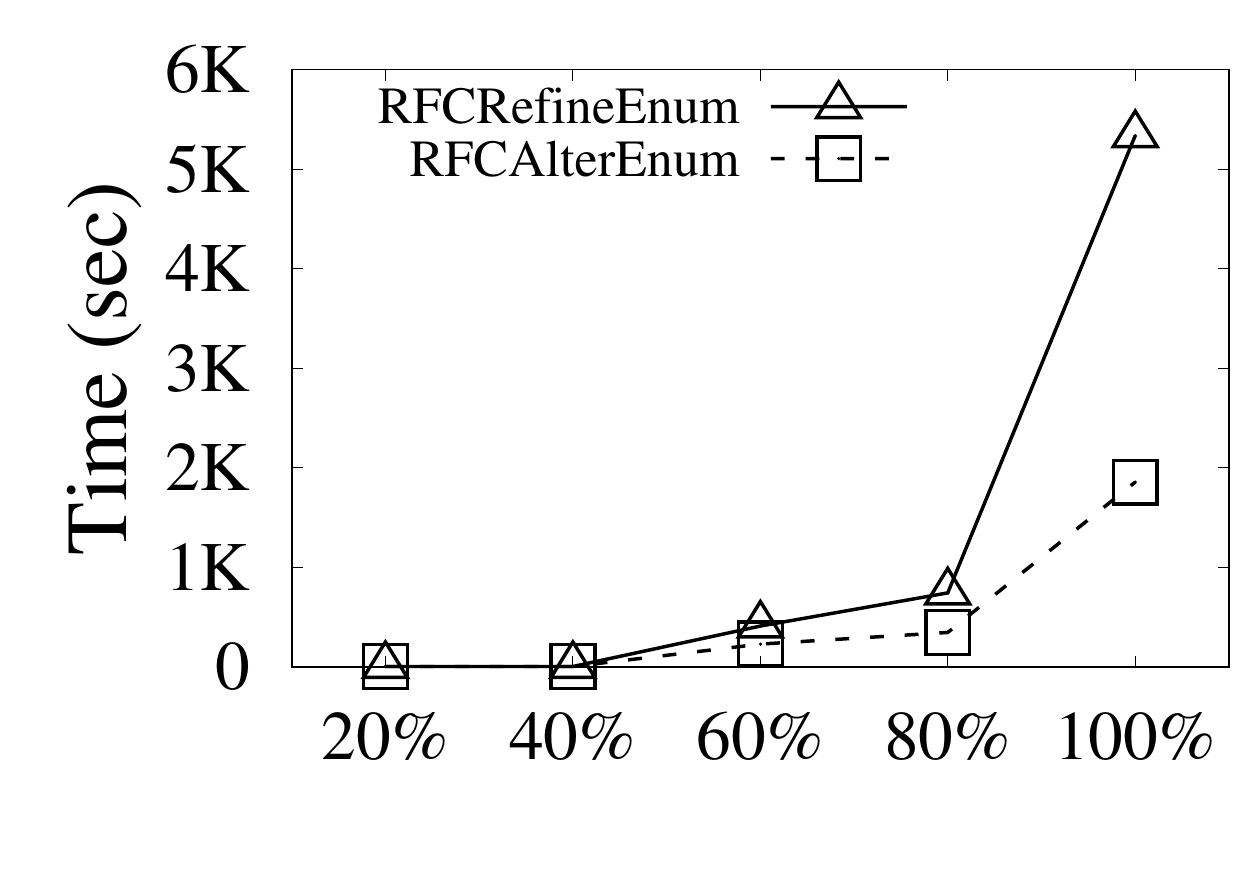}
			}
		\end{tabular}
	\end{center}
	\vspace*{-0.4cm}
	\caption{Scalability of \weak, \strong, \relativeweak and \relativestrong algorithms}
	\label{fig:exp-scalability-test}
	\vspace*{-0.4cm}
\end{figure}

\stitle{The number of fairness-aware cliques.} \figref{fig:exp:cliquenum} (a)-(d) show the numbers of weak fair cliques, strong fair cliques and relative fair cliques with different $k$. Clearly, there are significant numbers of fair cliques in each dataset. In general, the number of strong fair cliques is larger than that of relative fair cliques, and the number of relative fair cliques is larger than that of weak fair cliques. This finding is consistent with our analysis in \secref{sec:preliminaries}, since a weak fair clique often contains a set of relative fair cliques, and a relative fair clique includes a set of strong fair cliques. Additionally, we can see that the number of fair cliques decreases when $k$ increases. This is because with a larger $k$, both the fairness and clique constraints become stricter, thus resulting in fewer fair cliques. Similar results can also be observed when varying $d$ from \figref{fig:exp:cliquenum} (e)-(h).  \figref{fig:exp:cliquenum} (i)-(l) also illustrate the numbers of relative fair cliques with different $\delta$. As expected, the number of relative fair cliques decreases with increasing $\delta$. This is because a relative fair clique with larger $\delta$ often contains many relative fair cliques with smaller $\delta$ according to the maximality in Definition \ref{def:relativefairclique}. These results confirm that our relative fair clique model indeed achieves a great compromise between the weak fair clique and strong fair clique models by introducing the difference threshold $\delta$, which is consistent with our analysis in \secref{sec:preliminaries}.

\stitle{Scalability testing.} To evaluate the scalability of the proposed algorithms, we generate four subgraphs for each dataset by randomly picking 20\%-80\% of the edges, and evaluate the runtime of all the proposed algorithms. \figref{fig:exp-scalability-test} illustrates the results on \flixster. The results on the other datasets are consistent. In \figref{fig:exp-scala-vary-m-weak-flixster}, the runtime of \weak with \bfsorder and \idorder increases sharply as the graph size increases, while for \colororder, it increases smoothly with varying $m$. Moreover, the \colororder ordering performs much better than the other orderings with all parameter settings, which is consistent with our previous findings. Analogously, when varying $m$, the runtime of \strong with \bfsorder and \idorder increases sharply with respect to the graph size in \figref{fig:exp-scala-vary-m-strong-flixster}. However, for \strong with \frorder and \heurorder, the runtime increases smoothly with $m$ increases. From \figref{fig:exp-scala-vary-m-relative-flixster}, we can also see that for relative fair clique enumeration algorithms, the runtime of \relativestrong increases very smoothly with increasing $m$, while the runtime of \relativeweak increases more sharply. Again, \relativestrong is significantly faster than \relativeweak, which is consistent with our previous findings. These results demonstrate the high scalability of the proposed algorithms. 

\begin{figure}[t!]
	\begin{center}
		\begin{tabular}[t]{c}
			\subfigure[{\scriptsize \weak}]{
				\includegraphics[width=0.43\columnwidth, height=2.3cm]{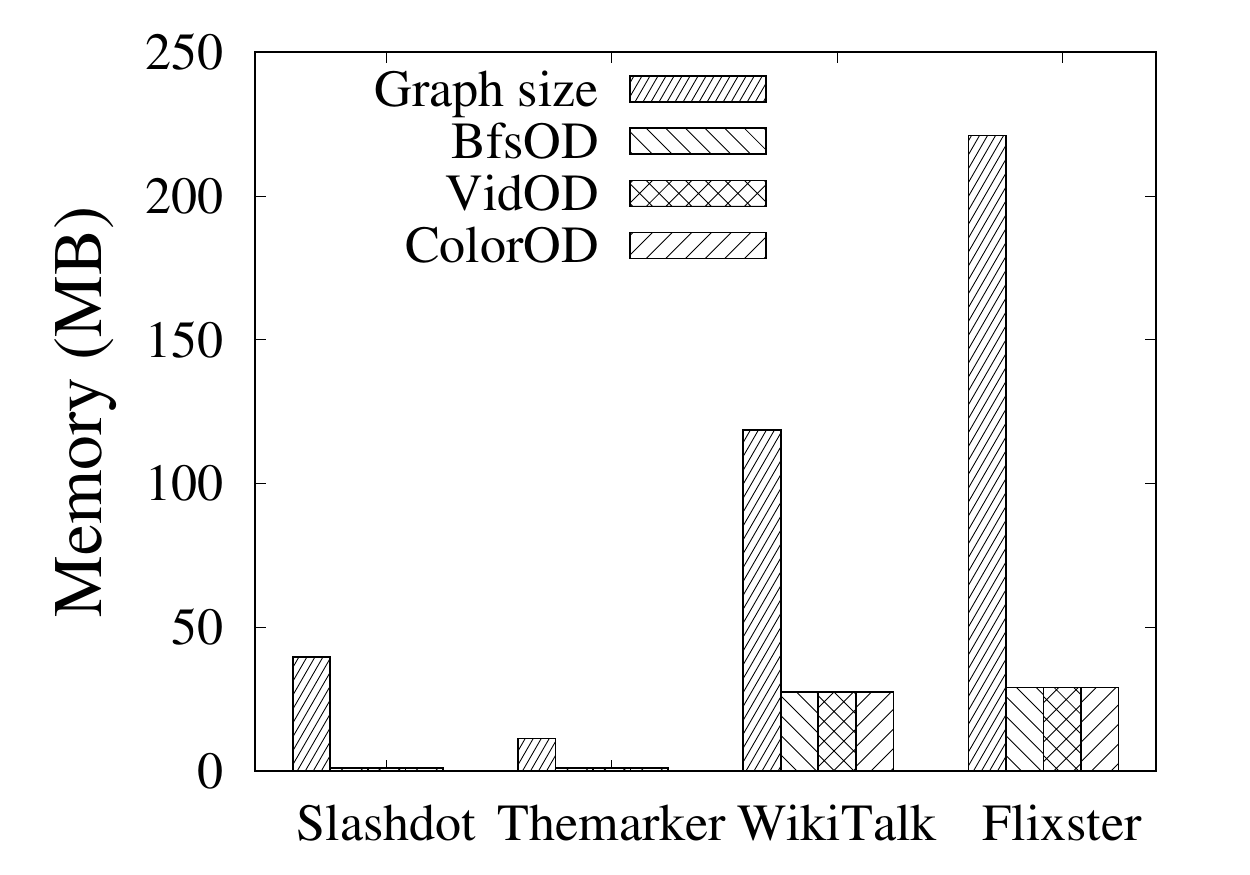}
			}
			\subfigure[{\scriptsize \strong}]{
				\includegraphics[width=0.43\columnwidth, height=2.3cm]{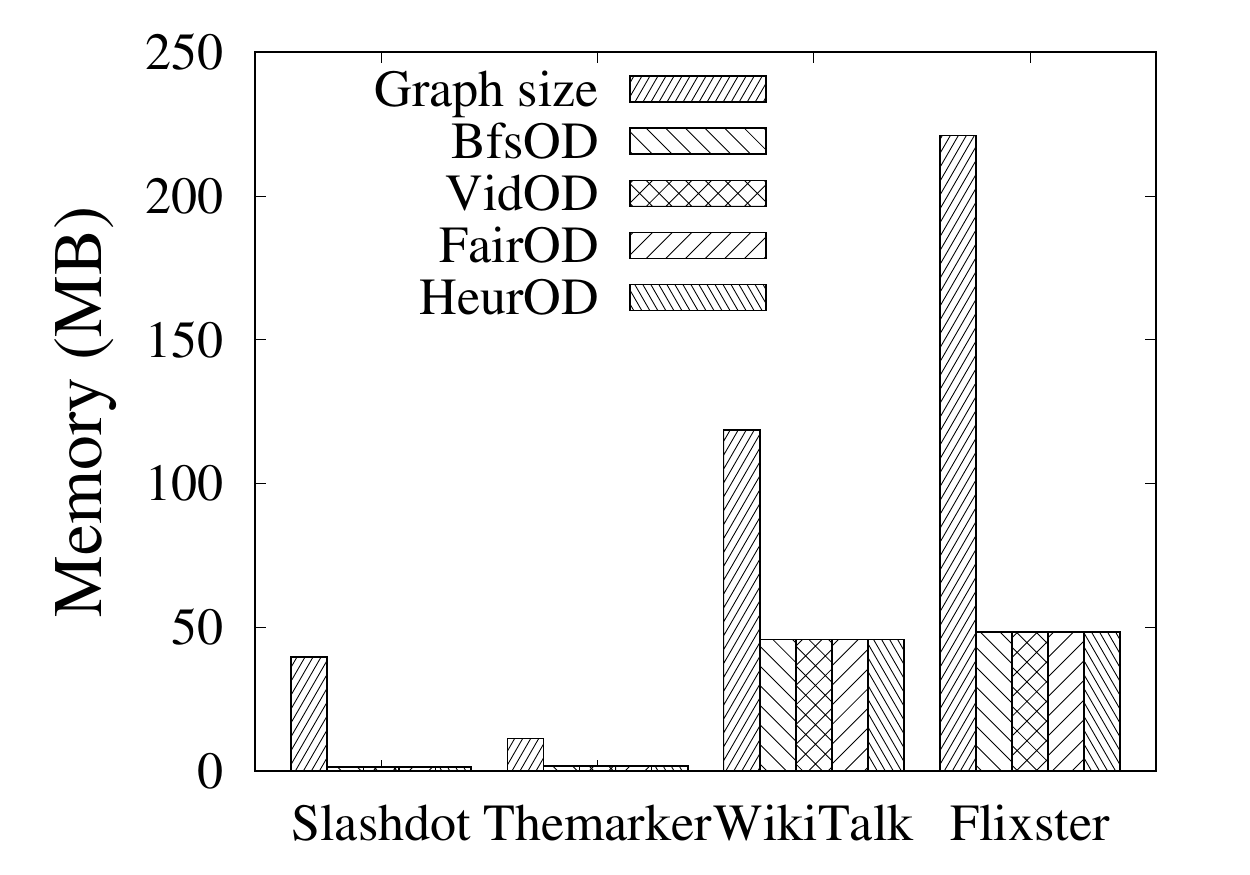}
			}\\
			
			\subfigure[{\scriptsize \relativeweak and \relativestrong}]{
				\includegraphics[width=0.43\columnwidth, height=2.3cm]{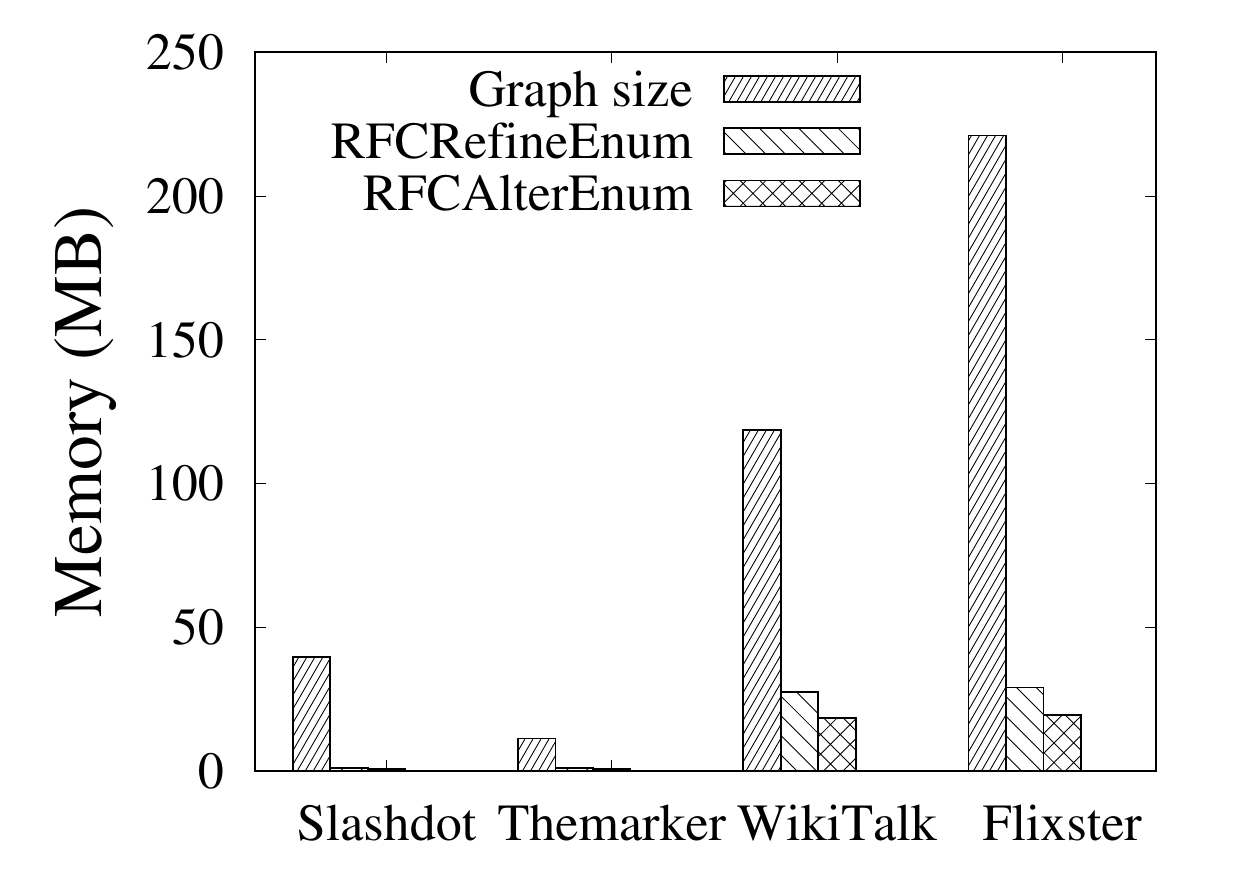}
			}
			\vspace*{-0.2cm}
		\end{tabular}
	\end{center}
	\vspace*{-0.2cm}
	\caption{Memory overhead}
	\vspace*{-0.4cm}
	\label{fig:exp-Memory-overhead}
\end{figure}

\stitle{Memory overhead.} \figref{fig:exp-Memory-overhead} shows the memory overheads of \weak, \strong, \relativeweak and \relativestrong algorithms on all datasets. Note that the memory costs of different algorithms do not include the size of the graph. From \figref{fig:exp-Memory-overhead} (a)-(b), we can see that the memory usages of \weak and \strong with different orderings are always smaller than the graph size. This is because both the \weak and \strong algorithms follow a depth-first manner, thus the space overhead is linear. Additionally, the memory overheads of \weak and \strong are robust with respect to different orderings. This is because the space usage in the enumeration procedure is mainly dominated by the depth of the enumeration tree. Since the tree depth is determined by the clique size, the space overhead is insensitive to different orderings. As can be seen from \figref{fig:exp-Memory-overhead} (c), the memory occupancy of \relativeweak and \relativestrong are also significantly smaller than the graph size since they also enumerate relative fair cliques in a depth-first manner like \weak and \strong. Compared with \relativeweak, the \relativestrong algorithm occupies less memory because the difference threshold $\delta$ can reduce the search space once the minimum number of nodes of an arbitrary attribute is determined based on the attribute-alternatively-selection strategy.

\subsection{Case study} \label{sec:casestudy}

We conduct a case study on a collaboration network \dblp to evaluate the effectiveness of our algorithms. The \dblp dataset is downloaded from \url{dblp.uni-trier.de/xml/}. We extract a subgraph \dbcase from \dblp which contains the authors who had published at least one paper in the database ($DB$), data mining ($DM$), and artificial intelligence ($AI$) related conferences. The \dbcase subgraph contains 52,106 vertices (authors) and 341,382 undirected edges. The attribute $A$ represents the author's main research area with $A_{val} = \{DB, DM, AI\}$. Each vertex has one attribute value selected from the set $A_{val}$. We set the attribute value for each vertex based on the maximum number of papers that the author published in the related conferences. For example, if an author has published 20 papers in $DB$ related conferences and 5 papers in $DM$ related conferences, we choose $DB$ as the author's attribute value.

\begin{figure}[t!]
	\begin{center}
		\begin{tabular}[t]{c}
            \subfigure[{\scriptsize a weak fair clique ($k=2$)}]{
            	\raisebox{0.05\height}{
					\includegraphics[width=0.6\columnwidth]{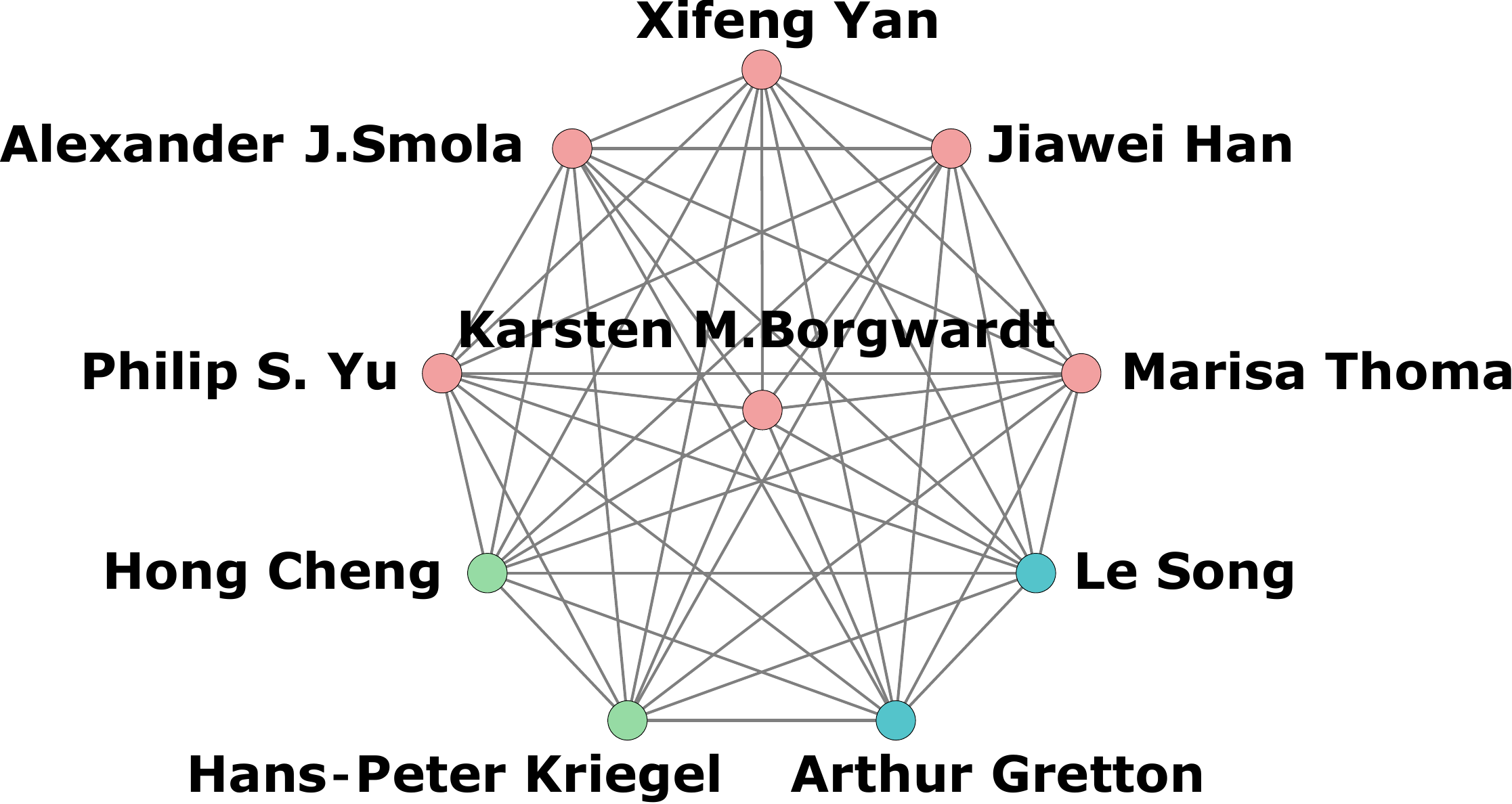}
				}
                \label{fig:exp-cs-3-weak}
			}\\
            \vspace*{0.2cm}

			\subfigure[{\scriptsize a strong fair clique ($k=2$)}]{
				\raisebox{0.05\height}{
					\includegraphics[width=0.45\columnwidth,]{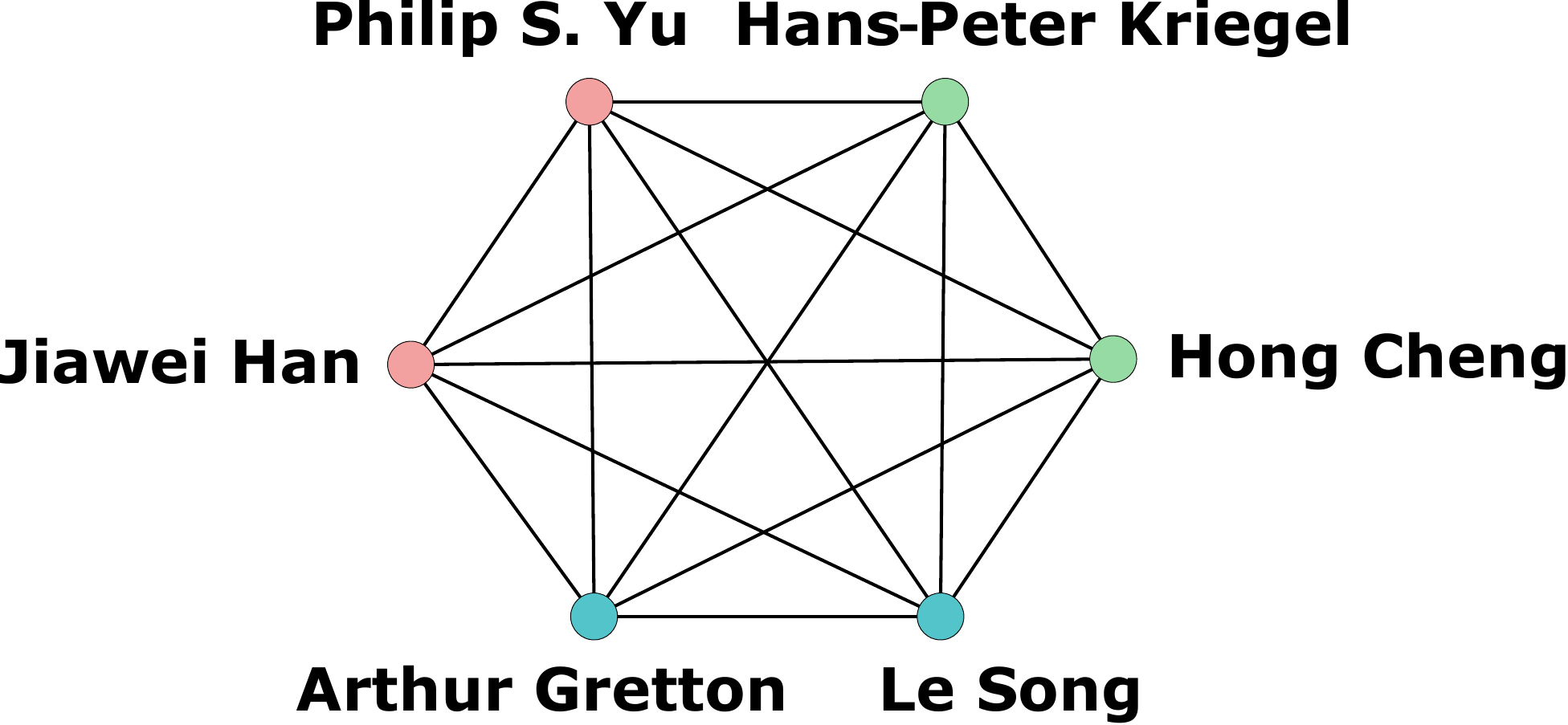}
				}
                \label{fig:exp-cs-3-strong1}
			}
			\subfigure[{\scriptsize a strong fair clique ($k=2$)}]{
				\raisebox{0.05\height}{
					\includegraphics[width=0.45\columnwidth,]{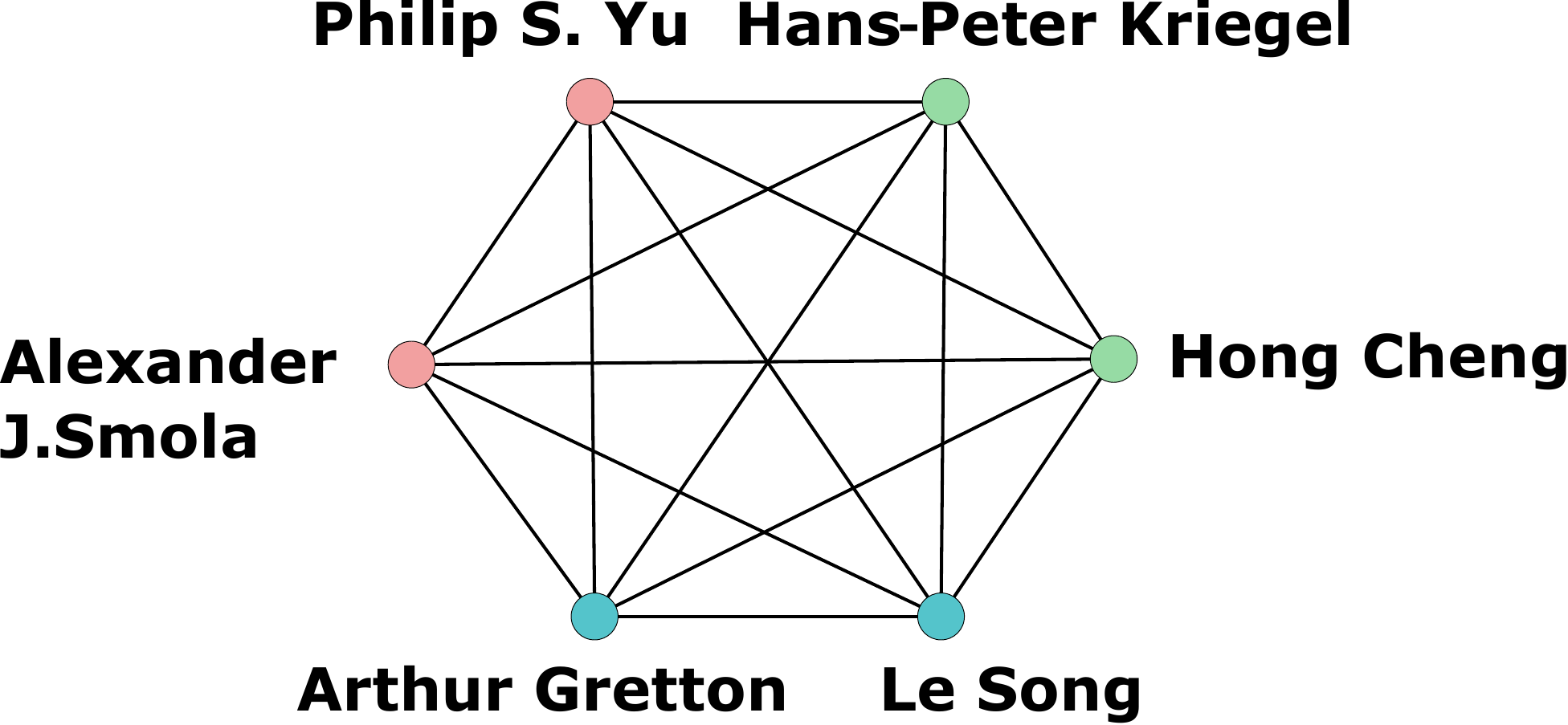}
                }
                \label{fig:exp-cs-3-strong2}
			}\\
		\end{tabular}
	\end{center}
	\vspace*{-0.3cm}
	\caption{Results of \weak and \strong on \dbcase with $A_{val} = \{DB, DM, AI\}$}
	\label{fig:exp:csattr3}
	\vspace*{-0.4cm}
\end{figure}

We perform the \weak, \strong and \relativeweak (\relativestrong) algorithms to find all weak fair cliques, strong fair cliques and relative fair cliques on \dbcase with $k = 2$ and $\delta=2, 3$. All algorithms apply \colorful to prune the unpromising vertices. The remaining graph after pruning by \colorful only has 61 vertices and 516 edges. \figref{fig:exp-cs-3-weak} shows a weak fair clique with size $10$, which involves 6 authors of $DB$, 2 authors of $DM$ and 2 authors of $AI$. We use different colors to represent the main research area of these authors, namely, $pink = DB$, $green = DM$, and $blue = AI$. Clearly, the number of vertices with different attribute values is no less than $k = 2$. These results indicate that \weak can find fair communities with diverse research areas. However, in \figref{fig:exp-cs-3-weak}, the weak fair clique is imbalanced (w.r.t.\ different attributes) due to the high percentage of authors with $DB$. \figref{fig:exp-cs-3-strong1} and \figref{fig:exp-cs-3-strong2} show two strong fair cliques which are also subgraphs of the clique in \figref{fig:exp-cs-3-weak}. This is consistent with the finding that a strong fair clique must be contained in a weak fair clique. As expected, the number of authors with different attribute values is exactly equal to $2$, thus it can avoid the \emph{attribute imbalance} problem in the weak fair clique. 

\begin{figure}[t!]
	\begin{center}
		\begin{tabular}[t]{c}
			\subfigure[{\scriptsize a relative fair clique ($\delta=2$)}]{
				\raisebox{0.08\height}{
					\includegraphics[width=0.45\columnwidth,]{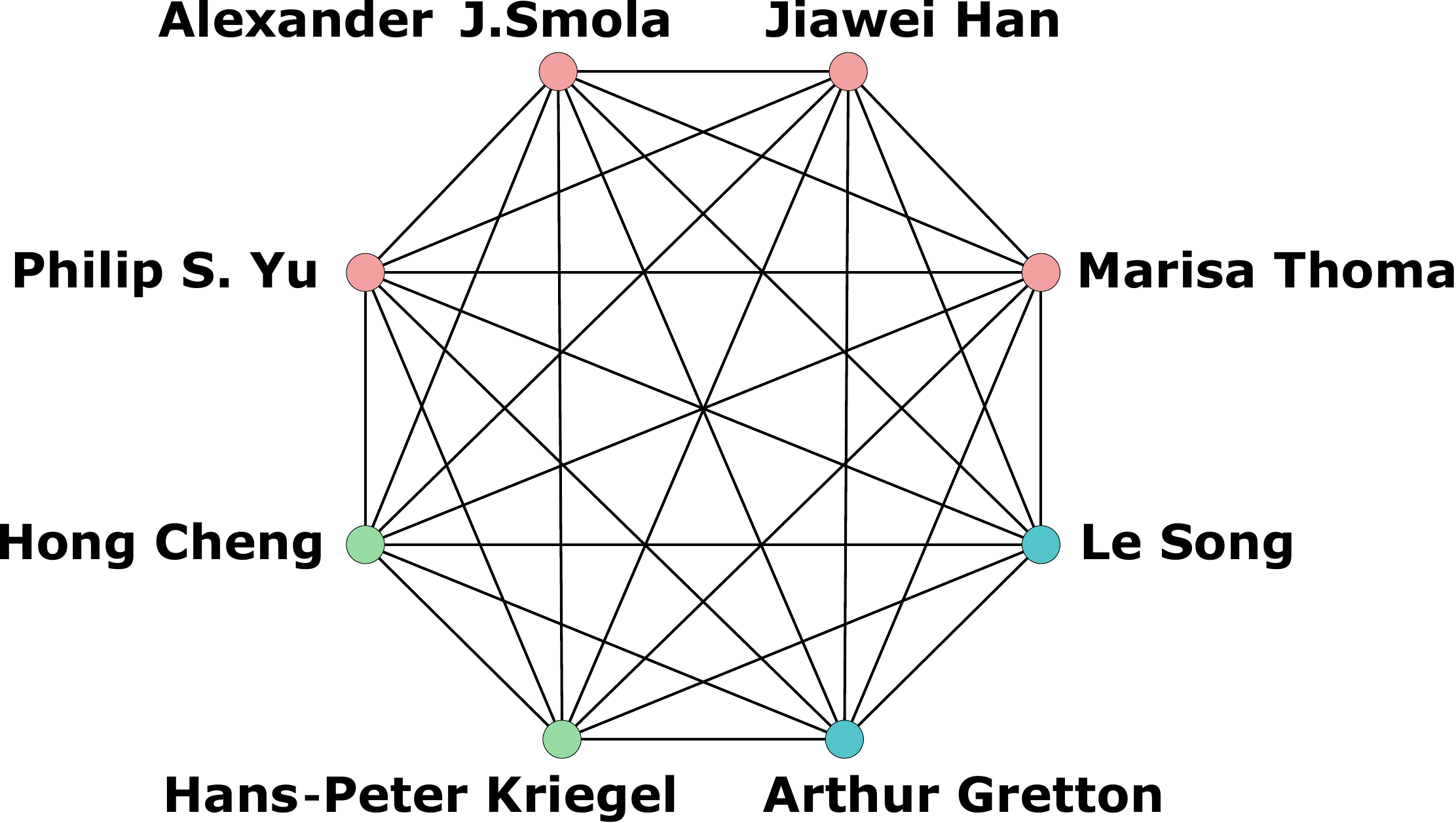}
				}
				\label{fig:exp-cs-2-relative1}
			}
			\subfigure[{\scriptsize a relative fair clique ($\delta=2$)}]{
				\raisebox{0.08\height}{
					\includegraphics[width=0.45\columnwidth,]{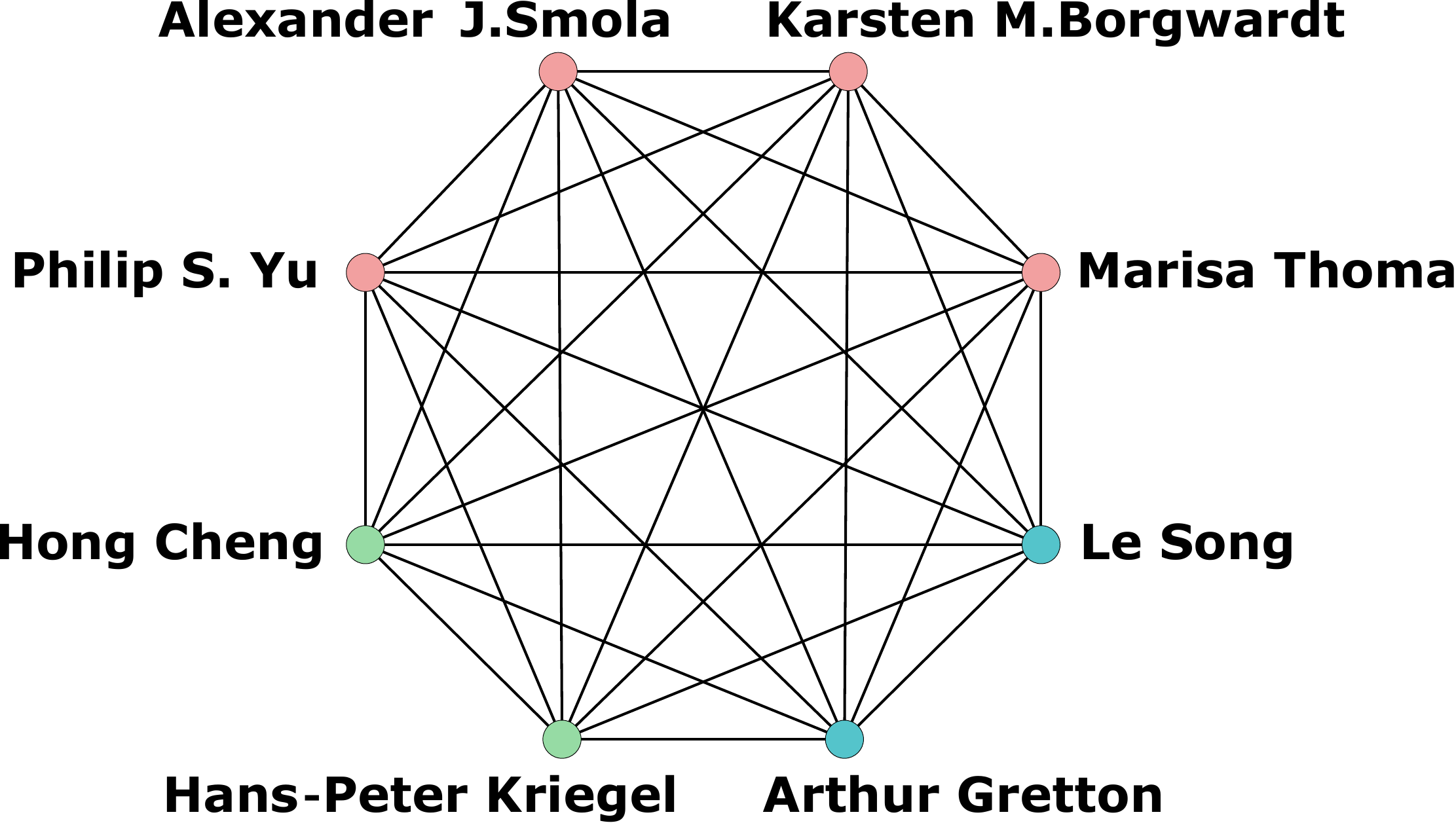}
				}
				\label{fig:exp-cs-2-relative2}
			}\\
			
			\vspace*{0.2cm}			
			\subfigure[{\scriptsize a relative fair clique ($\delta=3$)}]{
				\raisebox{0.1\height}{
					\includegraphics[width=0.45\columnwidth,]{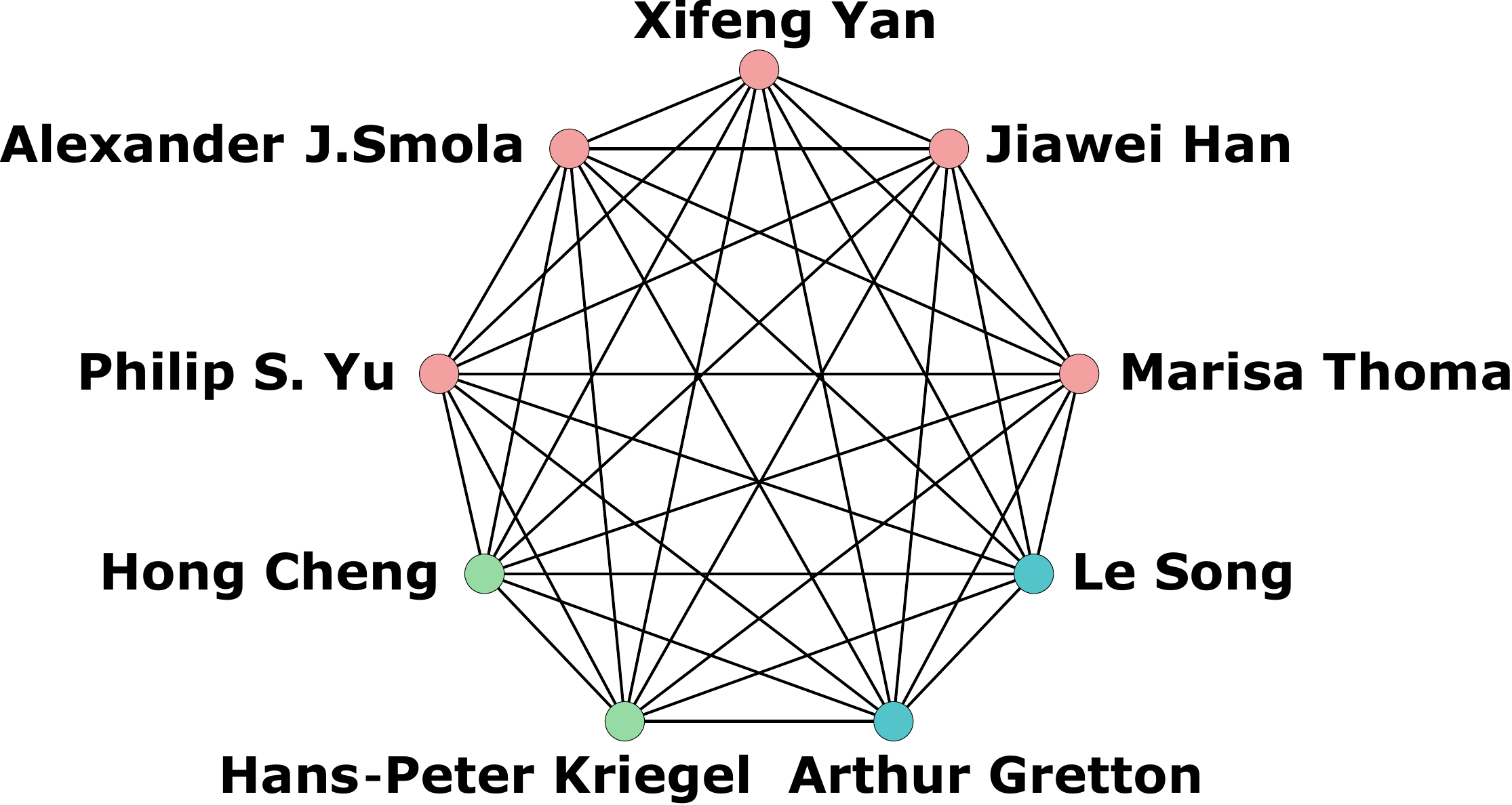}
				}
				\label{fig:exp-cs-3-relative1}
			}
			\subfigure[{\scriptsize a relative fair clique ($\delta=3$)}]{
				\raisebox{0.1\height}{
					\includegraphics[width=0.45\columnwidth,]{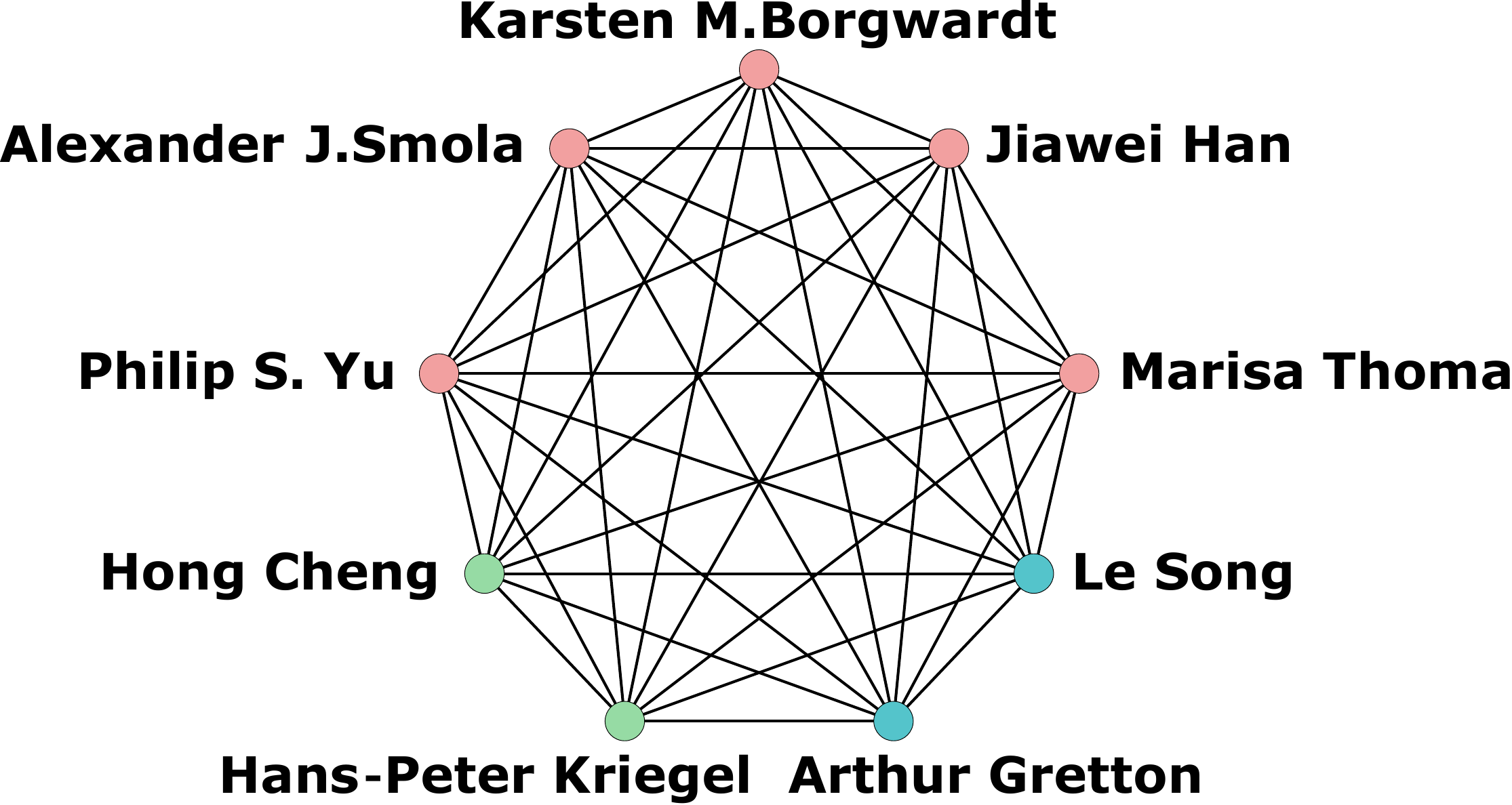}
				}
				\label{fig:exp-cs-3-relative2}
			}
		\end{tabular}
	\end{center}
	\vspace*{-0.3cm}
	\caption{Results of \relativeweak/\relativestrong on \dbcase with $A_{val} = \{DB, DM, AI\}$}
	\label{fig:exp:csattr_relative}
	\vspace*{-0.4cm}
\end{figure}

We also depict four relative fair cliques in \figref{fig:exp:csattr_relative}, which are related to the weak fair clique and strong fair cliques in \figref{fig:exp:csattr3}. \figref{fig:exp:csattr_relative} (a)-(b) and \figref{fig:exp:csattr_relative} (c)-(d) are the cliques for $\delta=2$ and $\delta=3$, respectively. As can be seen from \figref{fig:exp:csattr_relative} (a)-(b), the number of vertices with different attribute values is no less than $k = 2$ and the maximum difference in the number of vertices of those attributes is $2 \le \delta=2$. Moreover, these two relative cliques are also subgraphs of the clique in \figref{fig:exp-cs-3-weak} and they both contains the strong fair cliques shown in \figref{fig:exp-cs-3-strong1} and \figref{fig:exp-cs-3-strong2}. Similar results can also be found in \figref{fig:exp:csattr_relative} (c)-(d). By comparing the cliques with $\delta=2$ and $\delta=3$, we can find that the difference threshold $\delta$ does measure the balance between the attributes in a relative fair clique. A larger $\delta$ leads to finding a clique in which the number of nodes of each attribute varies greatly, and thus the result is closer to a weak fair clique. While for a smaller $\delta$, the enumerated relative fair cliques are closer to the model of strong fair clique. This finding reveals that our relative fair clique model is a good compromise between the weak fair clique and the strong fair clique models as described in \secref{sec:preliminaries}. 

All the results demonstrate that the \weak, \strong and \relativestrong/\relativeweak algorithms can be used to find fair communities with diverse attributes; \strong can further keep a balance over different attributes in the community; and \relativestrong and \relativeweak provide a more flexible way to find fair communities as a compromise by specifying the difference threshold $\delta$. In addition, this case study also indicates that the fairness-aware cliques show the scholars of different research areas who cooperate with each other, and further reflect the closeness of different research areas. That is, the closer these areas are, the larger fair cliques will be. If no fair clique can be found, then it means that at least one research area has no obvious connection to others. The fairness-aware clique models aim to find balance among different attributes, which are suitable to be used in cross-cutting areas.

\subsection{Discussions} \label{sec:select}
As shown in our experiments, seeking a suitable $k$ for our fair clique model is important for practical applications. Here we introduce a heuristic method to find an appropriate $k$. Since the sizes of fair cliques are clearly no larger than the maximum clique size of the graph, we can first compute the maximum clique size of a graph by using the state-of-the-art maximum clique search algorithms \cite{17maxclique,19kddmaxclique}. Suppose the size of a maximum clique is $C_{max}$. Then, the parameter $k$ in our fair clique models satisfies $k \le \lfloor \frac{C_{max}}{A_n} \rfloor$. Note that when the maximum clique size is hard to compute for some instances, an alternative solution is to compute an approximation of $C_{max}$ by using a linear-time greedy algorithm \cite{rossi2015parallel}. Therefore, for a particular application, we can use a binary search method to find an appropriate $k$ from the interval $[1, \lfloor \frac{C_{max}}{A_n} \rfloor]$ by invoking the proposed algorithms to compute the fairness-aware cliques.

\section{Related work} \label{sec:relatedwork}

\stitle{Attributed graph mining.} Our work is related to attributed graph mining which has attracted much attention in data mining due to the diverse applications \cite{li2018community, tong2007fast, fang2016effective, khan2020compact, pizzuti2018genetic, wu2014graph}. For example, Li \etal \cite{li2018community} proposed an embedding-based model to discover communities in attributed graphs. Tong \etal \cite{tong2007fast} studied the problem of finding subgraphs for given query patterns in attributed graphs. Fang \etal \cite{fang2016effective} investigated the attributed community search problem and developed an index structure, called CL-tree, to efficiently support attributed community search. Khan \etal \cite{khan2020compact} proposed an algorithm to mine subgraphs such that the vertices in the subgraph are closely connected and each vertex contains as many query keywords as possible. Pizzuti \etal \cite{pizzuti2018genetic} introduced a community mining algorithm for attributed graphs that considers both node similarity and structural connectivity. In this paper, we study a problem of mining fair communities (fair cohesive subgraph) in attributed graphs. To the best of our knowledge, our work is the first to study the fair community search problem in attributed networks.



\stitle{Fairness-aware data mining.} Our work is related to fairness-aware data mining which has been recognized as an important issue in data mining and machine learning. To measure fairness, many concepts have been proposed in the literature \cite{verma2018fairness}. Zehlike \etal \cite{zehlike2017fa} proposed a method to generate a ranking with a guaranteed group fairness, which can ensure the proportion of protected elements in the rank is no less than a given threshold. Serbos \etal \cite{serbos2017fairness} investigated a problem of fairness in package-to-group recommendation, and proposed a greedy algorithm to find approximate solutions. Beutel \etal \cite{beutel2019fairness} also studied fairness in recommendation systems and presented a set of metrics to evaluate algorithmic fairness. Another line of research on fairness was studied in classification algorithms. Some notable work includes demographic parity \cite{dwork2012fairness} and equality of opportunity \cite{hardt2016equality}. For instance, Hardt \etal \cite{hardt2016equality} proposed a framework that can optimally adjust any learned predictor to reduce bias. Compared to the existing studies, our definition of fairness which requires the equality of different attribute values in a group is different from those in the machine learning literature.


\stitle{Cohesive subgraph mining.} Our work is also related to cohesive subgraph mining. Clique is an important cohesive subgraph model and there are numerous studies that focus on clique mining. Finding maximum cliques, aiming to discover the cliques with the largest size, has attracted much attention. The algorithms for maximum clique search are mainly based on the branch-and-bound framework \cite{ostergaard2002fast}, \cite{konc2007improved}. Ostergard \etal \cite{ostergaard2002fast} presented a branch-and-bound algorithm with the vertex order taken from a coloring of the vertices. Konc \etal \cite{konc2007improved} proposed an approximate coloring algorithm and used it to provide bounds of the size of the maximum clique. Tomita et al.\ proposed a series of maximum clique algorithms, called MCQ \cite{03maxclique}, MCR \cite{09maxclique}, MCS \cite{10maxclique} and MCT \cite{16fawmaxclique,17maxclique}, based on the coloring technique. All these algorithms either use the coloring technique to obtain an upper bound of the maximum clique or apply the coloring heuristics to design a branching strategy. Moreover, all these algorithms are mainly tailored to non-attributed graphs. Different from these works, we use the coloring technique to develop a $k$-core based graph reduction approach; and our work aims to find fairness-aware cliques in attributed graphs.

Another research problem of clique mining is to enumerate maximal cliques. The well-known algorithm for enumerating all maximal cliques is the classic Bron-Kerbosch (BK) algorithm \cite{bron1973algorithm}. Tomita \etal \cite{tomita2006worst} proposed an algorithm, using a greedy pivoting technique, to find all maximal cliques. Eppsten \etal \cite{eppstein2011listing} further improved the BK algorithm based on a heuristic degeneracy ordering. In addition, some relaxed definitions of clique were also proposed, such as $n$-clique \cite{alba1973graph}, $n$-clan, $n$-club \cite{mokken1979cliques}, $k$-plex \cite{seidman1978graph, balasundaram2011clique}, quasi-clique \cite{pardalos1999maximum, abello2002massive}, $k$-core \cite{dorogovtsev2006k, khaouid2015k, montresor2012distributed}, and so on \cite{borgatti1990ls}. However, the solutions mentioned above are not tailored for attributed graphs, and thus cannot be directly used to solve our problems. In this work, we develop novel algorithms to compute maximal fair cliques in attributed graphs with several non-trivial pruning techniques.

\section{Conclusion} \label{sec:conclusion}
In this paper, we study a problem of enumerating fairness-aware cliques in attributed graphs. To this end, we propose a weak fair clique model, a strong fair clique model and a relative fair clique model. To enumerate all weak fair cliques, we first present a novel colorful $k$-core based pruning technique to prune unpromising vertices. And then we develop a backtracking algorithm with a carefully-designed ordering technique to enumerate all weak fair cliques in the pruned graph. To enumerate all strong fair cliques, we propose a new fairness $k$-core based pruning algorithm for the 2D case, and then develop a backtracking algorithm with a fairness $k$-core based ordering technique to enumerate all strong fair cliques. We also present a strong fair clique enumeration algorithm with a heuristic ordering for handling high-dimensional cases. To enumerate all relative fair cliques, we present two efficient algorithms based on a weak fair clique refinement strategy and an attribute-alternatively-selection strategy, respectively. We also design an enhanced colorful $k$-core based pruning technique for 2D attributes, which can also be applied to reduce the graph for weak fair clique enumeration. Extensive experiments are conducted using four large real-life graphs, and the results demonstrate the efficiency and effectiveness of the proposed algorithms.

There are several future directions that are deserved further investigation. First, the proposed models are based on the concept of clique which may be strict for some real-life applications. A promising direction is to relax the clique model used in our definitions, and apply other models (e.g., $k$-truss) to define the fairness-aware cohesive subgraphs. Second, the proposed pruning technique is mainly based on the colorful $k$-core. An interesting question is that can we develop a colorful $k$-truss based pruning technique? Since $k$-truss is often much denser than $k$-core, such a pruning technique may be more powerful than our colorful $k$-core based technique. Finally, it is also interesting to develop more efficient branching and ordering techniques to further speed up the backtracking enumeration procedure.

\comment{
\section{Acknowledgments}
This work was partially supported by (i) NSFC Grants 61772346, 61732003, U1809206, 61836005, 61672358; (ii) National Key R\&D Program of China 2018YFB1004402; (iii) Beijing Institute of Technology Research Fund Program for Young Scholars; (iv) ARC Discovery Project Grant DP160101513. Guoren Wang is the corresponding author of this paper.
}

{
\bibliography{fairclique_tkde}

\begin{thebibliography}{10}

\bibitem{li2015influential}
R.-H. Li, L.~Qin, J.~X. Yu, and R.~Mao, ``Influential community search in large
  networks,'' {\em PVLDB}, vol.~8, no.~5, pp.~509--520, 2015.

\bibitem{papadopoulos2012community}
S.~Papadopoulos, Y.~Kompatsiaris, A.~Vakali, and P.~Spyridonos, ``Community
  detection in social media,'' {\em Data Mining and Knowledge Discovery},
  vol.~24, no.~3, pp.~515--554, 2012.

\bibitem{huang2014querying}
X.~Huang, H.~Cheng, L.~Qin, W.~Tian, and J.~X. Yu, ``Querying k-truss community
  in large and dynamic graphs,'' in {\em SIGMOD}, 2014.

\bibitem{friedrich2015cliques}
T.~Friedrich and A.~Krohmer, ``Cliques in hyperbolic random graphs,'' in {\em
  INFOCOM}, 2015.

\bibitem{li2019improved}
R.-H. Li, Q.~Dai, G.~Wang, Z.~Ming, L.~Qin, and J.~X. Yu, ``Improved algorithms
  for maximal clique search in uncertain networks,'' in {\em ICDE}, 2019.

\bibitem{18tkdekpc}
L.~Yuan, L.~Qin, W.~Zhang, L.~Chang, and J.~Yang, ``Index-based densest clique
  percolation community search in networks,'' {\em {IEEE} Trans. Knowl. Data
  Eng.}, vol.~30, no.~5, pp.~922--935, 2018.

\bibitem{yu2006predicting}
H.~Yu, A.~Paccanaro, V.~Trifonov, and M.~Gerstein, ``Predicting interactions in
  protein networks by completing defective cliques,'' {\em Bioinformatics},
  vol.~22, no.~7, pp.~823--829, 2006.

\bibitem{boginski2006mining}
V.~Boginski, S.~Butenko, and P.~M. Pardalos, ``Mining market data: a network
  approach,'' {\em Computers \& Operations Research}, vol.~33, no.~11,
  pp.~3171--3184, 2006.

\bibitem{li2018community}
Y.~Li, C.~Sha, X.~Huang, and Y.~Zhang, ``Community detection in attributed
  graphs: An embedding approach,'' in {\em AAAI}, 2018.

\bibitem{tong2007fast}
H.~Tong, C.~Faloutsos, B.~Gallagher, and T.~Eliassi-Rad, ``Fast best-effort
  pattern matching in large attributed graphs,'' in {\em SIGKDD}, 2007.

\bibitem{fang2016effective}
Y.~Fang, R.~Cheng, S.~Luo, and J.~Hu, ``Effective community search for large
  attributed graphs,'' {\em VLDB}, vol.~9, no.~12, pp.~1233--1244, 2016.

\bibitem{khan2020compact}
A.~Khan, L.~Golab, M.~Kargar, {\em et~al.}, ``Compact group discovery in
  attributed graphs and social networks,'' {\em Information Processing \&
  Management}, vol.~57, no.~2, p.~102054, 2020.

\bibitem{pizzuti2018genetic}
C.~Pizzuti and A.~Socievole, ``A genetic algorithm for community detection in
  attributed graphs,'' in {\em EvoApplications}, pp.~159--170, 2018.

\bibitem{wu2014graph}
Y.~Wu, Z.~Zhong, W.~Xiong, and N.~Jing, ``Graph summarization for attributed
  graphs,'' in {\em ISEEE}, vol.~1, pp.~503--507, 2014.

\bibitem{yang2013community}
J.~Yang, J.~McAuley, and J.~Leskovec, ``Community detection in networks with
  node attributes,'' in {\em ICDM}, 2013.

\bibitem{xu2012model}
Z.~Xu, Y.~Ke, Y.~Wang, H.~Cheng, and J.~Cheng, ``A model-based approach to
  attributed graph clustering,'' in {\em SIGMOD}, 2012.

\bibitem{verma2018fairness}
S.~Verma and J.~Rubin, ``Fairness definitions explained,'' in {\em FairWare},
  2018.

\bibitem{hardt2016equality}
M.~Hardt, E.~Price, and N.~Srebro, ``Equality of opportunity in supervised
  learning,'' in {\em NIPS}, 2016.

\bibitem{dwork2012fairness}
C.~Dwork, M.~Hardt, T.~Pitassi, O.~Reingold, and R.~Zemel, ``Fairness through
  awareness,'' in {\em ITCS}, 2012.

\bibitem{zehlike2017fa}
M.~Zehlike, F.~Bonchi, C.~Castillo, S.~Hajian, M.~Megahed, and R.~Baeza-Yates,
  ``Fa* ir: A fair top-k ranking algorithm,'' in {\em CIKM}, 2017.

\bibitem{serbos2017fairness}
D.~Serbos, S.~Qi, N.~Mamoulis, E.~Pitoura, and P.~Tsaparas, ``Fairness in
  package-to-group recommendations,'' in {\em WWW}, 2017.

\bibitem{beutel2019fairness}
A.~Beutel, J.~Chen, T.~Doshi, {\em et~al.}, ``Fairness in recommendation
  ranking through pairwise comparisons,'' in {\em SIGKDD}, 2019.

\bibitem{matula1972graph}
D.~W. Matula, G.~Marble, and J.~D. Isaacson, ``Graph coloring algorithms,'' in
  {\em Graph theory and computing}, pp.~109--122, 1972.

\bibitem{jensen2011graph}
T.~R. Jensen and B.~Toft, {\em Graph coloring problems}, vol.~39.
\newblock 2011.

\bibitem{coresdecom2003}
V.~Batagelj and M.~Zaversnik, ``An o(m) algorithm for cores decomposition of
  networks,'' {\em CoRR}, vol.~cs.DS/0310049, 2003.

\bibitem{coresdecomMatulaB83}
D.~W. Matula and L.~L. Beck, ``Smallest-last ordering and clustering and graph
  coloring algorithms,'' {\em J. {ACM}}, vol.~30, no.~3, pp.~417--427, 1983.

\bibitem{mitchem1976various}
J.~Mitchem, ``On various algorithms for estimating the chromatic number of a
  graph,'' {\em The Computer Journal}, vol.~19, no.~2, pp.~182--183, 1976.

\bibitem{14spaacolororder}
W.~Hasenplaugh, T.~Kaler, T.~B. Schardl, and C.~E. Leiserson, ``Ordering
  heuristics for parallel graph coloring,'' in {\em SPAA}, 2014.

\bibitem{bron1973algorithm}
C.~Bron and J.~Kerbosch, ``Algorithm 457: finding all cliques of an undirected
  graph,'' {\em Communications of the ACM}, vol.~16, no.~9, pp.~575--577, 1973.

\bibitem{tomita2006worst}
E.~Tomita, A.~Tanaka, and H.~Takahashi, ``The worst-case time complexity for
  generating all maximal cliques and computational experiments,'' {\em
  Theoretical computer science}, vol.~363, no.~1, pp.~28--42, 2006.

\bibitem{17maxclique}
E.~Tomita, ``Efficient algorithms for finding maximum and maximal cliques and
  their applications,'' in {\em WALCOM} (S.~Poon, M.~S. Rahman, and H.~Yen,
  eds.), 2017.

\bibitem{19kddmaxclique}
L.~Chang, ``Efficient maximum clique computation over large sparse graphs,'' in
  {\em KDD} (A.~Teredesai, V.~Kumar, Y.~Li, R.~Rosales, E.~Terzi, and
  G.~Karypis, eds.), 2019.

\bibitem{rossi2015parallel}
R.~A. Rossi, D.~F. Gleich, and A.~H. Gebremedhin, ``Parallel maximum clique
  algorithms with applications to network analysis,'' {\em SIAM Journal on
  Scientific Computing}, vol.~37, no.~5, pp.~C589--C616, 2015.

\bibitem{ostergaard2002fast}
P.~R. {\"O}sterg{\aa}rd, ``A fast algorithm for the maximum clique problem,''
  {\em Discrete Applied Mathematics}, vol.~120, no.~1-3, pp.~197--207, 2002.

\bibitem{konc2007improved}
J.~Konc and D.~Janezic, ``An improved branch and bound algorithm for the
  maximum clique problem,'' {\em proteins}, vol.~4, no.~5, 2007.

\bibitem{03maxclique}
E.~Tomita and T.~Seki, ``An efficient branch-and-bound algorithm for finding a
  maximum clique,'' in {\em DMTCS} (C.~Calude, M.~J. Dinneen, and
  V.~Vajnovszki, eds.), 2003.

\bibitem{09maxclique}
E.~Tomita and T.~Kameda, ``An efficient branch-and-bound algorithm for finding
  a maximum clique with computational experiments,'' {\em J. Glob. Optim.},
  vol.~44, no.~2, p.~311, 2009.

\bibitem{10maxclique}
E.~Tomita, Y.~Sutani, T.~Higashi, S.~Takahashi, and M.~Wakatsuki, ``A simple
  and faster branch-and-bound algorithm for finding a maximum clique,'' in {\em
  WALCOM} (M.~S. Rahman and S.~Fujita, eds.), 2010.

\bibitem{16fawmaxclique}
E.~Tomita, K.~Yoshida, T.~Hatta, A.~Nagao, H.~Ito, and M.~Wakatsuki, ``A much
  faster branch-and-bound algorithm for finding a maximum clique,'' in {\em
  FAW} (D.~Zhu and S.~Bereg, eds.), 2016.

\bibitem{eppstein2011listing}
D.~Eppstein and D.~Strash, ``Listing all maximal cliques in large sparse
  real-world graphs,'' in {\em SEA}, pp.~364--375, 2011.

\bibitem{alba1973graph}
R.~D. Alba, ``A graph-theoretic definition of a sociometric clique,'' {\em
  Journal of Mathematical Sociology}, vol.~3, no.~1, pp.~113--126, 1973.

\bibitem{mokken1979cliques}
R.~J. Mokken {\em et~al.}, ``Cliques, clubs and clans,'' {\em Quality \&
  Quantity}, vol.~13, no.~2, pp.~161--173, 1979.

\bibitem{seidman1978graph}
S.~B. Seidman and B.~L. Foster, ``A graph-theoretic generalization of the
  clique concept,'' {\em Journal of Mathematical sociology}, vol.~6, no.~1,
  pp.~139--154, 1978.

\bibitem{balasundaram2011clique}
B.~Balasundaram, S.~Butenko, and I.~V. Hicks, ``Clique relaxations in social
  network analysis: The maximum k-plex problem,'' {\em Operations Research},
  vol.~59, no.~1, pp.~133--142, 2011.

\bibitem{pardalos1999maximum}
J.~Pardalos and M.~Resende, ``On maximum clique problems in very large
  graphs,'' {\em DIMACS series}, vol.~50, pp.~119--130, 1999.

\bibitem{abello2002massive}
J.~Abello, M.~G. Resende, and S.~Sudarsky, ``Massive quasi-clique detection,''
  in {\em Latin American symposium on theoretical informatics}, pp.~598--612,
  2002.

\bibitem{dorogovtsev2006k}
S.~N. Dorogovtsev, A.~V. Goltsev, and J.~F.~F. Mendes, ``K-core organization of
  complex networks,'' {\em Physical review letters}, vol.~96, no.~4, p.~040601,
  2006.

\bibitem{khaouid2015k}
W.~Khaouid, M.~Barsky, V.~Srinivasan, and A.~Thomo, ``K-core decomposition of
  large networks on a single pc,'' {\em VLDB}, vol.~9, no.~1, pp.~13--23, 2015.

\bibitem{montresor2012distributed}
A.~Montresor, F.~De~Pellegrini, and D.~Miorandi, ``Distributed k-core
  decomposition,'' {\em TPDS}, vol.~24, no.~2, pp.~288--300, 2012.

\bibitem{borgatti1990ls}
S.~P. Borgatti, M.~G. Everett, and P.~R. Shirey, ``Ls sets, lambda sets and
  other cohesive subsets,'' {\em Social networks}, vol.~12, no.~4,
  pp.~337--357, 1990.

\end{thebibliography}
}

%

\begin{IEEEbiography}[{\includegraphics[width=1in,height=1.25in,clip,keepaspectratio]{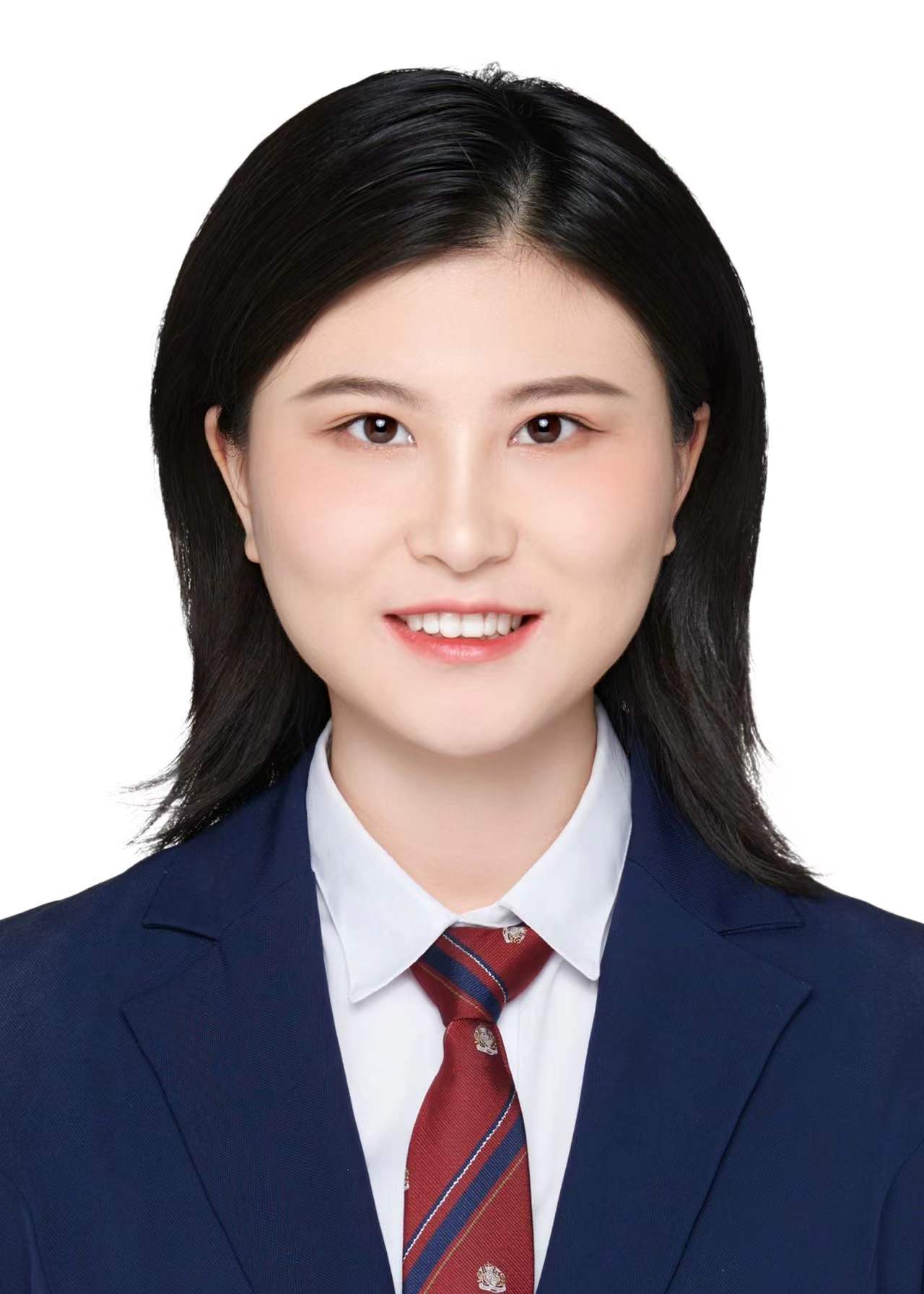}}]{Qi Zhang} is currently a Ph.D. Candidate in Beijing Institute of Technology, China. Her current research interests include social network analysis and data-driven graph mining.
\end{IEEEbiography}

\begin{IEEEbiography}[{\includegraphics[width=1in,height=1.25in,clip,keepaspectratio]{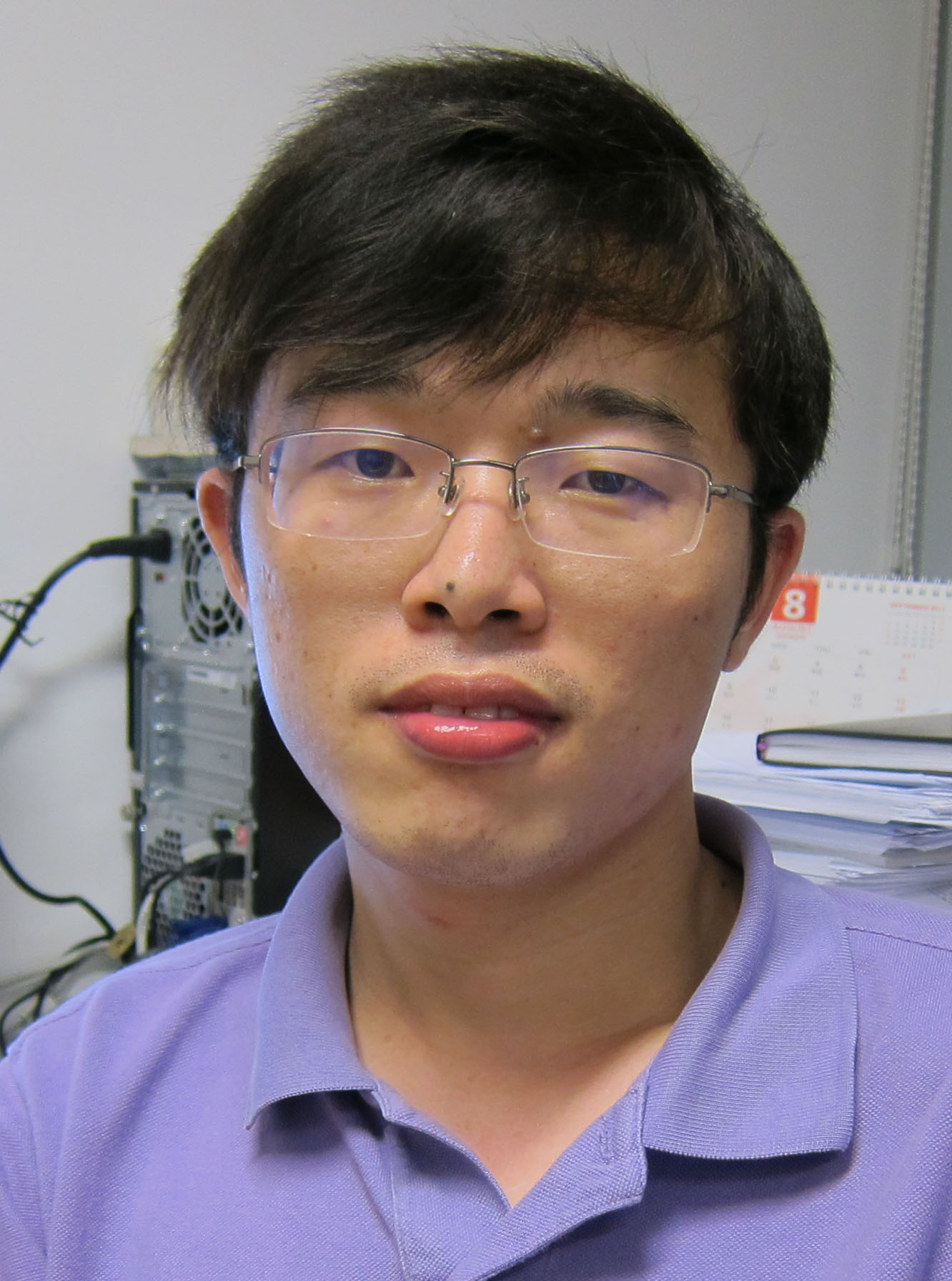}}] {Rong-Hua Li} received the Ph.D. degree from the Chinese University of Hong Kong in 2013. He is currently a Professor at Beijing Institute of Technology, Beijing, China. His research interests include graph data management and mining, social network analysis, graph computation systems, and graph-based machine learning.
\end{IEEEbiography}

\begin{IEEEbiography}[{\includegraphics[width=1in,height=1.30in,clip,keepaspectratio]{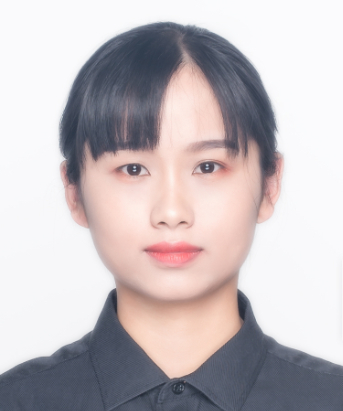}}]{Minjia Pan} is currently an under graduate student at Beijing Institute of Technology, China. She received the B.S. degree in computer science from Northeastern University, China in 2019. Her current research interests include social network analysis and graph mining.\end{IEEEbiography}

\begin{IEEEbiography}[{\includegraphics[width=1in,height=1.25in,clip,keepaspectratio]{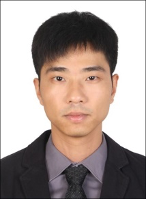}}]{Yongheng Dai} received his Ph.D. degree from The Chinese University of Hong Kong in 2011. Now he works as an R\&D engineer in the areas of domain modeling, knowledge formalization, and knowledge-driven machine learning in the China Academy of Electronics and Information Technology (CAEIT). From 2011 to 2013, he worked on optical fiber communication and digital signal processing in Huawei. After that, he worked on OFDM-based visible light communication and OCC-based indoor positioning until 2017 in CAEIT. Dr. Dai has published 33 papers in international journals and conferences, and holds 10 pending patents. He is also the recipient of China postdoctoral science foundation grant 2015, Beijing Science and Technology Award 2008, and IEEE Photonics Society (HK chapter) Best Paper Award 2008.
\end{IEEEbiography}

\begin{IEEEbiography}[{\includegraphics[width=1in,height=1.25in,clip,keepaspectratio]{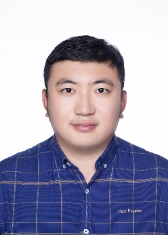}}]{Qun Tian} received his Master's Degree from Harbin Institute of Technology, is a senior engineer in artificial intelligence in Diankeyun Technologies Ltd, Beijing, China. His interests include complex network, graph neural network and knowledge inference.
\end{IEEEbiography}

\begin{IEEEbiography}[{\includegraphics[width=1in,height=1.25in,clip,keepaspectratio]{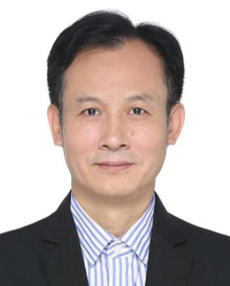}}]{Guoren Wang} received the BSc, MSc, and PhD degrees from the Department of Computer Science, Northeastern University, China, in 1988, 1991 and 1996, respectively. Currently, he is a Professor in the Department of Computer Science, Beijing Institute of Technology, Beijing, China. His research interests include XML data management, query processing and optimization, bioinformatics, high dimensional indexing, parallel database systems, and cloud data management. He has published more than 100 research papers.
\end{IEEEbiography}

\end{document}